\documentclass[aps,showpacs,preprintnumbers,twocolumn,superscriptaddress,floatfix,nofootinbib]{revtex4-2}

\usepackage[linesnumbered,lined,boxed,commentsnumbered,ruled,vlined]{algorithm2e}
\usepackage{algpseudocode}
\usepackage{amsfonts}
\usepackage{amsmath, amssymb, amsthm}
\usepackage{graphicx}
\usepackage{dcolumn}
\usepackage{bm,bbm, bbold}
\usepackage{xkcdcolors}
\usepackage{tabularx}
\usepackage{epstopdf}
\usepackage{xcolor}
\usepackage{tabularray} % for colors in tables
\usepackage{mathrsfs}
\usepackage{subcaption}
\usepackage{caption}
\usepackage{mathtools}
\usepackage{float}
\usepackage{verbatim}
\usepackage{comment}
\usepackage{ragged2e}
\usepackage{physics}
\usepackage{hyperref}
\usepackage{url}
\usepackage{cleveref}
\usepackage{tikz}
\usepackage{utfsym}
\usepackage{multirow}
\usepackage{scalerel}
\usepackage{amsthm}
\usepackage{booktabs}
\usepackage{placeins}
\usepackage[utf8]{inputenc}
\usepackage[T1]{fontenc}
\usepackage{enumitem}
\usepackage{caption}
\usepackage{subcaption}
\usepackage{natbib}

\newcommand{\bz}{\mathbf{z}}

\DeclareCaptionJustification{justified}{\justifying}
\hypersetup{colorlinks=true, citecolor=orange, urlcolor=blue, linkcolor=magenta}

\begin{document}

\title{Community detection in subject-subject networks from psychometrics data}

\author{Arianna Armanetti}
\affiliation{NETWORKS research unit, IMT School for Advanced Studies Lucca, P.zza San Francesco 19, 55100 Lucca (Italy)}

\author{Luca Cecchetti}
\affiliation{MOMILAB research group, IMT School for Advanced Studies Lucca, P.zza San Francesco 19, 55100 Lucca (Italy)}

\author{Eiko Fried}
\affiliation{Department of Clinical Psychology, Leiden University, Wassenaarseweg 52, 2333 AK Leiden, The Netherlands.}
\affiliation{Department of Methodology \& Statistics
Leiden University, Wassenaarseweg 52, Leiden 2333 AK, The Netherlands}

\author{Diego Garlaschelli}
\affiliation{NETWORKS research unit, IMT School for Advanced Studies Lucca, P.zza San Francesco 19, 55100 Lucca (Italy)}
\affiliation{INdAM-GNAMPA Istituto Nazionale di Alta Matematica `Francesco Severi', P.le Aldo Moro 5, 00185 Rome (Italy)}
\affiliation{Lorentz Institute for Theoretical Physics, University of Leiden, Einsteinweg 55, Leiden, 2333 NL, The Netherlands}

\author{Miguel Ibáñez-Berganza}
\affiliation{NETWORKS research unit, IMT School for Advanced Studies Lucca, P.zza San Francesco 19, 55100 Lucca (Italy)}
\affiliation{INdAM-GNAMPA Istituto Nazionale di Alta Matematica `Francesco Severi', P.le Aldo Moro 5, 00185 Rome (Italy)}
\date{\today}

\begin{abstract}
\noindent Identifying subgroups of respondents in psychometric data is traditionally addressed with Latent Class Analysis, which requires the number of classes to be specified \emph{a priori} and can perform poorly when strong inter-item correlations violate local independence assumptions.
We propose a network-theoretic alternative based on community detection in subject-subject similarity networks. To suppress the systematic artifacts induced by the factor structure of the items, the similarity is computed in a low-dimensional factor-score space and the null model for modularity maximisation is obtained by removing the leading (global) mode of the similarity matrix, rather than via the standard Newman--Girvan model.
The significance of a detected partition is then assessed against a column-wise resampling null through four complementary observables: the modularity, the differential entropy of the eigenvector point cloud at two neighbourhood scales, and the overlap of the within- and between-community similarity histograms. 
On a synthetic benchmark with controlled mixture signal, all four metrics correctly identify the homogeneous case as null-compatible---including the demanding regime of a dataset dominated by a single factor---and exhibit a graded departure from the null as the cluster separation grows. Applied to 14 widely used psychometric scales, the pipeline isolates a small group of datasets supporting a genuine and directly interpretable modular structure, while the remaining scales fall either in a mixed-signal regime or in one compatible with a single homogeneous community. The significance analysis is independent of the specific community-detection algorithm and provides an operational way to test for modular subject-level structure in questionnaire data.
\end{abstract}

\maketitle
%\tableofcontents
%\newpage

%%%%%%%%%%%%%%%%%%%%%%%%%%%%%%%%%%%%%%%%%%%%%%%%%%
\section*{Introduction}
%%%%%%%%%%%%%%%%%%%%%%%%%%%%%%%%%%%%%%%%%%%%%%%%%%
%\red{[I added some references, not sure if I found the \textit{must-cite}]} \\

\noindent Psychometrics has traditionally relied on latent variable models---in particular exploratory and confirmatory factor analysis---to describe the structure of questionnaire data \citep{Thurstone1947-gx,Fabrigar1999-ol}.
In this paradigm, observed item responses are treated as noisy linear projections of a small number of continuous latent traits, and individual differences are captured by positions along these dimensions.
Factor analytic models have proven effective for representing the covariance 
structure of items and for identifying the dominant dimensions of variation 
within a questionnaire \citep{Joreskog1969-my, Gorsuch1983-aw, McDonald2013-hd, Fabrigar1999-ol}.

\noindent A network-based paradigm has more recently emerged \citep{Borsboom2017-ke, Epskamp2018-yi, Borsboom2021-gt, Robinaugh2020-hu}, representing questionnaire items as nodes in a graph whose edges encode partial correlations.
This \emph{item network} perspective has proven productive for studying symptom clusters, feedback loops between constructs, and bridge symptoms in psychopathology \citep{Fried2017-cw, Robinaugh2020-hu}.
By shifting attention from latent variables to the relationships between observed indicators, item networks have opened new questions about how constructs are organised and how psychopathological symptoms and other features coexist and interact \citep{Borsboom2017-ke,Epskamp2018-yi,Olthof2023-fi,Scheffer2024-pq}.

\noindent Both paradigms are fundamentally \emph{variable-centred}: their primary object of study is the structure of the \emph{item space}---the covariance structure of items or the graph of their partial correlations---rather than the heterogeneity of the population of respondents \citep{Howard2018-ng}.
A complementary, \emph{person-centred} perspective asks whether the population is homogeneous or whether it decomposes into subgroups of individuals with systematically different item profiles. Identifying such subgroups is of direct clinical and psychometric interest, motivating the search for population-level structure in the respondent space.
The dominant tool in the psychometric literature for this purpose is Latent Class Analysis (LCA) \citep{lazarsfeld1968,goodman1974,Hagenaars2002, Collins2009}, which has several well-documented limitations.
First, the number of classes must be specified \emph{a priori}; selection via information criteria such as BIC and AIC is possible but in practice unreliable, and no consensus on the selection criterion has emerged \cite{Nylund2007}.
Second, LCA performs poorly when classes are unbalanced in size, tending to over-split large groups.
Third, LCA assumes that items are mutually independent given class membership, but this assumption is violated by design in the majority of psychometric datasets, which are constructed by selecting items with strong inter-item correlations \cite{Weller2020-nk}. A more detailed discussion can be found in Appendix \ref{app:lca}.

\noindent A network approach can also be used within this \emph{person-centred} perspective. A dataset can be viewed as a weighted bipartite adjacency matrix \cite{Asratian1998-vf}, from which two monopartite projections can be built: an \emph{item network} and a \emph{subject network}.
While the \emph{item network} has begun to attract attention, as noted above, the \emph{subject network} remains a largely unexplored projection.

\noindent In this work, we study population heterogeneity by leveraging the network representation of subjects to search for community structure.
In this representation, nodes correspond to respondents and edge weights encode pairwise similarity in response patterns; community structure in the subject network therefore corresponds directly to subgroups of individuals who share a distinctive response profile.

%- - - - - - - - - - - - - - - - - - - - - - - -
%\subsubsection{The problem of spurious community detection in the subject network}

\noindent The central methodological challenge is to distinguish \emph{genuine} subgroup structure from \emph{spurious} modular patterns that arise as artifacts of the data-generating process even when the population is homogeneous.
First, strong inter-item correlations induced by the latent factor structure, combined with the ordinal discretisation of response scales, can produce block-diagonal patterns in the similarity matrix that mislead standard community detection algorithms \cite{fortunato2010-os}, identifying clusters even when all respondents belong to the same population.
Second, a particularly difficult case arises when the item set has a dominant axis of variation: if not handled correctly, subjects are artificially split into two groups along this axis, regardless of any genuine subgroup structure, as illustrated in Fig.~\ref{fig:demo-comm}. %
%
%------------------------------------------------------ 
\begin{center}
    \textit{Outline of contributions }
\end{center}
\noindent This work aims to control, as far as possible, for the biases and artifacts described above, and proceeds in two complementary directions.
First, to suppress spurious communities in the detected partition, we design a pipeline that explicitly accounts for the confounding effects of factor structure: we propose a pairwise similarity metric based on factor scores rather than raw responses, combined with a spectral null model that removes the dominant eigenmode of the similarity matrix before modularity optimisation.
Second, to catch artifacts that may survive the first step, we introduce a set of complementary statistical tests that assess \textit{post-hoc} whether a detected partition reflects genuine population heterogeneity.
The \emph{post-hoc} validation framework is not tied to the specific pipeline choices made here and can in principle be applied to any community detection approach on psychometric data.\\
\noindent The pipeline is first validated on synthetic datasets designed to cover both homogeneous and clustered regimes. Once the behaviour of the pipeline and the sensitivity of the proposed metrics are established on known ground truth, we apply them to 14 real psychometric datasets and discuss the results.

%%%%%%%%%%%%%%%%%%%%%%%%%%%%%%%%%%%%%%%%%%%%%%%%%%
\section{Methods}
%%%%%%%%%%%%%%%%%%%%%%%%%%%%%%%%%%%%%%%%%%%%%%%%%%
\noindent We treat the questionnaire dataset as a weighted bipartite network between subjects and items, and project onto the subject layer to obtain a weighted subject-subject similarity network. Community detection algorithms are then applied to this network. Each design choice in the pipeline is motivated by the need to distinguish genuine subgroup structure from artifacts induced by the presence of inter-item covariances and ordinal discretisation.

\noindent Bipartite network projections have been used to identify opinion-based groups from survey data~\citep{MacCarron2020-dv, Dinkelberg2021-ug}, where respondent communities are found by thresholding co-agreement on individual items. The approach presented here differs in two key respects: similarity is computed in a low-dimensional factor-score space rather than on raw responses, and the null model for modularity is spectral rather than threshold-based, with the explicit goal of separating genuine population heterogeneity from artifacts induced by the presence of inter-item covariances.\\

\noindent The proposed community detection pipeline consists of three steps: 
\begin{itemize}
    \item[(i)]~\emph{factor-score projection} (Section~\ref{similarity}), in which a factor analysis model is fitted on the dataset and subjects are embedded in an $F$-dimensional score space;
    \item[(ii)]~\emph{similarity matrix construction}, based on the pairwise squared Euclidean distance in factor-score space (Section~\ref{similarity});
    \item[(iii)]~\emph{market mode removal and community detection} (Section~\ref{sec:market-mode}), the leading eigenvalue contribution is subtracted from the similarity matrix before modularity optimisation that is done using the Leiden algorithm.
\end{itemize}

Once a partition is found on the data, we propose a \emph{procedure to statistically validate the significance of such partition} (Section~\ref{sec:significance}).
The strength of the partition found is assessed with four complementary tests against a resampling null model. The observables used are (i) the modularity of the partition, (ii)-(iii) the differential entropy of the cloud points in the eigenvector space, estimated at two different scales of granularity, and (iv) the overlap of the within- and between-community similarity histograms.\\
%
% - - - - - - - - - - - - - - - - - - - - - - - - - - - 
\subsection*{Code availability}
\noindent The codes developed for this paper are available in the Python package \texttt{psycomm} \cite{psycomm}, together with tutorial notebooks on how to reproduce all the results in this paper. 

% - - - - - - - - - - - - - - - - - - - - - - - - - - - 
\subsection{Dataset structure and notation}
\noindent The datasets we study are $N \times M$ integer matrices $\mathbf{X}$, where each entry $x_{ij} \in \{0,\ldots,R-1\}$ is the response of subject $i$ to item $j$ on a Likert scale of $R$ values. Items are grouped into blocks corresponding to latent constructs; questionnaires in this work range from single-factor short scales ($M \sim 10\text{--}15$) to multi-factor inventories ($M \sim 50$, up to $M \sim 100\text{--}200$). See Table~\ref{tab:datasets-tab} for the full list.

\noindent We use both synthetic datasets -- to validate the pipeline against known ground truth -- and real datasets. The generative models for synthetic data are described in Appendix~\ref{app:synthetics}.
For all real datasets, a uniformly random subsample of $N=1000$ subjects is drawn from the respondents with complete responses on all items; missing values are not imputed. This subsampling keeps the pairwise similarity matrix tractable across heterogeneous dataset sizes and standardises the operating point of the cloud-entropy criteria (Section~\ref{sec:significance}).

%------------------------------------------------------
\subsection{A pipeline for community detection in subject networks}

% - - - - - - - - - - - - - - - - - - - - - - - - - - - 
\subsubsection{The similarity network: projection in a low-dimensional space}
\label{similarity}
\noindent A central concern is whether the modularity maximisation produces non-trivial partitions even on data drawn from a single Gaussian with no community structure.
The choice of similarity metric is critical: a poorly chosen metric either degrades the signal or introduces spurious structure. 
The first natural choice for the similarity measure is to compute the squared item-wise distance between pairs of subjects and take its opposite as similarity.
In this approach, the squared Euclidean distance between subjects $i$ and $j$ in item space is:
\begin{equation}
    d_{ij}^{(\mathrm{IS})} = \sum_{m=1}^{M} (x_{im} - x_{jm})^2,
    \label{eq:dist_zscore}
\end{equation}
where $x_{im}$ is the answer of subject $i$ to item $m$. The corresponding similarity is $S_{ij}^{(\mathrm{IS})} = - d_{ij}^{(\mathrm{IS})}$.
Similar approaches are used for example in \cite{Babeanu2021-rw}. \\
\noindent However, in the $M$-dimensional item space, $S^{(\mathrm{IS})}$ conflates genuine between-subject differences with the redundancy induced by items that share a latent factor: pairs of subjects who differ only in their loading on a single factor appear highly dissimilar across all $M$ items simultaneously, inflating the contrast between subjects on opposite sides of the latent axes.
This leads to a strong separation of subjects in two groups even in the case in which the dataset is well known to be generated from a single multivariate Gaussian distribution. 
The redundancy in the datasets is not a flaw, but a deliberate design property \citep{Clark1995-gw}, hence a characteristic that needs to be accounted for in the design of a community detection pipeline for this type of datasets. \\
\noindent The solution we propose to overcome the issue of over-accounting for the item redundancy is to \textit{compute the similarity} not in the item space, but \textit{in a low-dimensional space}, where subjects are projected by computing their scores along the axes found by a Factor Analysis (FA) model fitted on the dataset. \\

\noindent In this approach, the responses are first projected into an $F$-dimensional factor-score space, where $F \ll M$. A factor analysis model is fitted to $\mathbf{X}$.
Factor analysis (FA) is a latent-variable model that decomposes the observed item vector into a linear combination of $F$ latent factors plus item-specific noise 
\begin{align}
\label{eq:famodel}
&{\bf x}_i = {\bm\mu} +  \Lambda\,\hat{\bf z}_i+ {\bm\epsilon}_i,\\ & \hat{\bf z}_i \sim {\cal N}({\bf 0}, I_F), \qquad {\bm\epsilon}_s \sim {\cal N}({\bf 0}, \Psi),
\end{align}
where $\Lambda \in \mathbb{R}^{M\times F}$ is the matrix of factor loadings, $\Psi = \mathrm{diag}(\psi_1,\ldots,\psi_M)$ is the diagonal matrix of uniquenesses (item-specific variances not explained by the factors), and $\hat{\bf z}_i$ and ${\bm\epsilon}_i$ are independent. The marginal distribution of ${\bf x}_i$ under this model is ${\cal N}({\bm\mu},\,\Lambda\Lambda^\top + \Psi)$. The parameters $(\Lambda,\Psi)$ are estimated from the training data by maximising the Gaussian log-likelihood, which we implement via the EM algorithm as provided by \texttt{scikit-learn} \cite{scikit-learn} with no rotation applied to the factor scores. 
Given the estimated parameters $(\hat\Lambda, \hat\Psi)$ and an observed response vector ${\bf x}_i$, we compute the factor scores $\hat{\bf z}_i$ as the mean of the posterior distribution of the latent factors:
\begin{equation}
    \hat{\bf z}_i = \hat\Lambda^\top(\hat\Lambda\hat\Lambda^\top + \hat\Psi)^{-1}({\bf x}_i - \hat{{\bm\mu}}).
\end{equation}
The resulting score matrix $\hat{\mathbf{Z}} \in \mathbb{R}^{N \times F}$ is used as starting point to build the similarity matrix.
We use the maximum-likelihood factor-analysis implementation in scikit-learn's \texttt{FactorAnalysis} \cite{scikit-learn} . 
%Pearson correlations are used as input; we have not assessed sensitivity to the use of polychoric correlations, which would in principle be more appropriate for ordinal data.
The similarity becomes:
\begin{equation}
    S_{ij}^{(\mathrm{FS})} = - \|\hat{\bz}_i - \hat{\bz}_j\|_2^2,
    \label{eq:sim_fs}
\end{equation}
where $\hat{\bz}_i \in \mathbb{R}^F$ is the factor score of subject $i$.

\noindent The number of FA components $F$ is in principle arbitrary.
A good rule of thumb that we propose is to select $F \in [N_{\mathrm{constr}}-1,\, N_{\mathrm{constr}}+2]$, where $N_{\mathrm{constr}}$ is the number of latent constructs that the scale aims to investigate, paying attention to setting $F$ at 2--3 if the scale investigates only one construct. Nevertheless, it is worth pointing out that the sensitivity analysis reported in Appendix~\ref{app:F-sensitivity} shows that for small datasets with 10--20 items or datasets with a small number of items per construct, the similarity matrix tends to be more noisy, leading to less reliable results on the modular structure.

\noindent The factor-score projection removes inter-item covariance by collapsing correlated items onto shared latent axes, avoiding the inflation of similarity due to repeated measurement of the same construct. 
This representation in some cases restores the discriminative power of detection methods and avoids the effect produced by IS-similarity where homogeneous datasets tend to be split in two communities, simply separating the dataset into the positive and negative halves along the main axis of variation. 
See Appendix~\ref{app:item-space-art} for a detailed description of these results.

\begin{figure*}[t]
    \centering
    \includegraphics[width=0.6\linewidth]{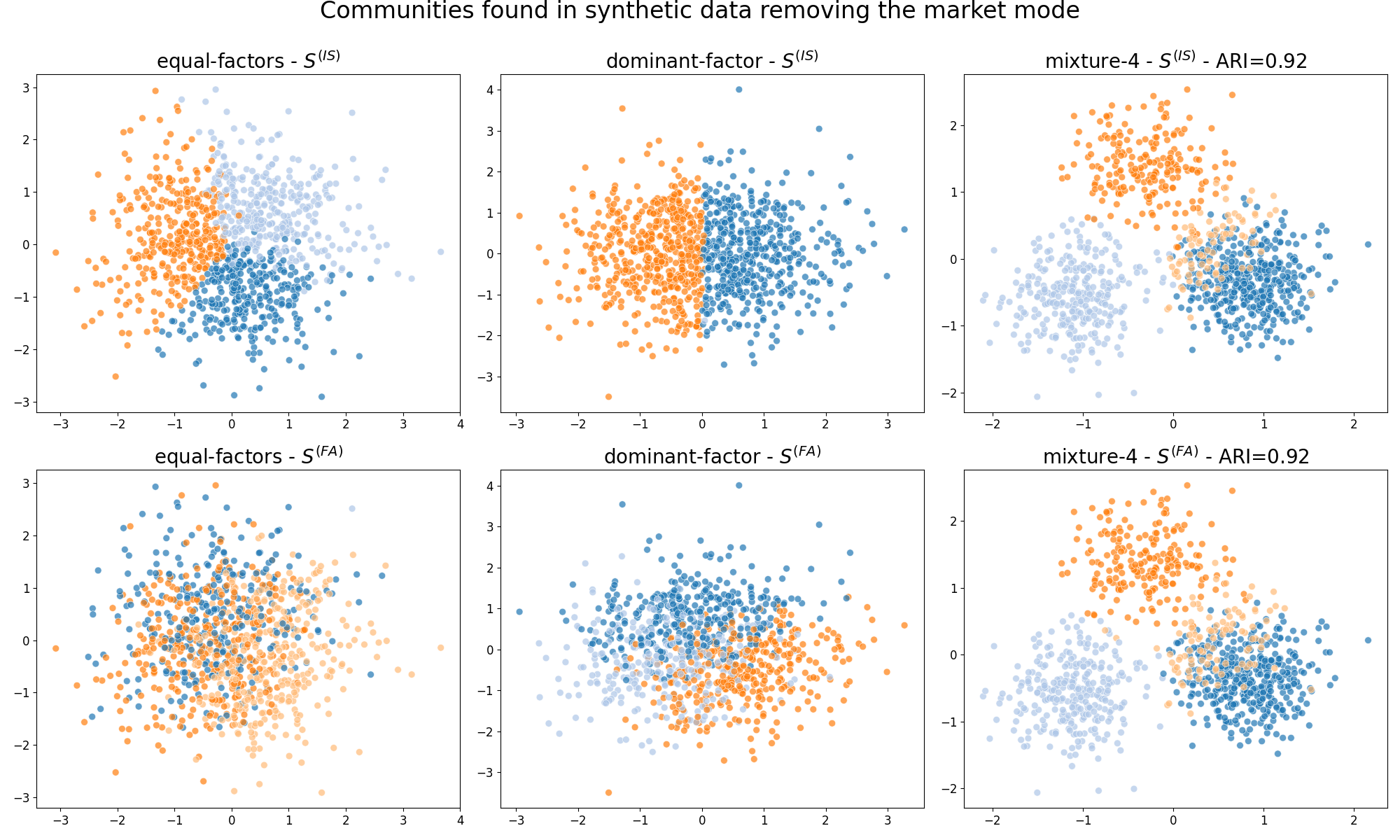}
    \caption{Two-dimensional PCA projections of subject similarity matrices for three synthetic datasets, coloured by the communities detected by the community detection algorithm. The \textit{equal-factors} and \textit{dominant-factor} datasets are drawn from a single homogeneous population (no planted community structure); \textit{mixture-4} contains four equally-sized subgroups with distinct response profiles. 
    \emph{Top row}: similarity computed directly in the $M$-dimensional item-response space.   
    \textit{Bottom row}: similarity computed in the $F$-dimensional factor-score space. 
    Community detection is done in both cases maximising the modularity of the similarity matrix after removal of its leading eigenvector of the similarity matrix (market-mode removal), which captures the global response tendency shared across all subjects (See Sec. \ref{sec:market-mode}). 
    In the two null datasets, item-space similarity yields a spurious partition whose boundaries align with the principal axis of variation — an artifact of inter-item covariance structure rather than genuine subject heterogeneity. 
    Factor-score similarity suppresses this artifact: the resulting partition is confused and unstable, consistent with the absence of planted structure. For the well-separated mixture, both approaches recover the planted communities, confirming that the FA-space method does not discard genuine signal.}
    \label{fig:demo-comm}
\end{figure*}

% - - - - - - - - - - - - - - - - - - - - - - - - - - -
\subsubsection{Community detection: modularity maximisation with a Random Matrix Theory null model}
\label{sec:market-mode}

\noindent To detect communities in the subject network, we work within the modularity maximisation framework, which casts the problem as an optimisation against a null model for baseline connectivity. This allows different null model choices to be compared without altering the rest of the detection procedure.
As briefly stated in the introduction, the choice of the null model is the second critical aspect of the methodological set-up after the choice of the similarity measure. \\ 

\noindent Given the similarity matrix $S$, we seek the partition $\boldsymbol{\sigma} = \{\sigma_i\}_{i=1}^N$ that maximises the (unnormalised) modularity
\begin{equation}
    Q(\boldsymbol{\sigma}) = \sum_{i,j} B_{ij}\,\delta(\sigma_i,\sigma_j), \qquad B_{ij} = S_{ij} - S_{ij}^{\mathrm{null}},
    \label{eq:modularity}
\end{equation}
where $\sigma_i =1, \, .. \, , \, K$ is the label of subject $i$ marking the community it is assigned to, while $S_{ij}^{\mathrm{null}}$ is the expected similarity under the chosen null model and $B$ plays the role of the modularity matrix. The total number of communities $K$ is a free parameter of the maximisation. 
We drop the conventional $1/S_{\mathrm{tot}}$ prefactor because $S$ has signed entries here, which would make the normalisation either degenerate or sign-flipping. The optimisation is performed with the signed-weights variant of the Leiden algorithm \cite{Traag2019-ax, Traag2011-signed} applied directly to $B$, at default resolution.\\

\noindent The standard choice for $S^{\mathrm{null}}$ would be the Newman--Girvan model \citep{Newman2004-zp, newman2006-dt} for weighted networks:
\begin{equation}
    S_{ij}^{\mathrm{null}} = \frac{s_i\, s_j}{S_{\mathrm{tot}}},
    \label{eq:newman_null}
\end{equation}
where $s_i = \sum_j S_{ij}$ is the strength of node $i$. This preserves the expected strength sequence while assuming otherwise random connectivity.

\noindent However, for fully connected subject--subject similarity networks derived from Likert data, this null model is insufficient.
In Appendix~\ref{app:newman-failure} we show that the dominant axis of variation in $S$ -- driven by mean response level rather than subgroup membership -- is not fully absorbed by the NG null model.
This is particularly problematic for homogeneous datasets that have a strong principal axis of variation: even with the correct FA-space similarity, the NG null model does not prevent the partition from aligning with that axis.\\

\noindent We use \textit{Random Matrix Theory} (RMT) to define a more principled null model. The largest eigenvalue $\lambda_0$ of $S$ is associated with a quasi-uniform eigenvector and encodes the overall level of pairwise similarity across all subjects, meaning that it captures a global response tendency shared across all subjects. It represents the \emph{global mode} of the similarity matrix. In the closely related setting of correlation matrices, the global mode is often called the \emph{market mode} when the correlations are constructed from financial time series data \citep{Laloux1999-di, Leonidas2011, Plerou1999-xt, masuda2025introduction}. To emphasize the analogy, we will adopt the same terminology for $\lambda_0$ here.
Crucially, the market mode reflects a global response tendency and, when not fully removed, it can obscure the modular signal in the optimisation landscape, precisely as its counterpart in correlation matrices \cite{Bun2017-oi, garlaschelli-macmahon15}. \\
Removing this component is also insightful because it leaves a residual similarity matrix that, under the hypothesis of isotropy, carry no preferred direction, so that positive and negative values are equally likely.
We define the cleaned similarity matrix as:
\begin{equation}
    S^{\text{clean}}_{ij} = S_{ij} - \lambda_0\,v_{0,i}\,v_{0,j},
    \label{eq:market_mode_removal}
\end{equation}
where $\mathbf{v}_0$ is the eigenvector associated with $\lambda_0$. The modularity in Eq.~\eqref{eq:modularity} is then maximised using $B_{ij} = S^{\text{clean}}_{ij}$ directly. All subsequent spectral observables (Section~\ref{sec:significance}) are computed on $S^{\text{clean}}$. \\

\noindent A related approach to community detection based on RMT-derived null models has already been introduced for correlation matrices \cite{garlaschelli-macmahon15} and applied to financial \cite{garlaschelli-macmahon15, almog2015mesoscopic, Anagnostou2021-zq, zema2025mesoscopic}, neuronal \cite{buijink2016evidence, almog2019uncovering}, and gene expression \cite{mircea2022phiclust} data.
In that methodology, RMT provides an asymptotic exact result on the eigenvalue distribution of a correlation matrix computed on a set of uncorrelated finite size time series, the Marchenko--Pastur (MP) distribution \cite{MarchenkoPastur1967, Utsugi2004-ps, Potters2005-kx, Bun2017-oi, masuda2025introduction}.
This would lead to a modularity matrix cleaned of all the components whose eigenvalues are compatible with the MP bulk and -- if present -- of the market mode.\\
However, correlation matrices have substantial properties that differ from the similarity matrices we are working with. In particular, for the MP distribution to be a good descriptor of the noise bulk, the variables over which the correlations (or similarities) are computed need to be i.i.d.. This assumption is systematically violated in psychometric data, where items typically present a strong factor structure.
We explored alternative spectral null models accounting for the noise bulk, but found no compromise that suppressed spurious communities without discarding genuine signal (Appendix~\ref{app:spectral-nullmodel}).
We therefore clean the modularity matrix only of the market mode and assess the significance of the partitions \emph{post hoc} through the statistical framework described in Section~\ref{sec:significance}. %
%
%
%
%------------------------------------------------------
\FloatBarrier
\subsection{Assessing partition significance}
\label{sec:significance}
\noindent Detecting genuine community structure requires distinguishing partitions that reflect true subgroup organisation from those that arise spuriously from noise or factor structure alone.
Due to the intrinsic difficulty of the problem, it is not always possible for the community detection pipeline to reliably distinguish between a genuinely modular dataset and one compatible with a single homogeneous population.
The general problem of assessing whether a clustering reflects genuine population heterogeneity rather than chance variation has already been addressed in the psychometrics literature \cite{Tibshirani2001-gap, Mair2014-gop, Liu2008-sigclust, Rosenstrom2017-ruo}; these works typically rely on parametric (Gaussian or uniform) reference distributions.
Our contribution extends this line of work in three directions: (i) we adapt the significance question to the \emph{network} community-detection setting, enriching the set of test statistics with graph-partition observables; (ii) we use a non-parametric column-wise resampling null that preserves the marginal distribution of each factor (in expectation) and breaks the joint structure across factors, making no Gaussianity or i.i.d.\ assumption on the responses; and (iii) we combine four complementary observables---modularity, eigenvector cloud entropy at two neighbourhood scales, and the within-vs-between similarity histogram overlap---rather than relying on a single statistic, providing converging evidence against artefactual partitions that any single test would miss.
These observables probe different aspects of community structure: modularity evaluates the quality of the optimal partition of subjects; cloud entropy is calibrated at two scales — small scales and scales around the characteristic cluster size — capturing whether the distribution of subjects in factor-score space is genuinely concentrated; similarity overlap measures the contrast between within- and between-community weights in factor space, quantifying how much more similar subjects are to members of their own community than to others.
\noindent Our significance analysis proceeds as follows:

\begin{enumerate}
    \item From the FA scores matrix $\hat{\mathbf{Z}} \in \mathbb{R}^{N\times F}$ of the original dataset, generate $T$ randomised versions by independently resampling each factor column (see Section~\ref{sec:nullmodel}); the FA model is \emph{not} refitted on the randomised matrices.
    \item Compute the similarity matrix $S^{\mathrm{clean}}$ for each randomised sample.
    \item Run the community detection algorithm on each of the $T$ similarity matrices.
    \item Compute the observables on the partitions found. 
    For the cloud entropy -- that requires to specify the embedding dimension $d$ -- we keep it fixed to $d = K_{\mathrm{cons}}-1$, where $K_{\mathrm{cons}}$ is the number of communities detected in the majority of Leiden runs on the original data. This ensure that the resulting cloud entropies are formally comparable.
    \item Build the empirical distribution of the observables under the null hypothesis and test whether the values observed in the original dataset are compatible with the null.
\end{enumerate}

% - - - - - - - - - - - - - - - - - - - - - - - - - - -
\subsubsection{Resampling-based null distribution}
\label{sec:nullmodel}
\noindent The null distribution is constructed to make as few assumptions as possible about the generative mechanism of the data.
Since the similarity matrix is built in factor-score space (Section~\ref{similarity}), the randomisation is applied to the factor score matrix $\hat{\mathbf{Z}}$: for each factor $f$, the scores $\{\hat{z}_{if}\}_{i=1}^N$ are independently resampled across subjects with replacement,
\begin{equation}
    \forall i, \> \forall f
    \qquad \hat{z}_{if} \rightarrow \hat{z}_{\xi(i),f} \qquad \xi(i) \sim \mathrm{Cat}[1, \, \dots, \, N],
    \label{eq:permutation}
\end{equation}
where each $\xi(i)$ is drawn i.i.d.\ uniformly. This preserves the marginal distribution of each factor in expectation and destroys the joint structure across factors, that encodes within-subject coherence. We use i.i.d.\ resampling rather than a strict permutation for computational convenience; for $N\gg 1$ the two are statistically equivalent up to corrections of order $1/N$.
The FA model is the one fitted on the original data and is \emph{not} refitted on the randomised samples.\\

\noindent This randomisation procedure is used to generate a set of $T$ null versions of $\hat{\mathbf{Z}}$. On each of them we compute the similarity matrix $S^{\mathrm{clean}}$ and run the community detection pipeline.

\noindent To gain information on the presence or not of a clear underlying community structure we consider a set of complementary observables: (a) the modularity, (b) the point-cloud entropy in the eigenvector space, evaluated at two neighbourhood scales, (c) the overlap between within- and between-community similarity histograms.
These observables are individually introduced in the next paragraphs. \\

% - - - - - - - - - - - - - - - - - - - - - - - - - - -
\paragraph{Consensus partition.}
\noindent A single Leiden run can return slightly different partitions across initialisations. To stabilise the analysis we repeat the optimisation $B$ times with independent random seeds and aggregate the resulting partitions into a co-assignment matrix $\mathbf{C}\in[0,1]^{N\times N}$, where $C_{ij}$ is the fraction of runs in which subjects $i$ and $j$ are co-assigned. The consensus partition is obtained by applying average-linkage hierarchical clustering to the dissimilarity matrix $1-\mathbf{C}$ and cutting the dendrogram at $K$ clusters, where $K = K_{\mathrm{cons}}$ is the modal number of communities across the $B$ runs. Subjects with low co-assignment stability (below a fixed threshold) are flagged as unstable in the figures but retained in all observables. The same $K_{\mathrm{cons}}$ is used when computing the cloud-entropy embedding on both the original and the resampled data, so that the test statistic is evaluated in a fixed-dimensional space.\\

\noindent To quantify whether an observable $\mathcal{O}$ computed on the empirical dataset differs significantly from chance, we compare its distribution under repeated optimisation runs on the original data with the null distribution generated by the $T$ resampled datasets.
The observables on the empirical datasets are also computed multiple times because the Leiden algorithm is run $n_{\mathrm{boot}}$ times with different random seeds on the original $S^{\mathrm{clean}}$ (the subjects indices are not resampled).\\
Let $\{\mathcal{O}^{(b)}\}_{b=1}^{n_{\mathrm{boot}}}$ denote the values of the observable obtained by re-running Leiden, and $\{\mathcal{O}^{(t)}\}_{t=1}^{T}$ its values on the $T$ column-wise resampled datasets.
We adopt as primary significance criterion the \emph{run-integrated $p$-value}
\begin{align}
    p(\mathcal{O}) & \;=\; \frac{1}{n_{\mathrm{boot}}}\sum_{b=1}^{n_{\mathrm{boot}}}\,\frac{1}{T}\sum_{t=1}^{T}\,\mathbb{1}\!\left[\mathcal{O}^{(t)}\;\trianglerighteq_{\mathcal{O}}\;\mathcal{O}^{(b)}\right]
    \label{eq:pvalue}
\end{align}
where the binary relation $\trianglerighteq_{\mathcal{O}}$ stands for $\geq$ when $\mathcal{O}=Q$ (right-tail test: genuine structure produces modularity values larger than the null) and for $\leq$ when $\mathcal{O}\in\{\hat H_{k_{\mathrm{low}}}, \hat H_{k_{\mathrm{high}}}, \mathrm{OV}\}$ (left-tail test: genuine structure produces lower entropy / overlap than the null). Equivalently, $p(\mathcal{O})$ is the run-average of the null-tail probability evaluated at the run-specific empirical value.
By averaging the null-tail probability across the distribution of the empirical statistic instead of evaluating it at a single point estimate, the test becomes more conservative whenever the inter-run spread $\sigma_{\mathrm{obs}}$ of $\mathcal{O}^{\mathrm{obs}}$ is non-negligible, and reduces to a standard permutation $p$-value in the limit $\sigma_{\mathrm{obs}}\to 0$. A standard reference for permutation $p$-values with finite resampling sets is \cite{PhipsonSmyth2010}.
We declare a metric significant whenever $p(\mathcal{O}) < \alpha$ with $\alpha = 0.05$; throughout this work we use $T = n_{\mathrm{rep}} = 150$ resamplings and $n_{\mathrm{boot}} = 100$ Leiden re-runs.

\noindent As a secondary diagnostic we also report the standardised z-score
\begin{equation}
    z(\mathcal{O}) \;=\; \frac{\overline{\mathcal{O}}^{\mathrm{obs}} - \mu_0}{\sqrt{\sigma_0^2 + \sigma_{\mathrm{obs}}^2}}\,,
    \label{eq:zscore}
\end{equation}
where $\overline{\mathcal{O}}^{\mathrm{obs}}$ is the mean of $\{\mathcal{O}^{(b)}\}$ over the $n_{\mathrm{boot}}$ Leiden re-runs, $\mu_0,\sigma_0$ are the mean and standard deviation of the resampling null, and $\sigma_{\mathrm{obs}}$ is the standard deviation of $\{\mathcal{O}^{(b)}\}$. For observables that are deterministic given $S^{\mathrm{clean}}$ and the partition (e.g.\ the cloud entropy at fixed $K_{\mathrm{cons}}$) the inter-run spread $\sigma_{\mathrm{obs}}$ comes only from the variability of the partition itself across Leiden seeds.
A large $|z|$ provides additional evidence against the null hypothesis of no community structure, with the sign carrying the same directional information as the corresponding tail of Eq.~\eqref{eq:pvalue} ($z>0$ for modularity, $z<0$ for cloud entropy and overlap).
The two quantities are reported jointly: $p(\mathcal{O})$ drives the binary significance call, $z(\mathcal{O})$ summarises the size of the effect.

% - - - - - - - - - - - - - - - - - - - - - - - - - - -
\subsubsection{Modularity}
\noindent The first observable is the modularity $Q$ of the detected partition (Eq.~\eqref{eq:modularity}).
Modularity measures how much higher the intra-cluster connectivity is relative to the baseline expected under the null model.
The absolute value of $Q$ is not directly interpretable in isolation, but comparing it with the null distribution from resampled datasets provides a principled significance test.
If the empirical $Q$ systematically exceeds the null distribution, the partition is considered significant, as it captures structure beyond what is expected under marginal independence;
if it falls within the null bulk, the detected structure is indistinguishable from a chance artifact.

% - - - - - - - - - - - - - - - - - - - - - - - - - - -
\subsubsection{Entropy of the eigenvector point cloud}
\noindent Properties of the spectrum of $S^{\mathrm{clean}}$ provide additional evidence for or against modular structure. Under RMT, clustered data with $K$ communities are expected to produce $K-1$ eigenvalues above the Marchenko--Pastur bulk edge $\lambda_+$ \citep{Von_Luxburg2007-vc, Nadakuditi2012-uc, Decelle2011-yd}, as also discussed in Appendix~\ref{app:spectral-nullmodel}. However, this criterion is not sufficient on its own: homogeneous data with strong factor structure can produce a similar spectral profile \citep{Bun2017-oi}, since factor-induced correlations also lift eigenvalues above the bulk. We therefore rely on the \emph{eigenvectors} of $S^{\mathrm{clean}}$ rather than its eigenvalues alone.

\noindent For clustered data, the $K-1$ leading eigenvectors of $S^{\mathrm{clean}}$ encode the community partition: the embedding $\mathbf{y}(s) = (v_{s1}, \ldots, v_{s,K-1}) \in \mathbb{R}^{K-1}$ forms a cloud of points concentrated near $K$ distinct regions.
For homogeneous data the same eigenvectors show no such concentration. The contrast between the two regimes is quantified via the differential entropy of the point cloud.
The entropy of the \emph{joint} distribution of the cloud points is needed: the marginal entropy of each eigenvector, taken independently, is dominated by the smooth one-dimensional envelope of its components and is not sensitive enough to clustered structure in this regime. In Appendix~\ref{app:eigenvector-entropy} we provide direct evidence of this point.

Let $v_{sk}$ denote the $s$-th component of the $k$-th eigenvector of $S^{\mathrm{clean}}$, with $k$ ranging from 1 (the leading eigenvector \emph{after} market-mode removal) to $K_{\mathrm{cons}}-1$. Each subject $s$ is embedded as:
\begin{equation}
    \mathbf{y}(s) = \bigl(v_{s1}, \ldots, v_{s,K-1}\bigr) \in \mathbb{R}^{K-1}.
\end{equation}
The empirical points cloud density is:
\begin{equation}
    \rho(\mathbf{y}) = \frac{1}{N}\sum_{s=1}^{N} \delta\!\left(\mathbf{y} - \mathbf{y}(s)\right), \qquad \int d\mathbf{y}\,\rho(\mathbf{y}) = 1.
\end{equation}
Well-separated clusters correspond to a localised, low-entropy distribution; an unstructured cloud yields high entropy. We estimate the differential entropy $H = -\int d\mathbf{y}\,\rho(\mathbf{y})\ln\rho(\mathbf{y})$ using the $k$-nearest-neighbours ($k_{\text{NN}}$) Kozachenko--Leonenko estimator \cite{KozachenkoLeonenko1987, Lombardi2016}, which is asymptotically unbiased:
\begin{equation}
    \hat{H}_k = \frac{d}{N}\sum_{s=1}^{N} \ln \epsilon_{s,k}
                + \psi(N) - \psi(k)
                + \ln\!\left(\frac{\pi^{d/2}}{\Gamma(1+d/2)}\right),
    \label{eq:KL_entropy}
\end{equation}
where $d = K-1$, $\epsilon_{s,k}$ is the Euclidean distance from $\mathbf{y}(s)$ to its $k$-th nearest neighbour, and $\psi$, $\Gamma$ denote the digamma and gamma functions. \\
\noindent The informative range of $k$ depends on $N$ and on the size-weighted average community size $\overline{n}_{\mathrm{comm}}$: a loose lower bound scales as $k_{\mathrm{low}} \sim 0.1\,N$ and an upper bound as $k_{\mathrm{high}} \sim 0.8\,\overline{n}_{\mathrm{comm}}$. The factor 0.8 is chosen instead of the geometric upper limit 0.9 derived in Appendix~\ref{app:knn-sensitivity} to keep the operating point safely below the crossing-point transition, which would otherwise be too demanding for datasets with unbalanced community sizes (see Appendix~\ref{app:knn-sensitivity} for the full sensitivity study).

\noindent It is worth noting that this measure provides insightful information on the structure of the dataset even independently of a specific partition to test. Indeed, fixing the dimension of the eigenvector embedding, a study of this observable across the whole range of $k_{NN}$ could highlight if the dataset shows sign of structure and at which scale. 

%The properties of the spectrum and of the eigenvectors of the similairty matrix can be insightful on the presence of a modular structure. \\ RMT results show that clustered data with K communities should have a spectrum of the similarity matrix with K-1 eigenvalues separated from the bulk \blue{(check if true and who proved it in case)}. \\
%The problem is that also homogeneous data that have a strong factor structure show a similar spectrum \blue{(check if true and who proved it in case)}, this means that the eigenvalues alone are not enough to discriminate between the two scenarios. \\
%Nonetheless, a difference can be seen in the eigenvectors. \\ Indeed, the eigenvectors related to the first K-1 eigenvalues (given K is the number of clusters in the data) encode the community structure if present. 
%Meaning that for homogeneous datasets the components of the eigenvectors will be distributed more uniformly rather than for clustered dataset. \\ To quantify this effect we compute an approximation of the \textbf{entropy of the density distribution of the cloud points in the reduced eigenvectors space}. \\

% - - - - - - - - - - - - - - - - - - - - - - - - - - -
\subsubsection{Within--between community overlap of the similarity histogram}
\noindent A complementary, geometric criterion for partition quality is the degree of separation between
the within-community and between-community similarity distributions on the market-mode-cleaned matrix $S^{\mathrm{clean}}$. 
Given the partition $\{C_1,\ldots,C_K\}$ of $N$ subjects, we collect the within-community similarities 
$\mathcal{W} = \{S^{\mathrm{clean}}_{ij} : i,j \in C_k,\, i<j\}$ and the between-community similarities 
$\mathcal{B} = \{S^{\mathrm{clean}}_{ij} : i\in C_k,\,j\in C_{k'},\, k\neq k'\}$, and estimate their kernel densities 
$\hat{f}_\mathcal{W}$ and $\hat{f}_\mathcal{B}$. 
The overlap coefficient is
\begin{equation}
    \mathrm{OV} \;=\; \int \min\!\left(\hat{f}_\mathcal{W}(x),\, \hat{f}_\mathcal{B}(x)\right)\,dx \;\in\; [0,1],
    \label{eq:overlap}
\end{equation}
where $\hat{f}_\mathcal{W}(x)$ and $\hat{f}_\mathcal{B}(x)$ are kernel density estimates of the within- and between-community similarity distributions, and the integral is evaluated numerically on a regular grid.
$\mathrm{OV}=0$ corresponds to perfect separation and $\mathrm{OV}=1$ to identical distributions. 
A well-defined partition is expected to produce a smaller overlap than the same partition applied to a resampled version
of the dataset, so the resampling framework above is used in its lower-tail form for this observable.

%%%%%%%%%%%%%%%%%%%%%%%%%%%%%%%%%%%%%%%%%%%%%%%%%%
\FloatBarrier
\section{Results in synthetic datasets}
%%%%%%%%%%%%%%%%%%%%%%%%%%%%%%%%%%%%%%%%%%%%%%%%%%
\label{sec:res-synth}
%------------------------------------------------------
\subsection{Experimental setup}
\noindent We validate all the methods on three families of synthetic datasets.
The design of the generative models and their parameters are chosen to reproduce the properties of real datasets commonly available in psychometrics. \\
We work with ten datasets: two featuring homogeneous data with only one community, and eight featuring two or more communities with increasing separation.

\noindent All synthetic datasets share the following parameters: $N=1000$ subjects, $M=60$ items on a 6-point ordinal scale from 0 to 5, and $F=4$ latent constructs of 15 items each.\\

\noindent The two homogeneous datasets (referred to as \textit{equal-factors} and \textit{dominant-factor} from now on) are both generated from discretized multivariate Gaussian distributions with a low-rank covariance matrix that groups items into constructs with high internal coherence and low inter-group covariance.
The difference lies in the loading structure: the first dataset uses latent dimensions that contribute approximately evenly to the total variance, while the second is generated with a dominant factor that absorbs a higher fraction of the total variance (see Appendix~\ref{app:synthetics}).\\

%creating a polarization of the datasets along one direction, which induces a pronounced block-diagonal artifact in the similarity matrix and represents a genuinely difficult null case for modularity-based methods \\
\noindent The family of eight clustered datasets is generated from a mixture of binomial distributions.
The mixture component models the separation of subjects into different groups. To reproduce a latent structure of the items, they are assigned to $F$ groups and for each community $k$, the binomial probabilities $p_{ki}$ ($i = 1, \, .. \, , \, M$) are sampled so that items belonging to the same group will have similar $p_{ki}$. This allows us to model both a community-specific behaviour and a latent construct structure typical of psychometric questionnaires. 
The parameters of the generative function allow datasets with a varying level of overlap between the communities.
This lets us work with increasingly challenging datasets and explore the regions where clustered datasets are hard to distinguish from the Gaussian ones.\\
We refer to this set as \textit{mixture series}; each dataset is labelled by an index from 1 to 8 that grows with the signal-strength parameter and spans from an indistinguishable to a well-separated regime. Each dataset has 4 planted communities.
Full details of the generative models are given in Appendix~\ref{app:synthetics}.

\FloatBarrier
\subsection{Partition significance: validation against randomized versions of the datasets}

\begin{figure*}[t]
    \centering
    \includegraphics[width=0.8\linewidth]{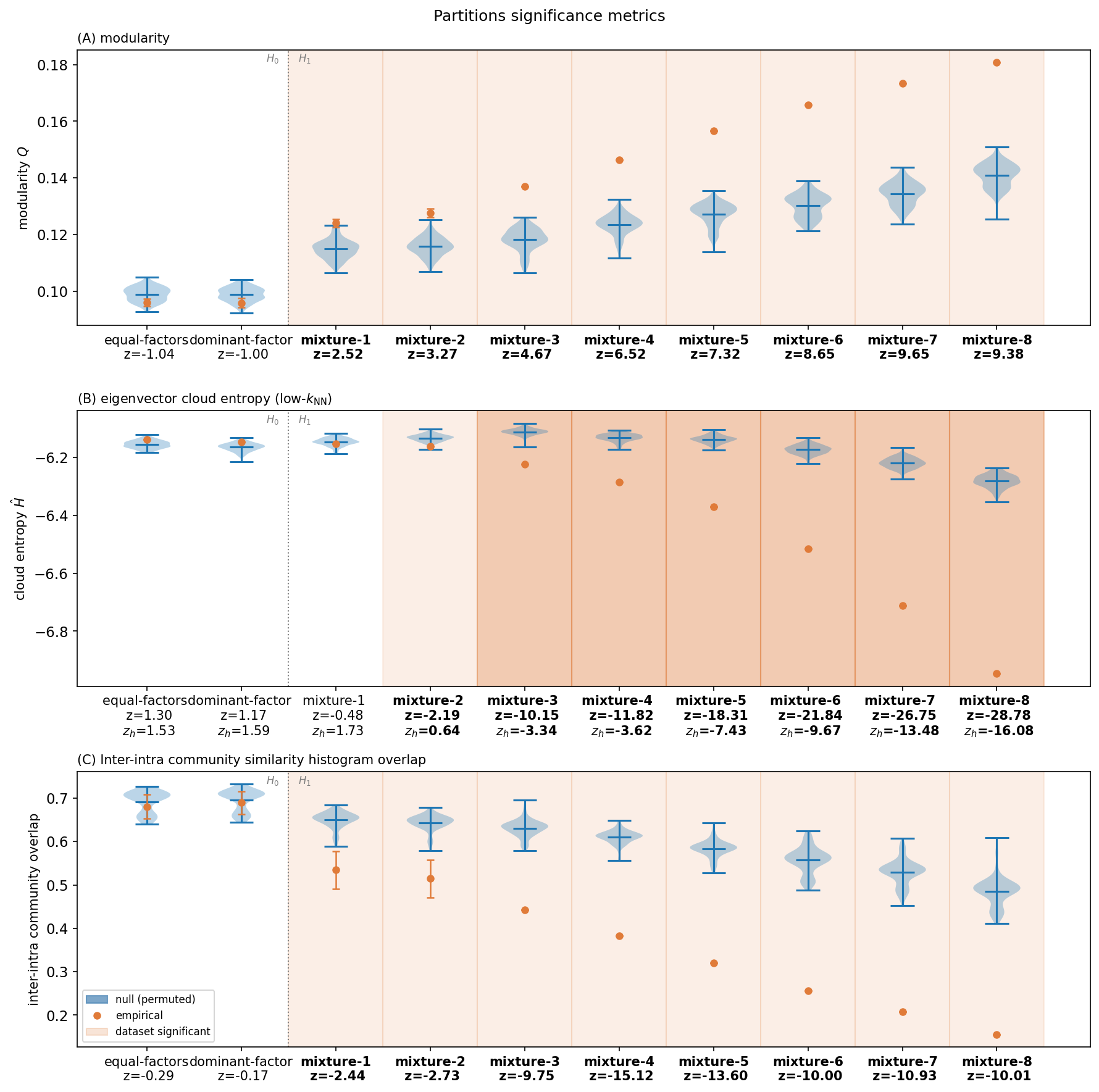}
    \caption{Significance metrics for the ten synthetic datasets.
        \textbf{Panel (A):} Modularity $Q$ (Eq.~\eqref{eq:modularity}).
        \textbf{Panel (B):} Differential entropy $\hat{H}$ of the $(K_{\mathrm{cons}}-1)$-dimensional eigenvector point cloud (Eq.~\eqref{eq:KL_entropy}) at $k_{\mathrm{NN}} = \lfloor 0.1\,N \rfloor = 100$ (loose).
        \textbf{Panel (C):} Within- vs.\ between-community similarity histogram overlap (Eq.~\eqref{eq:overlap}).
        \textbf{Panel (D):} Cloud entropy at $k_{\mathrm{NN}} \approx 0.8\,\overline{n}_{\mathrm{comm}}$ (strict). Across-dataset comparisons of the entropy value are not meaningful since the embedding dimension $K_{\mathrm{cons}}-1$ varies; the test is intra-dataset (empirical value vs.\ its own null).
        In every panel the shaded background marks datasets for which the run-integrated resampling $p$-value of Eq.~\eqref{eq:pvalue} satisfies $p(\mathcal{O}) < 0.05$ under the appropriate one-sided test (right tail for $Q$, left tail for the cloud entropy and the overlap); each violin is annotated with the standardised z-score of Eq.~\eqref{eq:zscore} as a secondary effect-size diagnostic.
        Calibration of $k_{\mathrm{NN}}$ is discussed in Appendix~\ref{app:knn-sensitivity}.}
    \label{fig:sint-validation}
\end{figure*}

\noindent Figure~\ref{fig:sint-validation} summarises the four significance metrics across the ten synthetic datasets. Each panel reports the empirical value of one metric (orange dot, with error bars over the $n_{\mathrm{boot}}$ Leiden re-runs) against the null distribution obtained by column-wise resampling of the factor-score matrix (blue violin).
For the modularity, the empirical value is expected to lie above the null; for both cloud-entropy panels and for the histogram overlap it is expected to lie below the null.
Columns are shaded when the run-integrated $p$-value of Eq.~\eqref{eq:pvalue} satisfies $p(\mathcal{O}) < 0.05$, i.e.\ when the metric flags the dataset as significant under the appropriate one-sided test; each violin is annotated with the standardised z-score of Eq.~\eqref{eq:zscore}, reported as a secondary diagnostic of the effect size.
The four metrics are complementary: the $k_{\mathrm{low}}$ cloud entropy detects fine-grained local concentration and is the most sensitive; modularity probes the global graph structure; the $k_{\mathrm{high}}$ cloud entropy is the conservative criterion targeting community-sized neighbourhoods; and the histogram overlap (panel C) measures the geometric separation of within-community vs.\ between-community similarities. A partition supported by all four criteria carries substantially stronger evidence than one flagged by a single metric.
%
% - - - - - - - - - - - - - - - - - - - - - - - - - - -
%
\paragraph{Modularity significance}
\noindent Both $H_0$ datasets yield empirical $Q$ values that fall within the null distribution, including the more demanding \textit{dominant-factor} case: the combination of FA-space projection and market-mode removal is sufficient to prevent the dominant factor from inducing a false positive.

\noindent Among the mixture datasets, interestingly, modularity is significant ($p(Q) < 0.05$) for all mixtures. 
The empirical $Q$ increases with cluster separation (parametrized with $\xi$) while the null distribution remains approximately constant across the mixture series, widening the gap as cluster separation grows. 
%
%
% - - - - - - - - - - - - - - - - - - - - - - - - - - -
\bigskip
\paragraph{Eigenvector cloud entropies}
The two operating points test different geometric scales. The loose criterion $k_{\mathrm{low}} = \lfloor 0.1\,N \rfloor$ is significant ($p(\hat H_{k_{\mathrm{low}}}) < 0.05$) from \textit{mixture-2}. The strict criterion $k_{\mathrm{high}} = \lfloor 0.8\,\overline{n}_{\mathrm{comm}} \rfloor$ -- chosen below the empirical $\sim 0.9\,\overline{n}_{\mathrm{comm}}$ saturation derived in Appendix~\ref{app:knn-sensitivity} to remain safe on unbalanced community sizes -- crosses the same $p < 0.05$ threshold at \textit{mixture-3}.
The two criteria therefore bracket the detectability transition: $k_{\mathrm{low}}$ is sensitive to local clumping even when the global graph signal is weak, whereas $k_{\mathrm{high}}$ requires that the concentration extends to community-sized neighbourhoods. Both $H_0$ datasets are compatible with the null under either setting.
\bigskip
% - - - - - - - - - - - - - - - - - - - - - - - - - - -
\paragraph{Within-vs-between similarity histogram overlap}
\noindent For each consensus partition we measure the overlap between the within-community and between-community similarity distributions on the market-mode-removed matrix $S^{\mathrm{clean}}$ via a kernel density estimator (Eq.~\eqref{eq:overlap}).
Lower overlap indicates better separation between within- and between-community pairs. The overlap is significant ($p(\mathrm{OV}) < 0.05$) already at \textit{mixture-1}, slightly anticipating the modularity threshold. Both $H_0$ datasets are compatible with the null.

%%%%%%%%%%%%%%%%%%%%%%%%%%%%%%%%%%%%%%%%%%%%%%%%%%
%\FloatBarrier
\section{Results on real psychometric datasets}
%%%%%%%%%%%%%%%%%%%%%%%%%%%%%%%%%%%%%%%%%%%%%%%%%%
\label{sec:res-real}
%------------------------------------------------------
% DATASETS
\noindent We apply the pipeline to 14 real psychometric datasets, covering questionnaires on personality, mental disorders, political beliefs, and empathic sensitivity. A short description of each dataset, including the number of items, constructs and respondents, is reported in Table~\ref{tab:datasets-tab} together with the results discussed below.\\
For each dataset we run the same pipeline used for the synthetic validation: similarity computed in the space of factor scores, market-mode removal, and the (signed) Leiden algorithm. To keep computational cost comparable across datasets, all real datasets are first subsampled to a uniformly random subset of $N=1000$ respondents with complete responses (no imputation).
The datasets differ in the number of constructs they explore, ranging from one (single-scale surveys) up to 20 (multi-scale instruments).
For each multi-scale dataset, we set the number of latent factors $F$ for the similarity calculation equal to the number of constructs that the questionnaire explores.
For single-scale or two-scale datasets (RWAS, GCBS, PWE, MACH, CFCS) we set $F=3$ following the rule of thumb in Section~\ref{similarity}. A more detailed discussion of the choice of $F$ is provided in Appendix~\ref{app:F-sensitivity}.
%
% Merged table: dataset descriptions + significance outcomes
\begin{table*}[t]
\centering
\small
\begin{tblr}{
    colspec  = {|c|p{2.3cm}|p{3.8cm}|r|r|r|c|c|c|c|r|p{1.2cm}|},
    hlines,
    column{2,3, 12} = {cmd=\RaggedRight},
    row{1}     = {font=\bfseries},
    row{2,6,9,10}    = {bg=gray!25},   % 3 metrics significant
    row{4,5}         = {bg=gray!15},   % 2 metrics significant
    row{7}           = {bg=gray!8},    % 1 metric significant
    row{12,13,14,15} = {bg=gray!45},   % 4 metrics significant
    colsep = 4pt,
}
Acronym & Full name & Domain \& subscales & $N$ & $M/F$ & $K$ & $Q$ & $\hat{H}_{k_{\rm low}}$ & $\hat{H}_{k_{\rm high}}$ & OV & \#sig & Access \\
BIG5   & Big Five Inventory                          & personality traits. Subscales: Openness, Conscientiousness, Extraversion, Agreeableness, Neuroticism             & 873173 & 50/5  & 4 & $\circ$ & $\bullet$ & $\bullet$ & $\bullet$ & 3 & open \cite{goldberg1992bigfive} \\
HEXACO & HEXACO Personality Inventory                & personality traits. Subscales: Honesty-Humility, Emotionality, eXtraversion, Agreeableness, Conscientiousness, Openness & 22727  & 240/6   & 4 & $\circ$ & $\circ$   & $\circ$   & $\circ$   & 0 & open \cite{lee2004hexaco, openpsychometrics_rawdata}\\
SD3    & Short Dark Triad  & dark triad personality traits: Machiavellianism, Narcissism, Psychopathy                               & 17738  & 27/3    & 3 & $\bullet$ & $\circ$ & $\circ$   & $\bullet$ & 2 & open \cite{jones2014sd3, openpsychometrics_rawdata}\\
MACH   & Mach-IV Machiavellianism Scale  & Machiavellianism (single scale; views of human nature, manipulative tactics, abstract morality as facets) & 47090  & 20/1  & 3 & $\circ$   & $\bullet$ & $\circ$ & $\bullet$ & 2 & open \cite{christie1970mach, openpsychometrics_rawdata}\\
ACME   & Affective \& Cognitive Measure of Empathy  & multi-scale empathy measure; subscales on cognitive empathy, affective resonance, affective dissonance   & 1262   & 36/3    & 3 & $\bullet$ & $\bullet$ & $\circ$ & $\bullet$ & 3 & upon request \cite{vachon2016fixing} \\
IRI    & Interpersonal Reactivity Index  & multi-scale empathy measure; Perspective Taking, Fantasy, Empathic Concern, Personal Distress           & 1262   & 28/4      & 3 & $\circ$   & $\circ$   & $\circ$ & $\bullet$ & 1 & upon request \cite{albiero2006iri}\\
EI     & Empathy Index Scale   & two-scale empathy measure; empathy subscale and the behavioral contagion subscale  & 1262   & 14/2   & 4 & $\circ$   & $\circ$   & $\circ$ & $\circ$   & 0 & upon request \cite{jordan2002ei}\\
DASS   & Depression Anxiety Stress Scales   & mental health. Subscales on Depression, Anxiety, Stress                                                      & 3550   & 42/3   & 4 & $\bullet$ & $\bullet$ & $\circ$ & $\bullet$ & 3 & upon request \cite{lovibond1995dass}\\
CFCS   & Consideration of Future Consequences Scale  & temporal orientation; single scale                                                              & 14814  & 12/1   & 3 & $\bullet$ & $\bullet$ & $\circ$ & $\bullet$ & 3 & open \cite{strathman1994cfcs, openpsychometrics_rawdata}\\
HSNS   & Hypersensitive Narcissism Scale \& Dirty Dozen & narcissism (HSNS, 10 items) and dark triad (Dirty Dozen, 12 items)                         & 50483  & 22/2   & 4 & $\circ$   & $\circ$   & $\circ$ & $\circ$   & 0 & open \cite{hendin1997hsns, openpsychometrics_rawdata} \\
MSSCQ  & Multidimensional Sexual Self-Concept Questionnaire & sexual self-concept; 20 subscales (e.g.\ esteem, anxiety, motivation, assertiveness)  & 12789  & 100/20  & 4 & $\bullet$ & $\bullet$ & $\bullet$ & $\bullet$ & 4 & open \cite{snell1995msscq, openpsychometrics_rawdata}\\
RWAS   & Right-Wing Authoritarianism Scale           & political attitudes; single scale                                                               & 9680   & 22/1  & 3 & $\bullet$ & $\bullet$ & $\bullet$ & $\bullet$ & 4 & open \cite{altemeyer1981rwas, openpsychometrics_rawdata}\\
GCBS   & Generic Conspiracist Beliefs Scale          & conspiracist beliefs; single scale                                                              & 1283   & 15/1   & 3 & $\bullet$ & $\bullet$ & $\bullet$ & $\bullet$ & 4 & open \cite{brotherton2013gcbs, openpsychometrics_rawdata}\\
PWE    & Protestant Work Ethic Scale & work values, single scale & 892    & 19/1  & 3 & $\bullet$ & $\bullet$ & $\bullet$ & $\bullet$ & 4 & open \cite{mirels1971pwe, openpsychometrics_rawdata}\\
\end{tblr}
\caption{Summary of the 14 psychometric datasets used in the analysis.
\textit{Domain \& subscales}: psychological domain investigated and subscale structure; ``single scale'' indicates a unidimensional instrument.
$N$: total number of respondents available (the pipeline operates on a uniformly random subsample of $N=1000$ with complete responses).
$M\,/\,F$: total items / number of latent constructs in the original scale (the FA projection in the pipeline is set to $\max(F,3)$ following the rule of thumb in Section~\ref{similarity}, so it differs from $F$ only for single-construct scales such as RWAS, GCBS, PWE, MACH, CFCS).
$K$: number of communities found by the community detection pipeline.
\textit{Significance metrics}: $\bullet$ = run-integrated $p<0.05$; $\circ$ = $p\geq 0.05$; \#sig counts the number of metrics simultaneously significant; row shading is proportional to it.}
\label{tab:datasets-tab}
\end{table*}
%
%
%
%
%------------------------------------------------------
\subsection{Partition significance}
\begin{figure*}[t]
    \centering
    \includegraphics[width=0.95\linewidth]{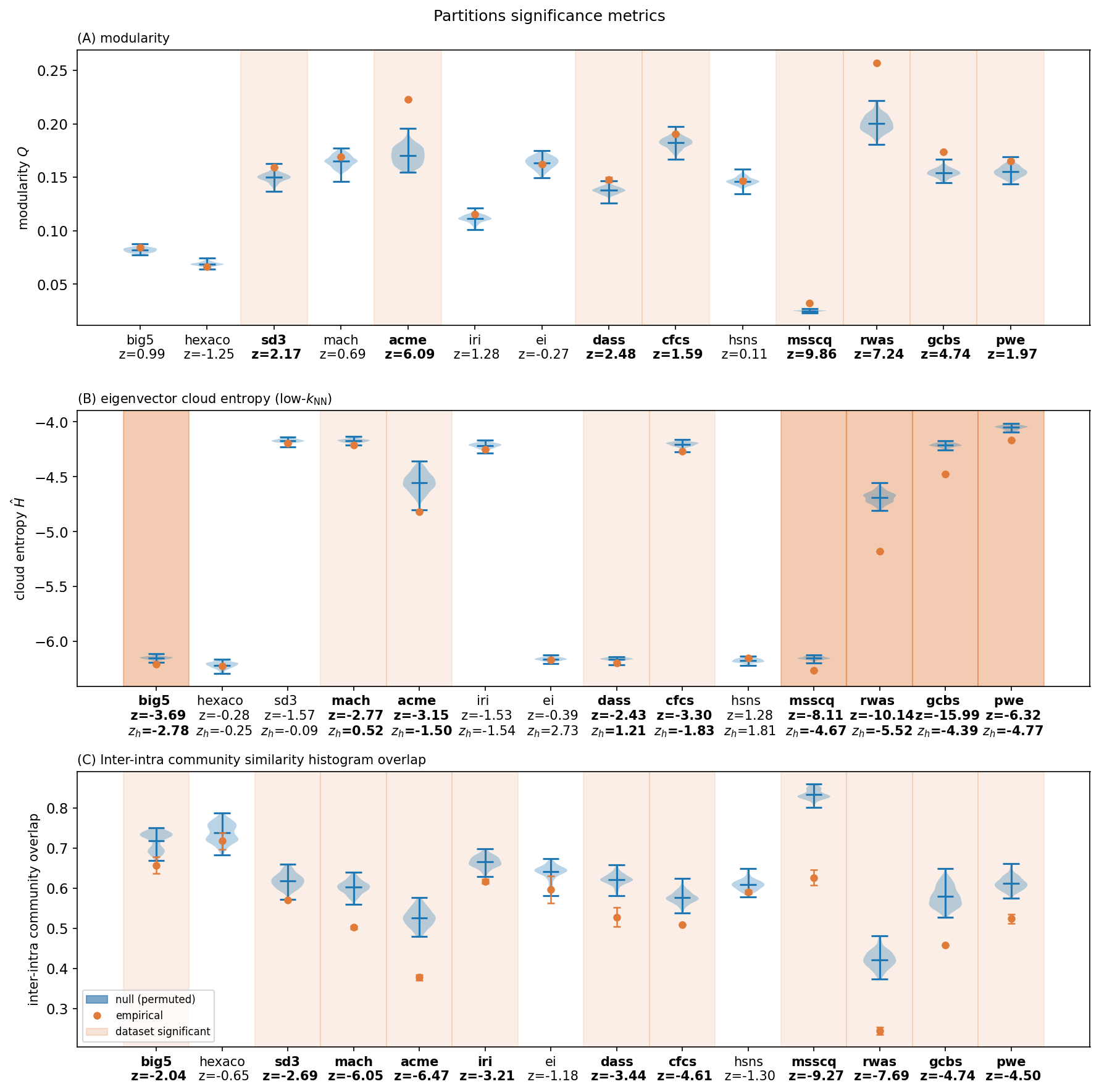}
    \caption{Significance metrics for the 14 real datasets, structured as in Fig.~\ref{fig:sint-validation}. Panels: (A) modularity $Q$, (B) cloud entropy at $k_{\mathrm{low}}$, (C) within- vs.\ between-community similarity histogram overlap, (D) cloud entropy at $k_{\mathrm{high}}$.
    Shaded background marks datasets for which the run-integrated resampling $p$-value of Eq.~\eqref{eq:pvalue} satisfies $p(\mathcal{O}) < 0.05$ under the appropriate one-sided test (right tail for $Q$, left tail for the cloud entropy and the overlap); the z-score annotated on each violin is the secondary effect-size diagnostic of Eq.~\eqref{eq:zscore}. Cloud-entropy values are not comparable across datasets since the embedding dimension $K_{\mathrm{cons}}-1$ varies; the test is intra-dataset.}
    \label{fig:real-validation}
\end{figure*}

Significance metrics for the real data are computed exactly as on the synthetic benchmark and reported in Figure~\ref{fig:real-validation}; we again use the run-integrated $p$-value of Eq.~\eqref{eq:pvalue} at $\alpha=0.05$ as the binary significance criterion, and the z-score of Eq.~\eqref{eq:zscore} as the effect-size diagnostic. Table~\ref{tab:datasets-tab} aggregates the outcomes together with the items-per-factor ratio $r=M/F$, the number of communities $K$ returned by the pipeline, and the count of metrics simultaneously flagging each dataset as significantly clustered.
Three groups emerge from the table.\\
\textit{Strong-signal datasets} (RWAS, GCBS, MSSCQ and PWE) are marked as significantly clustered by all four metrics. \\
\textit{Mixed-signal datasets} (BIG5, ACME, DASS, CFCS) reach significance on three metrics, while SD3 and MACH reach significance on two metrics. \\
\textit{Null-compatible datasets} (HEXACO, IRI, EI and HSNS) reach significance on at most one metric. The HEXACO case is particularly informative: with 240 items distributed over 6 broadband personality dimensions, none of the four metrics provides evidence of a clustered structure, suggesting that the 6-factor Big-Five-like representation does not yield strong discrete subgroups of respondents but rather a broad distribution over the trait space.\\
A caveat applies to IRI, EI, HSNS and CFCS, and — with a slightly lower $r$ — also to MACH and PWE.
These scales have items-per-factor ratios $r = M/F_{\mathrm{true}} \lesssim 12$ which, according to the sensitivity analysis of Appendix~\ref{app:F-sensitivity}, places them in a hard regime: a low items-per-factor ratio combined with a short Likert range limits the exploration of nuanced answer patterns, inducing non-negligible inhomogeneities in the factor-score space even when subjects belong to a single population.
Qualitatively, this effect stems from finite-sample discretization: with few items per factor and a short Likert range, the number of achievable discrete response patterns is small (roughly of order $R^{M/F}$ for $M$ items strongly loading on $F$ latent constructs with $R$ answer levels), and subjects concentrate on a grid of high-probability cells. This may produces a non-uniform, lattice-like distribution in the factor-score space that community detection algorithms may interpret as genuine subgroup structure.
Such spurious patterns are the hardest to identify, as clustering algorithms are prone to flag them as communities.
Any signature of modular structure emerging in datasets within this regime — such as the lower-than-null inter-to-intra community overlap observed for IRI — should therefore be interpreted with caution.

%------------------------------------------------------
\subsection{Answer patterns in clustered datasets}

\noindent Beyond the question of statistical significance, it is informative to examine what the detected communities look like in terms of response patterns. We focus as an example on two of the datasets with the strongest evidence in the significance framework: RWAS and GCBS.

\begin{figure*}[t]
    \centering
    \includegraphics[width=\linewidth]{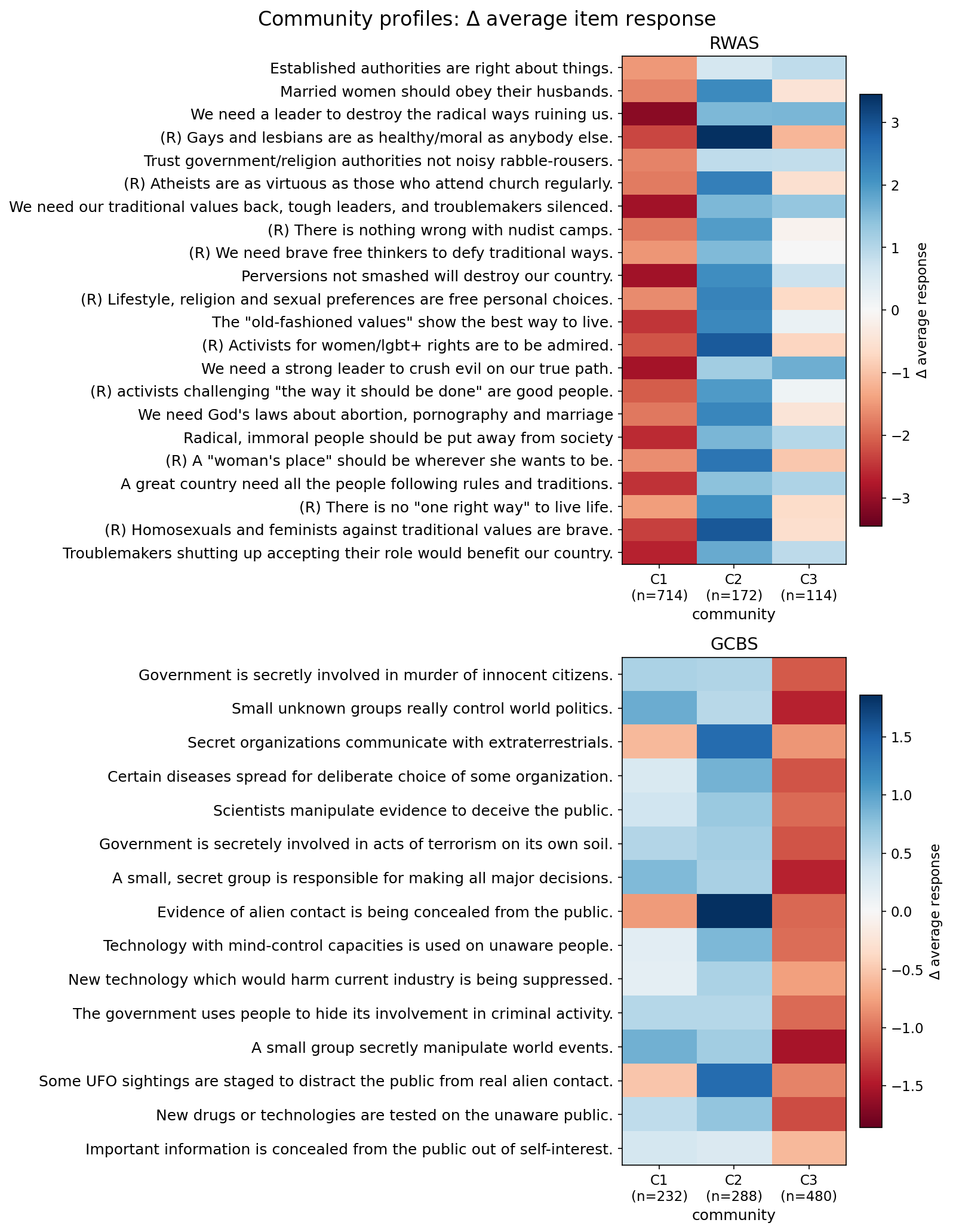}
    \caption{Per-item deviations from the global mean ($\Delta$ average item response) for each detected community, shown for RWAS and GCBS. Each column is a community; each row
    is an item. Cool colours indicate above-average responses, warm colours below-average.}
    \label{fig:comm-profiles}
\end{figure*}

\noindent RWAS (Fig.~\ref{fig:comm-profiles}, top panel) is a scale that investigates the tendency of a person toward right-wing authoritarian beliefs. In the literature, these beliefs are known to generate polarisation between those who fully embrace them and those who reject them \cite{altemeyer1981rwas}.
This separation is very strong between communities C1 and C2, which present opposite answer profiles with large deviations from the population mean: C2 groups the right-wing authoritarians, while C1 groups the left-wing libertarians. C3 groups respondents with centrist attitudes, leaning slightly left or right depending on the item.
The items indicated by (R) have been reverse-scored for visualisation purposes, so that all answers in blue correspond to a right-wing answer style.

\noindent GCBS is the General Conspiracy Beliefs Scale and it is another example of a scale investigating a polarizing topic. In this scale, all the items are conspiratory statements. 
In Fig.~\ref{fig:comm-profiles}, bottom panel), the three communities separate along the overall level of conspiracy belief. C1 ($n=478$) lies below the global mean on all items, identifying low conspiracy believers; C2 ($n=288$) lies above the mean on all items, identifying high believers; C3 ($n=234$) shows small and mixed deviations, consistent with intermediate or selective endorsement across the conspiracy theories probed by the scale.
Item 3, 8 and 13 stand out as particularly strongly endorsed in the group of high believers, interestingly, these three statements are the only claiming the evidence of alien contact is being kept secret from the public. 
% item 3: Secret organizations communicate with extraterrestrials, but keep this fact from the public.
% item 8: Evidence of alien contact is being concealed from the public.
% item 13: Some UFO sightings and rumours are planned or staged in order to distract the public from real alien contact.
These conspiracy beliefs are the most extremest conspiracy theory included in the survey, and indeed the group who believes in them is the group that is believing to conspiracies in general, being suspicious of governments, scientists and technology. \\
The third community is also interesting, since it is less extremist than C2, agreeing that extraterrestrial contacts are not being kept secret from the public, but they present a moderate level of skepticism towards political institutions and who detains the real power in the world (e.g. item 12 reads: \emph{Certain significant events have been the result of the activity of a small group who secretly manipulate world events}).

\clearpage
\section{Discussion}
\label{sec:discussion}
%%%%%%%%%%%%%%%%%%%%%%%%%%%%%%%%%%%%%%%%%%%%%%%%%%
From the results illustrated so far, we can distill various key considerations.\\

\paragraph{The importance of designing a similarity projection and a community detection algorithm specifically for psychometric data.}
Psychometric data are constrained to a discrete ordinal scale, items within the same construct are deliberately correlated, and the number of items is typically small. These properties serve interpretability and measurement reliability but have direct consequences for pairwise similarity measures: naive community detection on the raw item space produces spurious community partitions even on homogeneous data (Fig.~\ref{fig:demo-comm}). FA-space similarity and market-mode removal jointly suppress this artifact, ensuring the pipeline behaves correctly on null data before any genuine signal is present.\\
\paragraph{The importance of statistically validating the detected partition.}
Community detection algorithms return a partition regardless of whether a meaningful one exists, so statistical validation is a necessary, not optional, step. The four observables introduced here probe complementary facets of the partition (graph cohesion, fine- and community-scale concentration in the eigenvector embedding, and within- vs.\ between-community geometric separation), and their joint signature is more informative than any single test. Disagreement between metrics is itself diagnostic: local eigenvector concentration without global modular cohesion, for instance, points to a mildly clustered landscape rather than to well-separated communities.
The four metrics are not equivalent in strictness: in general, in dataset with good item-per-factor ration, modularity emerged as the most liberal, followed by cloud entropy evaluated at a short neighborhood scale, similarity overlap, and lastly cloud entropy at a broad neighborhood scale, which imposes the most stringent criterion for detecting community structure. 
The framework is independent of the specific community detection algorithm used and can be applied to assess the output of any alternative pipeline.\\
\paragraph{Limitations.}
The central limitation is shared by any clustering method on pairwise distances at finite $N$: when community separation is weak, the data are indistinguishable from a homogeneous population and the partition cannot be reliably recovered (this is a statistical limit, not a failure specific to the pipeline). On real data this affects most intermediate cases, where subpopulations may exist but are not sharply separated; the significance framework provides an operational criterion but cannot resolve borderline cases. A partial characterisation of the detectable regime as a function of $(M, F, K)$ is given in Appendices~\ref{app:F-sensitivity}, \ref{app:F-sensitivity-real}, and~\ref{app:Kstar}.\\
\noindent
Constraints intrinsic to the questionnaire format -- short instruments, narrow ordinal scales, response-style heterogeneity (acquiescence, extreme responding), and sampling bias -- can additionally blur community boundaries with respect to the underlying latent structure. A quantitative account of these effects is left for future work.\\
\paragraph{Future directions.}
Three directions stand out. First, extending the synthetic benchmark to more complex generative models would enable a systematic type-I/power characterisation under varied data-generating processes. Second, mixed-signal datasets may be better described as core-periphery structures \citep{Borgatti2000-corepe, Zhang2015-corepe} -- where small, well-defined subgroups coexist with a broader population sharing traits with more than one core -- as suggested by the qualitative nested organisation visible in some similarity heatmaps (Appendix~\ref{app:item-space-art}). Third, the framework extends naturally to longitudinal data, where the question shifts from a single-snapshot partition to the stability of community membership over time. As a complementary use, the significance test can serve as a preprocessing step for LCA, providing a data-driven criterion for whether a discrete-subgroup analysis is warranted before committing to a particular number of classes.

\section{Conclusions}
%%%%%%%%%%%%%%%%%%%%%%%%%%%%%%%%%%%%%%%%%%%%%%%%%%
\noindent
Community detection in psychometric data is a non-trivial task: response style artifacts and factor structure can generate apparent clusters in homogeneous populations, and available clustering significance tests are not well calibrated for this setting.
The pipeline introduced here addresses both problems through two targeted design choices -- FA-space similarity and market-mode removal -- and the permutation-based significance framework provides a principled way to decide whether a detected partition is genuine.
Among the metrics used to assess the statistical significance of the communities detected, we introduced the \emph{cloud entropy} as a novel measure.

\noindent
Applied to 14 real datasets, the approach shows that the community detection algorithm alone is not enough to distinguish communities reflecting genuine population heterogeneity from communities emerging as artifacts of the data collection and processing.
The partition-significance framework introduced here addresses this gap by quantifying how probable it is that the partition observed reflects substantial structure rather than confounding artifacts.
The framework is not tied to the specific pipeline used here; it can serve as a general-purpose test for whether a psychometric dataset warrants a discrete-subgroup analysis at all, complementing existing tools such as factor analysis and mixture models. \\
The datasets that show the strongest signs of clusterisation in the population are scales measuring polarising content, such as right-wing authoritarian political opinions (RWAS) and belief in conspiracy theories (GCBS). Other strong-signal scales such as PWE and MSSCQ also display sharp subgroup structure, although the interpretation in terms of polarisation is less direct. Personality inventories show more diffuse landscapes, consistent with continuous transitions between personality types rather than sharply defined subgroups: this is itself a finding about subject-level heterogeneity in the constructs studied, beyond what factor analysis alone can reveal.

\section{Acknowledgments}
We thank Giulio Costantini and Gabriele Limonta for the fruitful discussions and for pointing us to relevant references on the analysis of clusters in psychometrics.

%\section{Author Contributions}

\bibliography{references}

@misc{psycomm,
  author  = {Armanetti, Arianna},
  title   = {psycomm: community detection in psychometric data},
  year    = {2026},
  url     = {https://github.com/ariannaarmanetti/psycomm},
  doi     = {10.5281/zenodo.20386040},
}

@BOOK{Asratian1998-vf,
  title     = "Cambridge tracts in mathematics: Bipartite graphs and their
               applications series number 131",
  author    = "Asratian, Armen S and Denley, Tristan M J and Haggkvist, Roland",
  publisher = "Cambridge University Press",
  month     =  jul,
  year      =  1998,
  address   = "Cambridge, England"
}

@ARTICLE{Robinaugh2020-hu,
  title     = "The network approach to psychopathology: a review of the
               literature 2008-2018 and an agenda for future research",
  author    = "Robinaugh, Donald J and Hoekstra, Ria H A and Toner, Emma R and
               Borsboom, Denny",
  journal   = "Psychol. Med.",
  publisher = "Cambridge University Press (CUP)",
  volume    =  50,
  number    =  3,
  pages     = "353--366",
  month     =  feb,
  year      =  2020,
  language  = "en"
}

@ARTICLE{Olthof2023-fi,
  title     = "Complexity theory of psychopathology",
  author    = "Olthof, Merlijn and Hasselman, Fred and Oude Maatman, Freek and
               Bosman, Anna M T and Lichtwarck-Aschoff, Anna",
  journal   = "J. Psychopathol. Clin. Sci.",
  publisher = "American Psychological Association (APA)",
  volume    =  132,
  number    =  3,
  pages     = "314--323",
  month     =  apr,
  year      =  2023,
  language  = "en"
}

@ARTICLE{Scheffer2024-pq,
  title     = "A dynamical systems view of psychiatric disorders-theory: A
               review",
  author    = "Scheffer, Marten and Bockting, Claudi L and Borsboom, Denny and
               Cools, Roshan and Delecroix, Clara and Hartmann, Jessica A and
               Kendler, Kenneth S and van de Leemput, Ingrid and van der Maas,
               Han L J and van Nes, Egbert and Mattson, Mark and McGorry, Pat D
               and Nelson, Barnaby",
  journal   = "JAMA Psychiatry",
  publisher = "American Medical Association (AMA)",
  volume    =  81,
  number    =  6,
  pages     = "618--623",
  month     =  jun,
  year      =  2024,
  language  = "en"
}

@ARTICLE{Borsboom2021-gt,
  title     = "Network analysis of multivariate data in psychological science",
  author    = "Borsboom, Denny and Deserno, Marie K and Rhemtulla, Mijke and
               Epskamp, Sacha and Fried, Eiko I and McNally, Richard J and
               Robinaugh, Donald J and Perugini, Marco and Dalege, Jonas and
               Costantini, Giulio and Isvoranu, Adela-Maria and Wysocki, Anna C
               and van Borkulo, Claudia D and van Bork, Riet and Waldorp,
               Lourens J",
  journal   = "Sci. Rep.",
  publisher = "Springer Science and Business Media LLC",
  volume    =  4,
  number    =  1,
  pages     = "5918",
  month     =  aug,
  year      =  2014,
  copyright = "https://creativecommons.org/licenses/by-nc-nd/4.0",
  language  = "en"
}

@ARTICLE{Borsboom2017-ke,
  title     = "A network theory of mental disorders",
  author    = "Borsboom, Denny",
  journal   = "World Psychiatry",
  publisher = "Wiley",
  volume    =  16,
  number    =  1,
  pages     = "5--13",
  month     =  feb,
  year      =  2017,
  copyright = "http://onlinelibrary.wiley.com/termsAndConditions\#vor",
  language  = "en"
}

@ARTICLE{Epskamp2018-yi,
  title     = "Estimating psychological networks and their accuracy: A tutorial
               paper",
  author    = "Epskamp, Sacha and Borsboom, Denny and Fried, Eiko I",
  journal   = "Behav. Res. Methods",
  publisher = "Springer Science and Business Media LLC",
  volume    =  50,
  number    =  1,
  pages     = "195--212",
  month     =  feb,
  year      =  2018,
  language  = "en"
}

@ARTICLE{Fried2017-cw,
  title     = "Moving forward: Challenges and directions for psychopathological
               network theory and methodology",
  author    = "Fried, Eiko I and Cramer, Ang{\'e}lique O J",
  journal   = "Perspect. Psychol. Sci.",
  publisher = "SAGE Publications",
  volume    =  12,
  number    =  6,
  pages     = "999--1020",
  month     =  nov,
  year      =  2017,
  language  = "en"
}

@ARTICLE{Howard2018-ng,
  title     = "Variable-centered, person-centered, and person-specific
               approaches",
  author    = "Howard, Matt C and Hoffman, Michael E",
  journal   = "Organ. Res. Methods",
  publisher = "SAGE Publications",
  volume    =  21,
  number    =  4,
  pages     = "846--876",
  month     =  oct,
  year      =  2018,
  language  = "en"
}

@ARTICLE{fortunato2010-os,
  title     = "Community detection in graphs",
  author    = "Fortunato, Santo",
  journal   = "Phys. Rep.",
  publisher = "Elsevier BV",
  volume    =  486,
  number    = "3-5",
  pages     = "75--174",
  month     =  feb,
  year      =  2010,
  language  = "en"
}

@ARTICLE{Newman2004-zp,
  title     = "Finding and evaluating community structure in networks",
  author    = "Newman, M E J and Girvan, M",
  journal   = "Phys. Rev. E Stat. Nonlin. Soft Matter Phys.",
  publisher = "American Physical Society (APS)",
  volume    =  69,
  number    = "2 Pt 2",
  pages     = "026113",
  month     =  feb,
  year      =  2004,
  copyright = "http://link.aps.org/licenses/aps-default-license",
  language  = "en"
}

@ARTICLE{newman2006-dt,
  title     = "Modularity and community structure in networks",
  author    = "Newman, M E J",
  journal   = "Proc. Natl. Acad. Sci. U. S. A.",
  publisher = "Proceedings of the National Academy of Sciences",
  volume    =  103,
  number    =  23,
  pages     = "8577--8582",
  month     =  jun,
  year      =  2006,
  language  = "en"
}

@article{garlaschelli-macmahon15,
  title = {Community Detection for Correlation Matrices},
  author = {MacMahon, Mel and Garlaschelli, Diego},
  journal = {Phys. Rev. X},
  volume = {5},
  issue = {2},
  pages = {021006},
  numpages = {34},
  year = {2015},
  month = {Apr},
  publisher = {American Physical Society},
  doi = {10.1103/PhysRevX.5.021006},
  url = {https://link.aps.org/doi/10.1103/PhysRevX.5.021006}
}

@ARTICLE{Traag2019-ax,
  title     = "From Louvain to Leiden: guaranteeing well-connected communities",
  author    = "Traag, V A and Waltman, L and van Eck, N J",
  journal   = "Sci. Rep.",
  publisher = "Springer Science and Business Media LLC",
  volume    =  9,
  number    =  1,
  pages     = "5233",
  month     =  mar,
  year      =  2019,
  copyright = "https://creativecommons.org/licenses/by/4.0",
  language  = "en"
}

@ARTICLE{Von_Luxburg2007-vc,
  title     = "A tutorial on spectral clustering",
  author    = "von Luxburg, Ulrike",
  journal   = "Stat. Comput.",
  publisher = "Springer Science and Business Media LLC",
  volume    =  17,
  number    =  4,
  pages     = "395--416",
  month     =  dec,
  year      =  2007,
  language  = "en"
}

@ARTICLE{Nadakuditi2012-uc,
  title     = "Graph spectra and the detectability of community structure in
               networks",
  author    = "Nadakuditi, Raj Rao and Newman, M E J",
  journal   = "Phys. Rev. Lett.",
  publisher = "American Physical Society (APS)",
  volume    =  108,
  number    =  18,
  pages     = "188701",
  month     =  may,
  year      =  2012,
  copyright = "http://link.aps.org/licenses/aps-default-license",
  language  = "en"
}

@ARTICLE{Decelle2011-yd,
  title     = "Asymptotic analysis of the stochastic block model for modular
               networks and its algorithmic applications",
  author    = "Decelle, Aurelien and Krzakala, Florent and Moore, Cristopher
               and Zdeborov{\'a}, Lenka",
  journal   = "Phys. Rev. E Stat. Nonlin. Soft Matter Phys.",
  publisher = "American Physical Society (APS)",
  volume    =  84,
  number    = "6 Pt 2",
  pages     = "066106",
  month     =  dec,
  year      =  2011,
  copyright = "http://link.aps.org/licenses/aps-default-license",
  language  = "en"
}

@ARTICLE{Bun2017-oi,
  title     = "Cleaning large correlation matrices: Tools from Random Matrix
               Theory",
  author    = "Bun, Jo{\"e}l and Bouchaud, Jean-Philippe and Potters, Marc",
  journal   = "Phys. Rep.",
  publisher = "Elsevier BV",
  volume    =  666,
  pages     = "1--109",
  month     =  jan,
  year      =  2017,
  copyright = "http://creativecommons.org/licenses/by-nc-nd/4.0/",
  language  = "en"
}

@article{almog2015mesoscopic,
  title={Mesoscopic community structure of financial markets revealed by price and sign fluctuations},
  author={Almog, Assaf and Besamusca, Ferry and MacMahon, Mel and Garlaschelli, Diego},
  journal={PloS one},
  volume={10},
  number={7},
  pages={e0133679},
  year={2015},
  publisher={Public Library of Science San Francisco, CA USA}
}

@ARTICLE{Anagnostou2021-zq,
  title     = "Uncovering the mesoscale structure of the credit default swap
               market to improve portfolio risk modelling",
  author    = "Anagnostou, I and Squartini, T and Kandhai, D and Garlaschelli,
               D",
  journal   = "Quant. Finance",
  publisher = "Informa UK Limited",
  volume    =  21,
  number    =  9,
  pages     = "1501--1518",
  month     =  sep,
  year      =  2021,
  language  = "en"
}

@article{zema2025mesoscopic,
  title={Mesoscopic structure of the stock market and portfolio optimization: SM Zema et al.},
  author={Zema, Sebastiano Michele and Fagiolo, Giorgio and Squartini, Tiziano and Garlaschelli, Diego},
  journal={Journal of Economic Interaction and Coordination},
  volume={20},
  number={2},
  pages={307--333},
  year={2025},
  publisher={Springer}
}

@article{buijink2016evidence,
  title={Evidence for weakened intercellular coupling in the mammalian circadian clock under long photoperiod},
  author={Buijink, M Renate and Almog, Assaf and Wit, Charlotte B and Roethler, Ori and Olde Engberink, Anneke HO and Meijer, Johanna H and Garlaschelli, Diego and Rohling, Jos HT and Michel, Stephan},
  journal={PloS one},
  volume={11},
  number={12},
  pages={e0168954},
  year={2016},
  publisher={Public Library of Science San Francisco, CA USA}
}

@article{almog2019uncovering,
  title={Uncovering functional signature in neural systems via random matrix theory},
  author={Almog, Assaf and Buijink, M Renate and Roethler, Ori and Michel, Stephan and Meijer, Johanna H and Rohling, Jos HT and Garlaschelli, Diego},
  journal={PLoS computational biology},
  volume={15},
  number={5},
  pages={e1006934},
  year={2019},
  publisher={Public Library of Science San Francisco, CA USA}
}

@article{mircea2022phiclust,
  title={Phiclust: a clusterability measure for single-cell transcriptomics reveals phenotypic subpopulations},
  author={Mircea, Maria and Hochane, Maz{\`e}ne and Fan, Xueying and Chuva de Sousa Lopes, Susana M and Garlaschelli, Diego and Semrau, Stefan},
  journal={Genome Biology},
  volume={23},
  number={1},
  pages={18},
  year={2022},
  publisher={Springer}
}

@ARTICLE{Babeanu2021-rw,
  title     = "A random matrix perspective of cultural structure: groups or
               redundancies?",
  author    = "B{\u a}beanu, Alexandru-Ionu{\c t}",
  journal   = "J. Phys. Complex.",
  publisher = "IOP Publishing",
  volume    =  2,
  number    =  2,
  pages     = "025008",
  month     =  jun,
  year      =  2021,
  copyright = "https://creativecommons.org/licenses/by/4.0/"
}

@ARTICLE{MacCarron2020-dv,
  title         = "Identifying opinion-based groups from survey data: a
                   bipartite network approach",
  author        = "MacCarron, P{\'a}draig and Maher, Paul J and Quayle, Michael",
  month         =  dec,
  year          =  2020,
  copyright     = "http://creativecommons.org/licenses/by/4.0/",
  archivePrefix = "arXiv",
  primaryClass  = "cs.SI",
  eprint        = "2012.11392",
  journal       = {...}
}

@ARTICLE{Dinkelberg2021-ug,
  title     = "Detecting opinion-based groups and polarization in survey-based
               attitude networks and estimating question relevance",
  author    = "Dinkelberg, Alejandro and O'sullivan, David J P and Quayle,
               Michael and Maccarron, P{\'a}draig",
  journal   = "Adv. Complex Syst.",
  publisher = "World Scientific Pub Co Pte Ltd",
  volume    =  24,
  number    =  02,
  month     =  mar,
  year      =  2021,
  language  = "en"
}

@ARTICLE{Laloux1999-di,
  title     = "Noise dressing of financial correlation matrices",
  author    = "Laloux, Laurent and Cizeau, Pierre and Bouchaud, Jean-Philippe
               and Potters, Marc",
  journal   = "Phys. Rev. Lett.",
  publisher = "American Physical Society (APS)",
  volume    =  83,
  number    =  7,
  pages     = "1467--1470",
  month     =  aug,
  year      =  1999,
  copyright = "http://link.aps.org/licenses/aps-default-license"
}

@ARTICLE{Plerou1999-xt,
  title     = "Universal and nonuniversal properties of cross correlations in
               financial time series",
  author    = "Plerou, Vasiliki and Gopikrishnan, Parameswaran and Rosenow,
               Bernd and Nunes Amaral, Lu{\'\i}s A and Stanley, H Eugene",
  journal   = "Phys. Rev. Lett.",
  publisher = "American Physical Society (APS)",
  volume    =  83,
  number    =  7,
  pages     = "1471--1474",
  month     =  aug,
  year      =  1999,
  copyright = "http://link.aps.org/licenses/aps-default-license"
}

@article{Leonidas2011,
author = {Jr, Leonidas and Franca, Italo},
year = {2011},
month = {02},
pages = {},
title = {Correlation of financial markets in times of crisis},
volume = {391},
journal = {Physica A: Statistical Mechanics and its Applications},
doi = {10.1016/j.physa.2011.07.023}
}

@ARTICLE{Utsugi2004-ps,
  title     = "Random matrix theory analysis of cross correlations in financial
               markets",
  author    = "Utsugi, Akihiko and Ino, Kazusumi and Oshikawa, Masaki",
  journal   = "Phys. Rev. E Stat. Nonlin. Soft Matter Phys.",
  publisher = "American Physical Society (APS)",
  volume    =  70,
  number    = "2 Pt 2",
  pages     = "026110",
  month     =  aug,
  year      =  2004,
  copyright = "http://link.aps.org/licenses/aps-default-license",
  language  = "en"
}

@article{masuda2025introduction,
  title={Introduction to correlation networks: Interdisciplinary approaches beyond thresholding},
  author={Masuda, Naoki and Boyd, Zachary M and Garlaschelli, Diego and Mucha, Peter J},
  journal={Physics reports},
  volume={1136},
  pages={1--39},
  year={2025},
  publisher={Elsevier}
}

@ARTICLE{Potters2005-kx,
  title         = "Financial applications of Random Matrix Theory: Old laces
                   and new pieces",
  author        = "Potters, M and Bouchaud, J P and Laloux, L",
  month         =  jul,
  year          =  2005,
  archivePrefix = "arXiv",
  primaryClass  = "physics.data-an",
  eprint        = "physics/0507111",
  journal       = {...}
}

@article{Lombardi2016,
  title = {Nonparametric $k$-nearest-neighbor entropy estimator},
  author = {Lombardi, Damiano and Pant, Sanjay},
  journal = {Phys. Rev. E},
  volume = {93},
  issue = {1},
  pages = {013310},
  numpages = {12},
  year = {2016},
  month = {Jan},
  publisher = {American Physical Society},
  doi = {10.1103/PhysRevE.93.013310},
  url = {https://link.aps.org/doi/10.1103/PhysRevE.93.013310}
}

@BOOK{Lazarsfeld1968,
  title   = "Latent Structure Analysis",
  author  = "Lazarsfeld, P F and Henry, N W",
  year    =  1968,
  address = "Houghton Mifflin; New York",
  publisher = {...}
}

@book{Hagenaars2002,
title = "Applied Latent Class Analysis",
editor = "J.A.P. Hagenaars and A.L. McCutcheon",
note = "Pagination: 454",
year = "2002",
language = "English",
isbn = "0521594510",
publisher = "Cambridge University Press",
address = "United Kingdom",
}

@ARTICLE{Goodman1974,
  title     = "Exploratory latent structure analysis using both identifiable
               and unidentifiable models",
  author    = "Goodman, Leo A",
  journal   = "Biometrika",
  publisher = "Oxford University Press (OUP)",
  volume    =  61,
  number    =  2,
  pages     = "215--231",
  year      =  1974
}

@BOOK{Collins2009,
  title     = "Latent class and latent transition analysis",
  author    = "Collins, Linda M and Lanza, Stephanie T",
  publisher = "Wiley-Blackwell",
  series    = "Wiley Series in Probability and Statistics",
  month     =  nov,
  year      =  2009,
  address   = "Hoboken, NJ",
  language  = "en"
}

@ARTICLE{Nylund2007,
  title     = "Deciding on the number of classes in latent class analysis and
               growth mixture modeling: A Monte Carlo simulation study",
  author    = "Nylund, Karen L and Asparouhov, Tihomir and Muth{\'e}n, Bengt O",
  journal   = "Struct. Equ. Modeling",
  publisher = "Informa UK Limited",
  volume    =  14,
  number    =  4,
  pages     = "535--569",
  month     =  oct,
  year      =  2007
}

@INCOLLECTION{Vermunt2002-xh,
  title     = "Latent Class Cluster Analysis",
  booktitle = "Applied Latent Class Analysis",
  author    = "Vermunt, Jeroen K and Magidson, Jay",
  editor    = "Hagenaars, Jacques A and McCutcheon, Allan L",
  publisher = "Cambridge University Press",
  pages     = "89--106",
  month     =  jun,
  year      =  2002,
  address   = "Cambridge"
}

@ARTICLE{Oberski2016-fs,
  title     = "Beyond the number of classes: separating substantive from
               non-substantive dependence in latent class analysis",
  author    = "Oberski, D L",
  journal   = "Adv. Data Anal. Classif.",
  publisher = "Springer Science and Business Media LLC",
  volume    =  10,
  number    =  2,
  pages     = "171--182",
  month     =  jun,
  year      =  2016,
  language  = "en"
}

@ARTICLE{Sinha2021-pu,
  title     = "Practitioner's guide to latent class analysis: Methodological
               considerations and common pitfalls",
  author    = "Sinha, Pratik and Calfee, Carolyn S and Delucchi, Kevin L",
  journal   = "Crit. Care Med.",
  publisher = "Ovid Technologies (Wolters Kluwer Health)",
  volume    =  49,
  number    =  1,
  pages     = "e63--e79",
  month     =  jan,
  year      =  2021,
  language  = "en"
}

@ARTICLE{Weller2020-nk,
  title     = "Latent class analysis: A guide to best practice",
  author    = "Weller, Bridget E and Bowen, Natasha K and Faubert, Sarah J",
  journal   = "J. Black Psychol.",
  publisher = "SAGE Publications",
  volume    =  46,
  number    =  4,
  pages     = "287--311",
  month     =  may,
  year      =  2020,
  language  = "en"
}

@BOOK{Thurstone1947-gx,
  title     = "Multiple factor analysis",
  author    = "Thurstone, L L",
  publisher = "University of Chicago Press",
  year      =  1947,
  address   = "Chicago"
}

@ARTICLE{Fabrigar1999-ol,
  title     = "Evaluating the use of exploratory factor analysis in
               psychological research",
  author    = "Fabrigar, Leandre R and Wegener, Duane T and MacCallum, Robert C
               and Strahan, Erin J",
  journal   = "Psychol. Methods",
  publisher = "American Psychological Association (APA)",
  volume    =  4,
  number    =  3,
  pages     = "272--299",
  month     =  sep,
  year      =  1999,
  language  = "en"
}

@ARTICLE{Joreskog1969-my,
  title     = "A general approach to confirmatory maximum likelihood factor
               analysis",
  author    = "J{\"o}reskog, K G",
  journal   = "Psychometrika",
  publisher = "Cambridge University Press (CUP)",
  volume    =  34,
  number    =  2,
  pages     = "183--202",
  month     =  jun,
  year      =  1969,
  language  = "en"
}

@BOOK{Gorsuch1983-aw,
  title     = "Factor Analysis",
  author    = "Gorsuch, Richard L",
  publisher = "Lawrence Erlbaum Associates",
  edition   =  2,
  month     =  nov,
  year      =  1983,
  address   = "Mahwah, NJ"
}

@BOOK{McDonald2013-hd,
  title     = "Test theory",
  author    = "McDonald, Roderick P",
  publisher = "Psychology Press",
  month     =  jun,
  year      =  2013,
  address   = "London, England"
}

@ARTICLE{Clark1995-gw,
  title     = "Constructing validity: Basic issues in objective scale
               development",
  author    = "Clark, Lee Anna and Watson, David",
  journal   = "Psychol. Assess.",
  publisher = "American Psychological Association (APA)",
  volume    =  7,
  number    =  3,
  pages     = "309--319",
  month     =  sep,
  year      =  1995,
  language  = "en"
}

@ARTICLE{Liu2008-sigclust,
  title     = "Statistical Significance of Clustering for High-Dimension, Low-Sample Size Data",
  author    = "Liu, Yufeng and Hayes, David Neil and Nobel, Andrew and Marron, J. S.",
  journal   = "Journal of the American Statistical Association",
  volume    =  103,
  number    =  483,
  pages     = "1281--1293",
  year      =  2008,
  publisher = "Taylor \& Francis",
  doi       = "10.1198/016214508000000454"
}

@ARTICLE{Tibshirani2001-gap,
  title     = "Estimating the number of clusters in a data set via the gap statistic",
  author    = "Tibshirani, Robert and Walther, Guenther and Hastie, Trevor",
  journal   = "Journal of the Royal Statistical Society: Series B (Statistical Methodology)",
  volume    =  63,
  number    =  2,
  pages     = "411--423",
  year      =  2001,
  publisher = "Wiley",
  doi       = "10.1111/1467-9868.00293"
}

@ARTICLE{Mair2014-gop,
  title     = "The grand old party -- a party of values?",
  author    = "Mair, Patrick and Rusch, Thomas and Hornik, Kurt",
  journal   = "SpringerPlus",
  volume    =  3,
  pages     = "697",
  year      =  2014,
  publisher = "Springer",
  doi       = "10.1186/2193-1801-3-697"
}

@ARTICLE{Rosenstrom2017-ruo,
  title     = "A Parsimonious Explanation of the Resilient, Undercontrolled, and Overcontrolled Personality Types",
  author    = "Rosenstr{\"o}m, Tom and Jokela, Markus",
  journal   = "European Journal of Personality",
  volume    =  31,
  number    =  6,
  pages     = "658--668",
  year      =  2017,
  publisher = "SAGE",
  doi       = "10.1002/per.2117"
}

@misc{openpsychometrics_rawdata,
  author       = {{Open-Source Psychometrics Project}},
  title        = {Raw data from online personality tests},
  howpublished = {\url{https://openpsychometrics.org/_rawdata/}},
  year         = {2019},
  note         = {Last updated November 2019}
}

@article{goldberg1992bigfive,
  author    = {Goldberg, Lewis R.},
  title     = {The development of markers for the {Big-Five} factor structure},
  journal   = {Psychological Assessment},
  volume    = {4},
  number    = {1},
  pages     = {26--42},
  year      = {1992},
  publisher = {American Psychological Association},
  doi       = {10.1037/1040-3590.4.1.26}
}

@article{lee2004hexaco,
  author    = {Lee, Kibeom and Ashton, Michael C.},
  title     = {Psychometric properties of the {HEXACO} personality inventory},
  journal   = {Multivariate Behavioral Research},
  volume    = {39},
  number    = {2},
  pages     = {329--358},
  year      = {2004},
  publisher = {Taylor \& Francis},
  doi       = {10.1207/s15327906mbr3902_8}
}

@article{jones2014sd3,
  author    = {Jones, Daniel N. and Paulhus, Delroy L.},
  title     = {Introducing the short dark triad ({SD3}): A brief measure of dark personality traits},
  journal   = {Assessment},
  volume    = {21},
  number    = {1},
  pages     = {28--41},
  year      = {2014},
  publisher = {SAGE Publications},
  doi       = {10.1177/1073191113514105}
}

@book{christie1970mach,
  author    = {Christie, Richard and Geis, Florence L.},
  title     = {Studies in {Machiavellianism}},
  publisher = {Academic Press},
  address   = {New York},
  year      = {1970}
}

@article{strathman1994cfcs,
  author    = {Strathman, Alan and Gleicher, Faith and Boninger, David S. and Edwards, C. Scott},
  title     = {The consideration of future consequences: Weighing immediate and distant outcomes of behavior},
  journal   = {Journal of Personality and Social Psychology},
  volume    = {66},
  number    = {4},
  pages     = {742--752},
  year      = {1994},
  publisher = {American Psychological Association},
  doi       = {10.1037/0022-3514.66.4.742}
}

@article{hendin1997hsns,
  author    = {Hendin, Holly M. and Cheek, Jonathan M.},
  title     = {Assessing hypersensitive narcissism: A reexamination of {Murray's} narcissism scale},
  journal   = {Journal of Research in Personality},
  volume    = {31},
  number    = {4},
  pages     = {588--599},
  year      = {1997},
  publisher = {Elsevier},
  doi       = {10.1006/jrpe.1997.2204}
}

@incollection{snell1995msscq,
  author    = {Snell, William E., Jr.},
  title     = {The multidimensional sexual self-concept questionnaire},
  booktitle = {Sexuality-Related Measures: A Compendium},
  editor    = {Davis, Clive M. and Yarber, William L. and Bauserman, Robert and Schreer, George and Davis, Sandra L.},
  edition   = {2nd},
  publisher = {Sage Publications},
  address   = {Thousand Oaks, CA},
  pages     = {509--513},
  year      = {1995}
}

@book{altemeyer1981rwas,
  author    = {Altemeyer, Bob},
  title     = {Right-wing authoritarianism},
  publisher = {University of Manitoba Press},
  address   = {Winnipeg},
  year      = {1981}
}

@article{brotherton2013gcbs,
  author    = {Brotherton, Robert and French, Christopher C. and Pickering, Alan D.},
  title     = {Measuring belief in conspiracy theories: The generic conspiracist beliefs scale},
  journal   = {Frontiers in Psychology},
  volume    = {4},
  pages     = {279},
  year      = {2013},
  publisher = {Frontiers Media SA},
  doi       = {10.3389/fpsyg.2013.00279}
}

@article{mirels1971pwe,
  author    = {Mirels, Herbert L. and Garrett, James B.},
  title     = {The {Protestant} ethic as a personality variable},
  journal   = {Journal of Consulting and Clinical Psychology},
  volume    = {36},
  number    = {1},
  pages     = {40--44},
  year      = {1971},
  publisher = {American Psychological Association},
  doi       = {10.1037/h0030477}
}

@article{lovibond1995dass,
  author    = {Lovibond, Peter F. and Lovibond, Sydney H.},
  title     = {The structure of negative emotional states: Comparison of the {Depression Anxiety Stress Scales (DASS)} with the {Beck Depression} and {Anxiety Inventories}},
  journal   = {Behaviour Research and Therapy},
  volume    = {33},
  number    = {3},
  pages     = {335--343},
  year      = {1995},
  publisher = {Elsevier},
  doi       = {10.1016/0005-7967(94)00075-U}
}

@article{KozachenkoLeonenko1987,
  author  = {Kozachenko, L. F. and Leonenko, N. N.},
  title   = {Sample estimate of the entropy of a random vector},
  journal = {Problems of Information Transmission},
  volume  = {23},
  number  = {2},
  pages   = {95--101},
  year    = {1987}
}

@article{MarchenkoPastur1967,
  author  = {Mar{\v c}enko, V. A. and Pastur, L. A.},
  title   = {Distribution of eigenvalues for some sets of random matrices},
  journal = {Mathematics of the USSR-Sbornik},
  volume  = {1},
  number  = {4},
  pages   = {457--483},
  year    = {1967},
  doi     = {10.1070/SM1967v001n04ABEH001994}
}

@article{PhipsonSmyth2010,
  author  = {Phipson, Belinda and Smyth, Gordon K.},
  title   = {Permutation p-values should never be zero: calculating exact p-values when permutations are randomly drawn},
  journal = {Statistical Applications in Genetics and Molecular Biology},
  volume  = {9},
  number  = {1},
  pages   = {Article 39},
  year    = {2010},
  doi     = {10.2202/1544-6115.1585}
}

@article{Borgatti2000-corepe,
  author  = {Borgatti, Stephen P. and Everett, Martin G.},
  title   = {Models of core/periphery structures},
  journal = {Social Networks},
  volume  = {21},
  number  = {4},
  pages   = {375--395},
  year    = {2000},
  doi     = {10.1016/S0378-8733(99)00019-2}
}

@article{Zhang2015-corepe,
  author  = {Zhang, Xiao and Martin, Travis and Newman, M. E. J.},
  title   = {Identification of core-periphery structure in networks},
  journal = {Physical Review E},
  volume  = {91},
  number  = {3},
  pages   = {032803},
  year    = {2015},
  doi     = {10.1103/PhysRevE.91.032803}
}

@article{scikit-learn,
  author  = {Pedregosa, F. and Varoquaux, G. and Gramfort, A. and Michel, V.
             and Thirion, B. and Grisel, O. and Blondel, M. and Prettenhofer, P.
             and Weiss, R. and Dubourg, V. and Vanderplas, J. and Passos, A.
             and Cournapeau, D. and Brucher, M. and Perrot, M. and Duchesnay, E.},
  title   = {Scikit-learn: Machine Learning in {P}ython},
  journal = {Journal of Machine Learning Research},
  volume  = {12},
  pages   = {2825--2830},
  year    = {2011}
}

@article{Traag2011-signed,
  author  = {Traag, V. A. and Bruggeman, Jeroen},
  title   = {Community detection in networks with positive and negative links},
  journal = {Physical Review E},
  volume  = {80},
  number  = {3},
  pages   = {036115},
  year    = {2009},
  doi     = {10.1103/PhysRevE.80.036115}
}

@article{albiero2006iri,
  title={Contributo all’adattamento italiano dell’Interpersonal Reactivity Index},
  author={Albiero, PAOLO and Ingoglia, SONIA and Lo Coco, Alida and others},
  journal={Testing Psicometria Metodologia},
  volume={13},
  number={2},
  pages={107--125},
  year={2006},
  publisher={Cises, Srl}
}

@article{jordan2002ei,
  title={Emotional intelligence as a moderator of emotional and behavioral reactions to job insecurity},
  author={Jordan, Peter J and Ashkanasy, Neal M and Hartel, Charmine EJ},
  journal={Academy of Management review},
  volume={27},
  number={3},
  pages={361--372},
  year={2002},
  publisher={Academy of Management Briarcliff Manor, NY 10510}
}

@article{vachon2016fixing,
  title={Fixing the problem with empathy: Development and validation of the affective and cognitive measure of empathy},
  author={Vachon, David D and Lynam, Donald R},
  journal={Assessment},
  volume={23},
  number={2},
  pages={135--149},
  year={2016},
  publisher={Sage Publications Sage CA: Los Angeles, CA}
}
 
%%%%%%%%%%%%%%%%%%%%%%%%%%%%%%%%%%%%%%%%%%%%%%%%%%%%%%%
% START OF THE APPENDIX
%%%%%%%%%%%%%%%%%%%%%%%%%%%%%%%%%%%%%%%%%%%%%%%%%%%%%%%
%\clearpage
\appendix

%============================================================
% Appendix: Generative Models for Synthetic Datasets
%============================================================

\section{Generative models for synthetic data}
\label{app:synthetics}
\noindent This appendix describes the three generative procedures used to produce the synthetic datasets in the numerical experiments.  All three follow a common design principle: a ground-truth latent structure is specified first, and observed ordinal responses are derived from it through a controlled stochastic process. Parameters such as the number of subjects $N$, items $M$, latent factors $q$, and ordinal response levels are kept as free arguments so that the same procedure can be used across all simulation conditions.

% -----------------------------------------------------------
\subsection{Simple-Structure Factor Model}
% -----------------------------------------------------------
The first generative model implements a \emph{simple-structure} linear factor model. Items are partitioned into $F$ non-overlapping groups of equal size, each associated with exactly one latent factor. \\
The first thing to generate is the \emph{loading matrix} $\mathbf{W} \in \mathbb{R}^{M \times F}$. 
For each item $j$ belonging to factor group $k$:
\begin{itemize}
  \item the primary loading $W_{jk}$ is drawn uniformly from a specified
        interval $[a_{\mathrm{prim}},\, b_{\mathrm{prim}}]$ (e.g.\ $[0.6, 1.0]$)
        and its sign is randomised (predominantly positive);
  \item the cross-loadings $W_{j,k'}$, $k' \neq k$, are drawn uniformly
        from a near-zero interval $[a_{\mathrm{cross}},\, b_{\mathrm{cross}}]$
        (e.g.\ $[-0.2, 0.2]$).
\end{itemize}

\noindent Item-specific noise variances (\emph{unique variances} are drawn from a log-normal distribution:
\[
  \psi_j = \bar{\psi}\,\exp\!\bigl(\sigma_\psi \,\varepsilon_j\bigr),
  \qquad \varepsilon_j \sim \mathcal{N}(0,1),
\]
where $\bar{\psi}$ controls the average noise level and $\sigma_\psi$ its
spread across items. \\

\noindent To sample a dataset of N subjects:
\begin{enumerate}
  \item Sample latent scores: $\mathbf{Z} \sim \mathcal{N}(\mathbf{0},\,\mathbf{I}_F)$,
        independently for each subject.
  \item Compute continuous manifest variables:
        $\mathbf{X}^* = \mathbf{Z}\mathbf{W}^\top + \boldsymbol{\varepsilon}$,
        where $\varepsilon_{ij} \sim \mathcal{N}(0, \psi_j)$.
  \item Discretise each column of $\mathbf{X}^*$ into $q$ ordered categories via thresholds defined in subsection \ref{sub-app:discretisation}, yielding ordinal responses $X_{ij} \in \{0,1,\ldots,q-1\}$.
\end{enumerate}

% ----------------------------------------------------------
\subsection{Bifactor Model with a Dominant General Factor}
% ----------------------------------------------------------
\noindent The second model generates data from a \emph{bifactor} (or hierarchical) structure in which one latent dimension --- the \emph{general factor} --- loads on all items, while $F-1$ group-specific factors each load on a disjoint subset of items.\\
A key design parameter is $r \in (0,1)$, the proportion of total variance explained by the general factor:
\[
  r = \frac{\sum_j W_{j,\mathrm{dom}}^2}
           {\sum_j \sum_k W_{jk}^2 + \sum_j \psi_j}.
\]
Rather than choosing loadings heuristically, the magnitudes are \emph{analytically calibrated} to hit the target $r$ exactly.\\
First we sample $\tilde{\mathbf{w}}_{\mathrm{dom}} \in \mathbb{R}^M$ and $\widetilde{\mathbf{W}}_{\mathrm{spec}} \in \mathbb{R}^{M \times (F-1)}$ similarly to the equal-factors case.  
Then, we define:
\[
  A = \|\tilde{\mathbf{w}}_{\mathrm{dom}}\|^2,\qquad
  B = \|\widetilde{\mathbf{W}}_{\mathrm{spec}}\|_f^2,\qquad
  P = \sum_j \psi_j;
\]
such that:
\[
  r = \frac{\alpha A}
           {\alpha A + \beta B + P}.
\]
Setting the specific-factor scale to $\beta = 1$ and solving for the dominant-factor scale $\alpha$ yields:
\[
  \alpha = \sqrt{\frac{r\,(B + P)}{(1-r)\,A}},
\]
so that $\mathbf{w}_{\mathrm{dom}} = \alpha\,\tilde{\mathbf{w}}_{\mathrm{dom}}$ and $\mathbf{W}_{\mathrm{spec}} = \widetilde{\mathbf{W}}_{\mathrm{spec}}$.\\

\noindent The sampling step and the ordinal discretisation are identical to the equal-factors case.

% ----------------------------------------------------------
\subsection{Ordinal Discretisation}
\label{sub-app:discretisation}
% ----------------------------------------------------------
\noindent The continuous-to-ordinal mapping used in the FA-based models follows a fixed-threshold scheme that preserves the bell-shaped marginals expected under Gaussian latent variables. 
Given a continuous variable $x^*$ and $q$ ordinal categories, the thresholds $\tau_1 < \tau_2 < \cdots < \tau_{q-1}$ are placed at equally spaced positions over the interval $[\bar{x} - h\hat{\sigma},\, \bar{x} + h\hat{\sigma}]$, where $\bar{x}$ and $\hat{\sigma}$ are the empirical mean and standard deviation of all continuous values, and $h$ (typically $h = 2.5$) controls how many standard deviations the threshold range spans.
The ordinal response is then:
\[
  X_{ij} = \sum_{\ell=1}^{q-1} \mathbf{1}[x^*_{ij} > \tau_\ell].
\]
This design choice produces response distributions that are approximately normal-shaped rather than uniform, which is more realistic for psychometric data.

% -----------------------------------------------------------
\subsection{Mixture Model with Community Structure}
% -----------------------------------------------------------

\noindent The third generative model is designed for settings where both a \emph{community structure} over subjects and a \emph{factor structure} over items are present simultaneously. 
It does not rely on a linear Gaussian factor model; instead, it uses a discrete-mixture / binomial generative mechanism.\\

\noindent There are $K$ communities (latent subject clusters) with mixture weights $\boldsymbol{\theta} = (\theta_1,\ldots,\theta_K)$, and $F$ item groups (factors) that partition the $M$ items. 
The key quantity is the $(K \times F)$ mean-response matrix $\boldsymbol{\mu}$, where $\mu_{kf}$ is the expected response of a community-$k$ subject on an item from factor group $f$, expressed on the scale $\{0,\ldots,q-1\}$.\\

\noindent  If $\boldsymbol{\mu}$ is not provided by the user, a \emph{cyclic dominance} pattern is constructed: community $k$ is dominant on factor group $(k \bmod F)$ and suppressed on the remaining groups,
\[
  \mu_{kf} =
  \begin{cases}
    \mu_0 + \delta & \text{if } f = k \bmod F, \\[4pt]
    \mu_0 - \dfrac{\delta}{F-1} & \text{otherwise},
  \end{cases}
\]
where $\mu_0$ is a global baseline and $\delta$ is the signal strength.
A mixing parameter $\rho \in [0,1]$ interpolates between this structured matrix and a flat (all-equal) baseline: $\boldsymbol{\mu} \leftarrow (1-\rho)\,\boldsymbol{\mu} + \rho\,\mu_0\,\mathbf{1}$. \\

\noindent The success probability for item $j$ (belonging to group $f$) under community $k$ is:
\[
  \nu_{kj} = \frac{\mu_{kf}}{q-1}.
\]
Optional heterogeneity at the item level is introduced by perturbing $\nu_{kj}$ in logit space:
\[
  \tilde{\nu}_{kj} = \sigma\!\left(\mathrm{logit}(\nu_{kj}) + \eta_{kj}\right),
  \qquad \eta_{kj} \sim \mathcal{N}(0,\sigma_{\mathrm{item}}^2).
\]
Similarly, within-community subject variation is added per subject in the same logit space with scale $\sigma_{\mathrm{subj}}$. \\

\noindent The data generation process goes as follow:
\begin{enumerate}
  \item Assign community: $c_i \sim \mathrm{Categorical}(\boldsymbol{\theta})$.
  \item For each subject $i$, compute personalised probabilities
        $p_{ij} = \sigma(\mathrm{logit}(\nu_{c_i,j}) + \xi_{ij})$,
        $\xi_{ij} \sim \mathcal{N}(0,\sigma_{\mathrm{subj}}^2)$.
  \item Draw responses: $X_{ij} \sim \mathrm{Binomial}(q-1,\, p_{ij})$.
\end{enumerate}

\section{Latent Class Analysis limitations}
\label{app:lca}
\noindent Latent Class Analysis (LCA) \citep{lazarsfeld1968,goodman1974,Hagenaars2002,
Collins2009} is the standard statistical tool for clustering respondents in psychometrics. 
It models the observed response vector $\mathbf{x}_i \in \{0,\ldots,R-1\}^M$ as drawn from a finite mixture of $K$ discrete latent classes:
\begin{equation}
    P(\mathbf{x}_i) = \sum_{k=1}^{K} \pi_k \prod_{j=1}^{M}
    P(x_{ij} \mid c_i = k),
    \label{eq:lca}
\end{equation}
where $\pi_k$ is the prevalence of class $k$ and the within-class independence assumption factorizes the within-class response distribution across items.

\noindent Despite being the most established tool for class separation in psychometrics, it suffers from several well-documented limitations.\\

\noindent First, the number of classes $K$ must be specified \emph{a priori}; selection via information criteria such as BIC and AIC can be informative, but in practice it has proved not fully reliable. Monte Carlo studies have shown that AIC systematically over-extracts classes, while BIC, although generally superior, becomes less reliable when the true number of classes is small, especially under population imbalance \citep{Nylund2007}.
The Bootstrap Likelihood Ratio Test (BLRT) is another popular selection criterion: it proved to be more reliable than BIC, although it is computationally more expensive and, being model-dependent, does not provide a principled way to distinguish whether additional classes reflect genuine population heterogeneity or artifacts of model misspecification \citep{Nylund2007, Weller2020-nk}.\\
\noindent Second, the assumption that items are mutually independent given class membership is violated by design in psychometric data: items loading on shared latent factors remain correlated even after conditioning on class membership \citep{Vermunt2002-xh}. This residual dependence biases the estimate of class prevalences, item-response probabilities, and class memberships \citep{Sinha2021-pu}. Critically, unmodelled local dependence manifests as global misfit, which practitioners often address by increasing $K$ rather than modelling the dependence directly \citep{Oberski2016-fs}.\\
\noindent Third, LCA is sensitive to class-size imbalance and to extreme item profiles: when $K$ is overspecified, the algorithm tends to fragment large homogeneous groups into spurious subclasses, yielding solutions that are statistically identifiable but substantively uninterpretable \citep{Sinha2021-pu,Weller2020-nk}. \\

\begin{figure*}[t]
    \centering
    \includegraphics[width=0.9\linewidth]{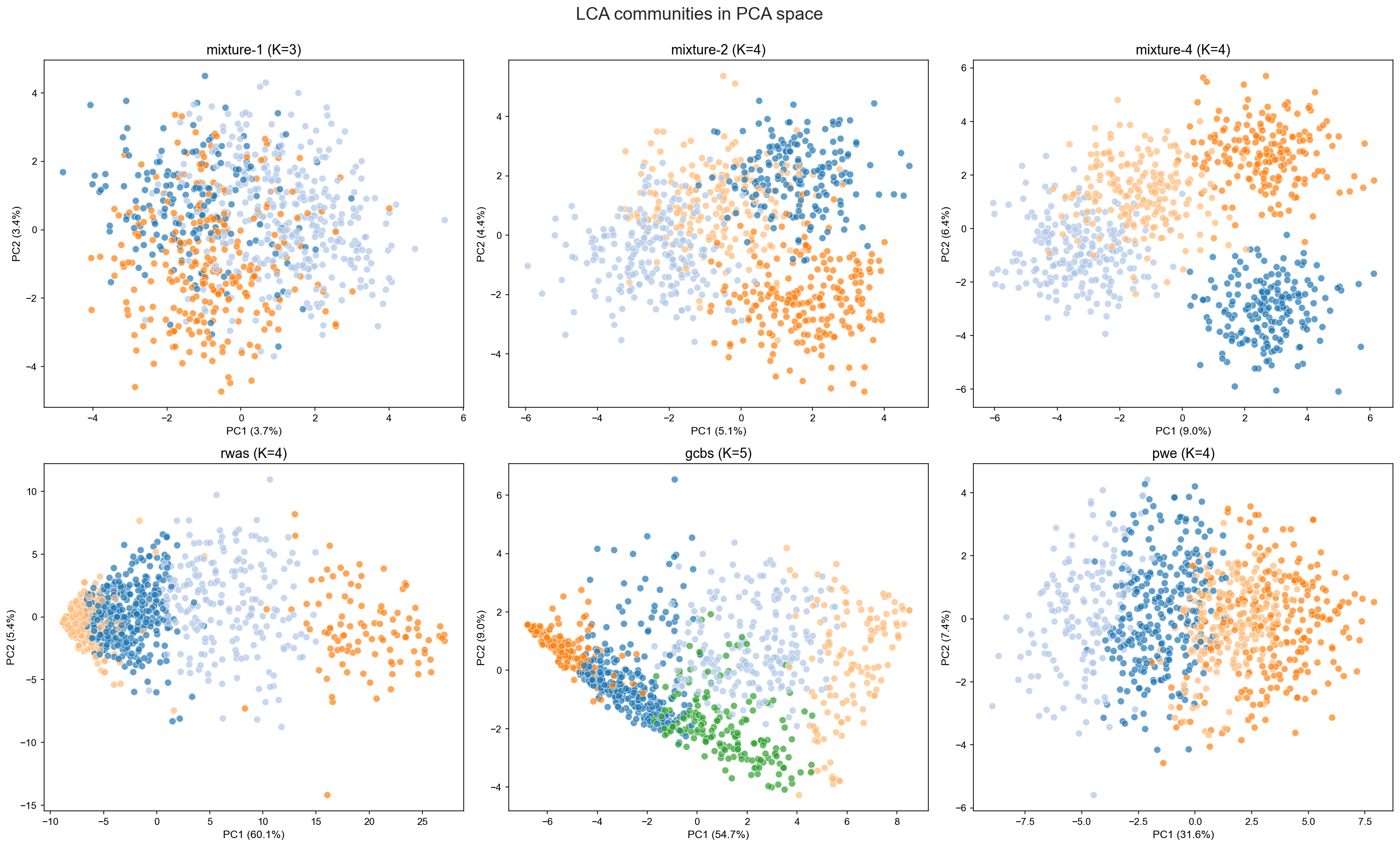}
    \caption{LCA partitions projected onto the first two principal components (PCA) for six datasets used in the comparison of Figure~\ref{fig:lca-vs-cd}.
    \textbf{Top row}: three synthetic mixtures generated with the item-response probability model (Section~\ref{app:synthetics}, $N=800$, $M=60$, $K=4$, $q=6$) at increasing signal strengths ($\xi \approx 0.1$, $0.35$, $0.85$ for mixture-1, mixture-2, mixture-4 respectively).
    \textbf{Bottom row}: three real datasets (RWAS, GCBS, PWE), each subsampled to $N \leq 1000$ subjects.
    In all panels, $K$ is selected via Parametric Bootstrap Likelihood Ratio Test (BLRT) and each colour represents one latent class.
    On synthetic data, LCA recovers coherent partitions only at high signal strength; at low-to-intermediate signal the class boundaries are arbitrary.
    On real datasets, LCA systematically segments subjects along the first principal component---the dominant axis of response-level variation---rather than identifying distinct profile subgroups.}
    \label{fig:lca-demo}
\end{figure*}

\begin{figure*}
    \centering
    \includegraphics[width=0.9\linewidth]{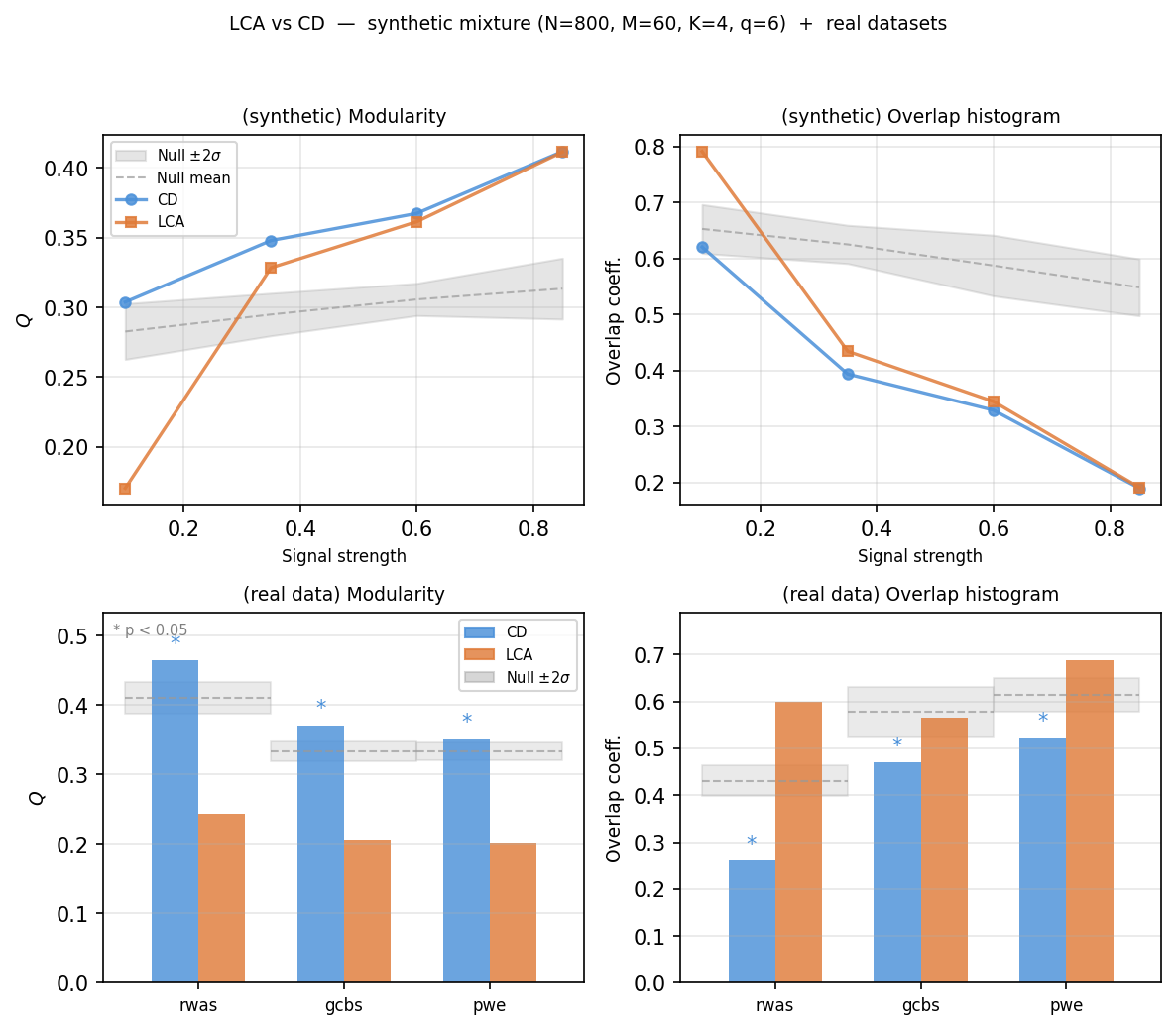}
    \caption{Comparison of LCA and community detection (CD) partitions on synthetic and real datasets, evaluated with the paper's validation metrics. \textbf{Top row}: modularity $Q$ (left) and KDE overlap coefficient between within- and between-community similarity distributions (right), as a function of the mixture signal strength~$\xi$.
    Synthetic data are generated with $N=800$ subjects, $M=60$ items, $K=4$ communities, and $q=6$ Likert levels using the item-response probability model of Section~\ref{app:synthetics}.
    CD uses FA scores (4 factors) $\to$ negative squared Euclidean similarity $\to$ market mode removal $\to$ signed Leiden; LCA uses StepMix with categorical measurement and $K$ selected by BLRT. Both partitions are evaluated on the same similarity matrix.
    The gray band shows the mean $\pm 2\sigma$ of $n_\mathrm{rep}=20$ resampled null realizations (column-wise resampling of FA scores, full pipeline rerun).
    \textbf{Bottom row}: same metrics on three real datasets (RWAS, GCBS, PWE), each subsampled to $N \leq 1000$ subjects; the gray band is the per-dataset resampling null. Asterisks mark partitions with $p < 0.05$ against the null (one-tailed).}
    \label{fig:lca-vs-cd}
\end{figure*}

\noindent Figure~\ref{fig:lca-demo} shows the partitions found by LCA on six of the seven datasets used in the quantitative comparison of Figure~\ref{fig:lca-vs-cd}: three (of four) synthetic mixtures with increasing cluster separation and three real datasets (RWAS, GCBS, PWE). The number of classes $K$ is selected via BLRT in all cases.\\
\noindent On synthetic data, the quality of the recovered partition tracks the signal strength. At low signal the four classes are almost entirely overlapping in PCA space, and the assignment is largely arbitrary. At intermediate signal, partial separation begins to emerge. Only at high signal does LCA recover a partition that is visually coherent, with classes occupying reasonably distinct regions of the PC1--PC2 plane. This confirms that LCA can recover mixture structure, but only when communities are strongly separated.\\
\noindent The pattern on real datasets is qualitatively different. For all three datasets, LCA partitions the subjects into classes that follow the main axis of variation (PC1), producing assignments that resemble a segmentation of the PC1 range into roughly equal-density intervals. \\
%The resulting partition has low modularity and high within-/between-community overlap, as confirmed by the quantitative metrics in Figure~\ref{fig:lca-vs-cd}.

\noindent Figure~\ref{fig:lca-vs-cd} provides a quantitative comparison between LCA and the proposed community detection (CD) pipeline on the same data, using as example two validation metrics adopted throughout the paper: the modularity $Q$ and the overlap coefficient between the within- and between-community similarity distributions (Section~\ref{sec:significance}).
On synthetic data, both methods are evaluated as a function of the mixture signal strength~$\xi$; on real datasets, results are shown as bar charts. In both cases, the resampling null band is superimposed.
At weak signal ($\xi \lesssim 0.4$), the proposed community detection algorithm (CD) already returns partitions with $Q$ significantly above the resampling null, while the LCA partition has substantially lower modularity---close to or within the null band---and higher overlap between the within- and between-community similarity distributions.
The two methods become comparable only when communities are strongly separated ($\xi \gtrsim 0.6$), at which point the mixture structure is detectable by essentially any clustering algorithm.
The overlap histogram shows the same ordering across all tested signal levels: CD consistently achieves better within/between separation than LCA. \\
In other words, on datasets in which population heterogeneity is expected, LCA does not recover the same partition structure as CD, and the partition it does find is less well-separated according to the network-level metrics.

%------------------------------------------------------
\section{The item-space artifact}
\label{app:item-space-art}

\begin{figure*}[t]
    \centering
    \includegraphics[width=0.9\linewidth]{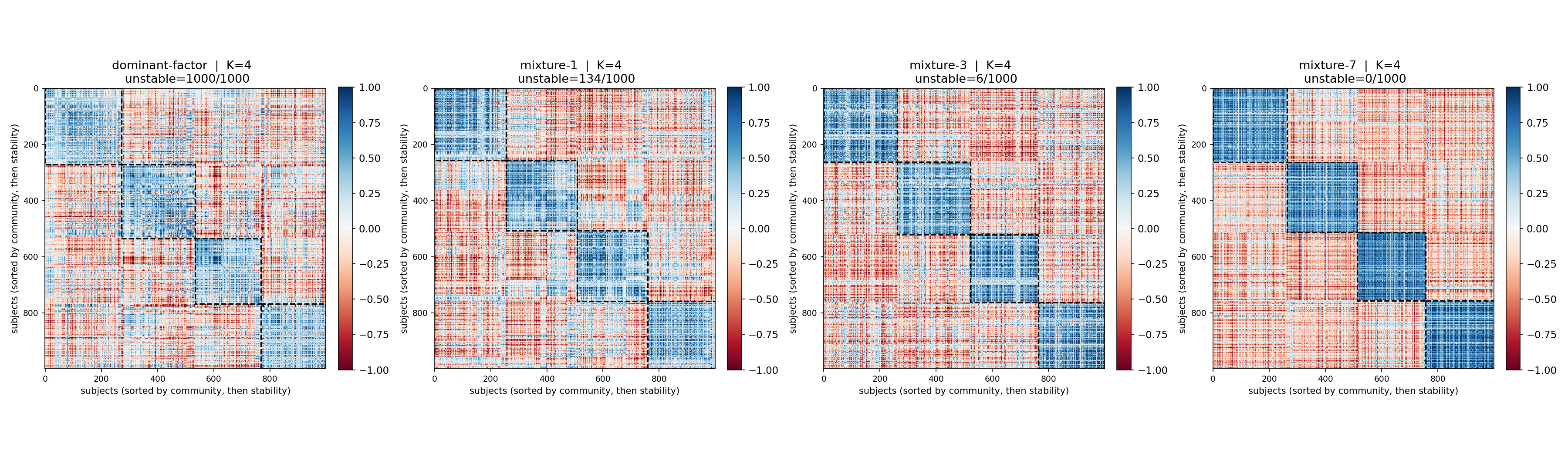}
    \includegraphics[width=0.9\linewidth]{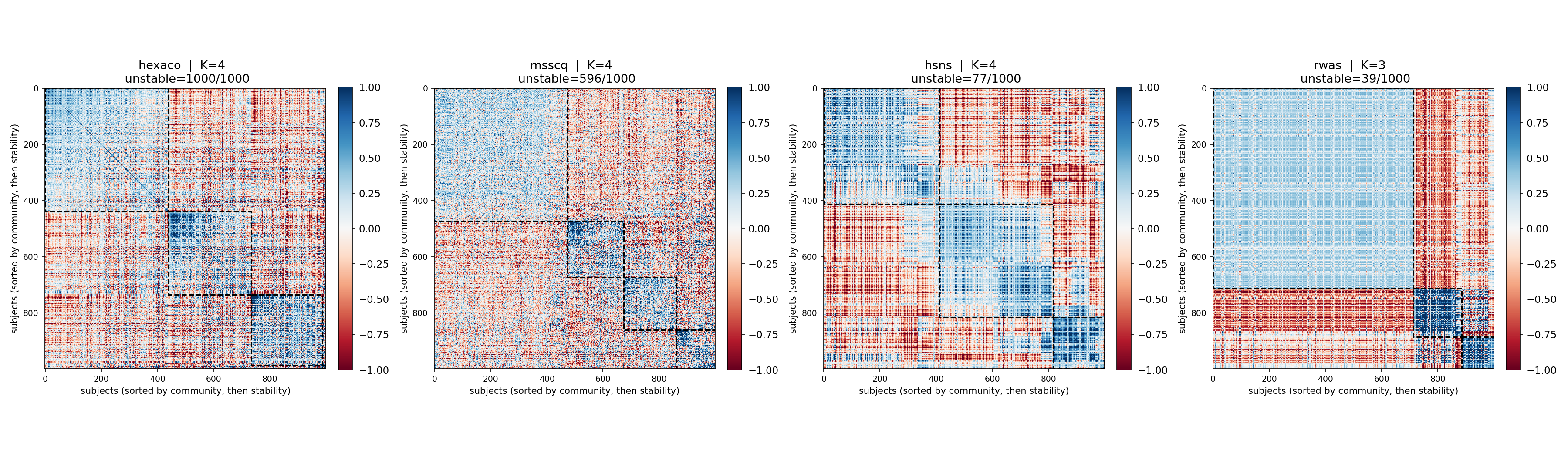}
    \vspace{-5mm}
    \caption{Similarity networks cleaned of the market mode and normalized for the max value.
    Nodes are grouped by consensus labels and within communities they are ranked from most to least stable.
    Blue (red) entries indicates positive (negative) similarities.
    The borders of the clusters detected are highlighted by black dashed squares.
    Fraction of unstable assignments is shown in each panel title.
    Top panel: synthetic datasets. Bottom panel: real datasets.}
    \label{fig:sim-heat}
\end{figure*}

\noindent A central concern is whether modularity maximisation produces non-trivial partitions even on data drawn from a single Gaussian with no community structure.\\
The choice of similarity representation has a decisive effect on both false-positive suppression and recovery power.
In the $M$-dimensional item space, $S^{(\mathrm{IS})}$ conflates genuine between-subject
differences with the redundancy induced by items that share a latent factor: pairs of subjects who differ only in their loading on a single factor appear highly dissimilar across all $M$ items simultaneously, inflating the inter-cluster contrast and distorting the noise bulk of the spectrum.\\
This leads to a strong separation of subjects into two groups even when the dataset is known to be generated from a single multivariate Gaussian distribution.\\
Plotting the dataset in the plane of the first two principal components (2-PC plane), the separation occurs along the principal axis of variation, splitting the dataset into the positive and negative halves along this axis (see Figure~\ref{fig:demo-comm}).\\
The apparent block-diagonal structure in the reordered similarity matrix is an artifact of the latent factor correlation structure, not of genuine subject-level subgroups, as visible in Figure~\ref{fig:sim-heat}.

\noindent The FA-space similarity $S^{(\mathrm{FS})}$ mitigates the redundancy in the full vectors by collapsing all highly correlated items onto a much lower number of coordinates in the transformed space.
This representation also tends to separate the two generative scenarios more cleanly: Gaussian datasets produce factor-score distributions that are approximately Gaussian along every axis, while mixture datasets produce multimodal or more structured marginal distributions.
As shown in Fig.~\ref{fig:demo-comm} (bottom row), the FA-space projection with market-mode removal successfully avoids the spurious two-community split in the $H_0$ datasets while recovering the planted partition in the well-separated mixture.

% =============================================================================
% APPENDIX: Sensitivity to the number of FA components $F$
% =============================================================================
\section{Sensitivity to the number of FA components and the items-per-factor ratio}
\label{app:F-sensitivity}
In the pipeline, one of the free parameter is $F$, the number of dimensions used to project subjects into the low-dimensional score space before computing the similarity.
A possible principled choice for $F$ could be the number of latent constructs explored in the questionnaire, but in practice other choices could be made.
We therefore ask two questions: (i) does the pipeline remain reliable when $F$ deviates from this value, identified with $F_{\mathrm{true}}$? and (ii) is there a structural property of the dataset that governs detectability independently of $F$?

\noindent We run the pipeline on a systematic grid of 30 synthetic structures, varying both the number of items $M \in \{10, 20, 30, 40, 50, 100, 150, 200\}$ and the number of planted latent constructs $F_{\mathrm{true}} \in \{1, \ldots, 20\}$.
For each structure we generate three datasets -- \textit{dominant-factor} ($H_0$), \textit{mix-low} ($\xi=0.2$, $H_1$ weak), and \textit{mix-high} ($\xi=0.9$, $H_1$ strong) -- and test $F \in [F_{\mathrm{true}}-2,\, F_{\mathrm{true}}+2]$ (minimum 1).

\begin{figure*}[t]
    \centering
    \includegraphics[width=\linewidth]{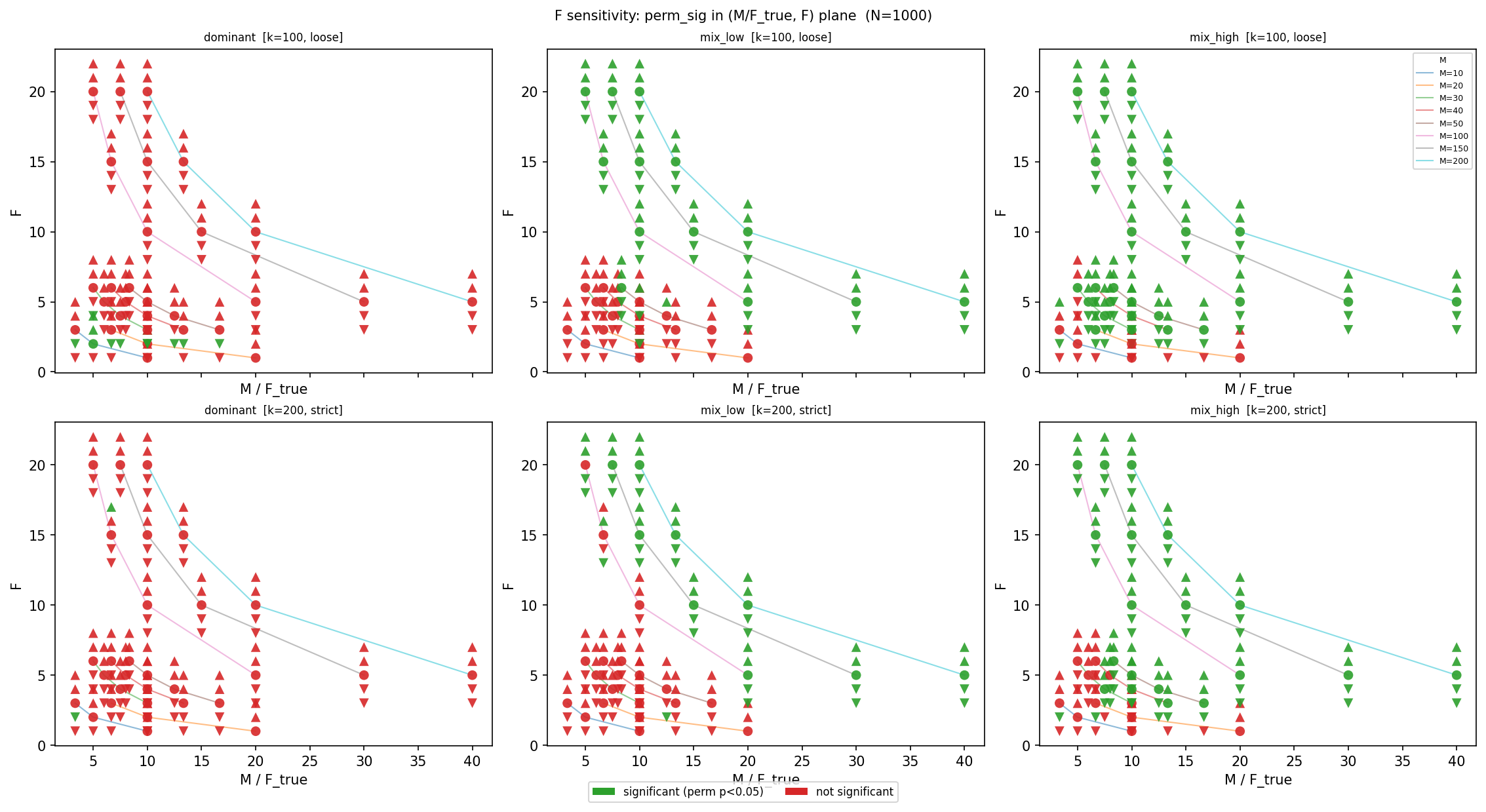}
    \caption{Each point is a tested $(M, F_{\mathrm{true}}, F)$ combination, plotted at coordinates $(M/F_{\mathrm{true}},\, F)$ and coloured by permutation significance. Marker shape indicates $F$ relative to $F_{\mathrm{true}}$: $\triangledown$ underfit, $\circ$ exact, $\triangle$ overfit. Lines connect the $F = F_{\mathrm{true}}$ points for the same $M$. The strict criterion (bottom row) eliminates all false positives for the dominant-factor dataset.}
    \label{fig:F-scatter}
\end{figure*}

\noindent To avoid reproducing the whole corpus of results in the paper, we limit the study of the clustered structure of the dataset to the significance of the cloud entropy alone.
We evaluate the cloud entropy at two values of $k$ corresponding to the loose ($k_{\mathrm{low}}$) and strict ($k_{\mathrm{high}}$) criteria defined in Eq.~\eqref{eq:k-rules} (Appendix~\ref{app:knn-sensitivity}). The operating points used here were calibrated for $N=1000$ and amount to $k_{\mathrm{low}}=0.1\,N=100$ and $k_{\mathrm{high}}=0.8\,\overline{n}_{\mathrm{comm}} = 200$, computed using $\overline{n}_{\mathrm{comm}}$ on the ground-truth partition of the mixtures.
The significance is assessed by permutation test ($n_{\mathrm{rep}}=30$). 
Results are shown in Figure~\ref{fig:F-scatter}. \\

\noindent The governing quantity turns out to be the ratio $r = M/F_{\mathrm{true}}$, i.e.\ the number of items per latent factor.
Three regimes emerge consistently across all structures:
\begin{itemize}
    \item $r < 7$: the pipeline does not detect mix-high reliably, and the dominant-factor dataset produces false positives under the loose criterion. This regime should be avoided.
    \item $7 \leq r < 15$: mix-high is detectable with the loose criterion; the strict criterion still misses it. The dominant-factor dataset can produce occasional false positives with the loose criterion.
    \item $r \geq 15$: both criteria give reliable detection of mix-high and no false positives for dominant-factor. Mix-low remains hard to detect unless $r \geq 20$--30.
\end{itemize}
Within each regime, varying $F$ by $\pm 2$ around $F_{\mathrm{true}}$ does not change the outcome: when a structure is in the detectable zone, the entire row $\Delta F \in \{-2,\ldots,+2\}$ is significant; when it is not, no choice of $F$ recovers the signal.
The strict criterion ($k=180$) produces no false positives for the dominant-factor dataset across all 30 structures, making it the safer choice when the goal is to avoid spurious detection.

\noindent These results sharpen the message about questionnaire length: it is not the total number of items $M$ that governs detectability, but the ratio $M/F_{\mathrm{true}}$.
A questionnaire with 50 items measuring 5 constructs ($r=10$) sits in the ambiguous regime, while one with 50 items measuring 3 constructs ($r\approx 17$) is safely detectable.
As a practical guideline, we recommend $M/F \geq 15$ for reliable detection with the strict criterion, and $M/F \geq 7$ as an absolute minimum for the loose criterion.
The number of FA components $F$ can be chosen by standard methods (parallel analysis or scree plot) without fine-tuning: any value within two units of the estimate is equivalent.

% - - - - - - - - - - - - - - - - - - - - - - - 
\subsection{F-sensitivity on real datasets}
\label{app:F-sensitivity-real}

\noindent The synthetic sweep above varies $F$ around a known $F_{\mathrm{true}}$. On real data the ``true'' number of latent constructs is unknown, so we ran a complementary check: for a panel of five real datasets (RWAS, GCBS, ACME, HEXACO, BIG5) we re-ran the full pipeline at $F \in \{F_{\mathrm{def}}-2, \ldots, F_{\mathrm{def}}+2\}$ (minimum $F=2$), where $F_{\mathrm{def}}$ is the value used in the main analysis, and tracked: (i) the consensus number of communities $K_{\mathrm{cons}}$; (ii) the inter-run ARI; (iii) the ARI between the consensus partition at $F$ and the consensus partition at $F_{\mathrm{def}}$ (label alignment on the common subjects); (iv) the four significance metrics.

\noindent Results are summarised in Table~\ref{tab:F-sens-real}. Three observations are robust across the panel:
\begin{itemize}
    \item $K_{\mathrm{cons}}$ is essentially insensitive to $F$ in the validated range: RWAS, GCBS, ACME and BIG5 retain the same $K_{\mathrm{cons}}$ at all tested $F$; HEXACO retains $K_{\mathrm{cons}}=4$ for $F\in\{4,\ldots,8\}$.
    \item The consensus partition at $F\neq F_{\mathrm{def}}$ agrees with the partition at $F_{\mathrm{def}}$ with ARI $\gtrsim 0.7$--$0.9$ for $|F-F_{\mathrm{def}}|\leq 1$, and stays above $\sim 0.8$ even at $|F-F_{\mathrm{def}}| = 2$ for the strongly-detected datasets (RWAS, GCBS, ACME). This is direct evidence that the recovered partition is not a trivial artifact of the projection dimension.
    \item Significance flags are stable: datasets that fire under the strict criterion at $F_{\mathrm{def}}$ keep firing across all tested $F$; datasets that do not fire keep not firing. The only borderline case is GCBS at $F=5$, where $K_{\mathrm{cons}}$ jumps from 3 to 4 and the modularity flag drops, while cloud entropy and overlap remain significant.
\end{itemize}

\begin{table*}[t]
\centering
\renewcommand{\arraystretch}{1}
\setlength{\tabcolsep}{6pt}
\begin{tabular}{l c c c c c c c c}
\toprule
Dataset & $F_{\mathrm{def}}$ & $F$ & $K_{\mathrm{cons}}$ & ARI$_{\mathrm{inter}}$ & ARI vs $F_{\mathrm{def}}$ & $Q$ & $\hat H_{k_{\mathrm{low}}}$ & $\hat H_{k_{\mathrm{high}}}$ \\
\midrule
RWAS   & 3 & 2 & 3 & 0.94 & 0.87 & $\bullet$ & $\bullet$ & $\bullet$ \\
       &   & 3 & 3 & 0.96 & 1.00 & $\bullet$ & $\bullet$ & $\bullet$ \\
       &   & 4 & 3 & 0.92 & 0.91 & $\bullet$ & $\bullet$ & $\bullet$ \\
       &   & 5 & 3 & 0.95 & 0.82 & $\bullet$ & $\bullet$ & $\bullet$ \\
\midrule
GCBS   & 3 & 2 & 3 & 1.00 & 0.87 & $\bullet$ & $\bullet$ & $\bullet$ \\
       &   & 3 & 3 & 1.00 & 1.00 & $\bullet$ & $\bullet$ & $\bullet$ \\
       &   & 4 & 3 & 1.00 & 0.97 & $\bullet$ & $\bullet$ & $\bullet$ \\
       &   & 5 & 4 & 0.88 & 0.83 & $\circ$   & $\bullet$ & $\bullet$ \\
\midrule
ACME   & 3 & 2 & 3 & 0.96 & 0.87 & $\bullet$ & $\bullet$ & $\circ$   \\
       &   & 3 & 3 & 0.92 & 1.00 & $\bullet$ & $\bullet$ & $\bullet$ \\
       &   & 4 & 3 & 0.90 & 0.87 & $\bullet$ & $\bullet$ & $\bullet$ \\
       &   & 5 & 3 & 0.91 & 0.78 & $\bullet$ & $\bullet$ & $\bullet$ \\
\midrule
HEXACO & 6 & 4 & 4 & 0.72 & 0.62 & $\circ$ & $\circ$ & $\circ$ \\
       &   & 5 & 4 & 0.59 & 0.69 & $\circ$ & $\bullet$ & $\bullet$ \\
       &   & 6 & 4 & 0.71 & 1.00 & $\circ$ & $\bullet$ & $\bullet$ \\
       &   & 7 & 4 & 0.68 & 0.82 & $\circ$ & $\bullet$ & $\bullet$ \\
       &   & 8 & 4 & 0.65 & 0.77 & $\circ$ & $\bullet$ & $\bullet$ \\
\midrule
BIG5   & 5 & 3 & 4 & 0.56 & 0.40 & $\circ$ & $\circ$ & $\circ$ \\
       &   & 4 & 4 & 0.71 & 0.65 & $\circ$ & $\circ$ & $\circ$ \\
       &   & 5 & 4 & 0.72 & 1.00 & $\circ$ & $\circ$ & $\circ$ \\
       &   & 6 & 4 & 0.63 & 0.89 & $\circ$ & $\circ$ & $\circ$ \\
       &   & 7 & 4 & 0.59 & 0.82 & $\circ$ & $\circ$ & $\circ$ \\
\bottomrule
\end{tabular}
\caption{Sensitivity of $K_{\mathrm{cons}}$, partition stability and significance metrics to $F$ on five real datasets. ARI vs $F_{\mathrm{def}}$ measures how much the consensus partition at $F$ shifts with respect to the partition at $F_{\mathrm{def}}$. $\bullet$: significant under extreme-value separation; $\circ$: compatible with the null. $n_{\mathrm{rep}} = 50$, $n_{\mathrm{boot}} = 30$.}
\label{tab:F-sens-real}
\end{table*}

% - - - - - - - - - - - - - - - - - - - - - - - - - - 
\subsection{Maximum number of detectable communities $K^*$}
\label{app:Kstar}

\noindent The synthetic and real F-sensitivity analyses suggest that the recovered $K_{\mathrm{cons}}$ is not driven by $F$ in a trivial way. A complementary question is: \emph{what is the maximum $K$ that the pipeline can recover at a given $(M,F)$?}\\

\noindent We address it by planting mixtures of $K_{\mathrm{true}} \in \{2,\ldots,8\}$ communities, with slightly unequal sizes ($\theta_k$ linearly decreasing from 1 to 0.3), strong signal ($\xi=0.9$), $N=800$, and sweeping $(M, F) \in \{20,50,100,200\}\times\{2,3,5,10\}$.\\
For each configuration we record $K_{\mathrm{cons}}$ and $\mathrm{ARI}$ against the planted partition.
We define $K^*(M,F)$ as the largest $K_{\mathrm{true}}$ for which at least half of the replicates satisfy
$|K_{\mathrm{cons}} - K_{\mathrm{true}}| \approx 1$ and $\mathrm{ARI}_{\mathrm{gt}} > 0.5$.

\begin{figure}[h]
    \centering
    \includegraphics[width=\linewidth]{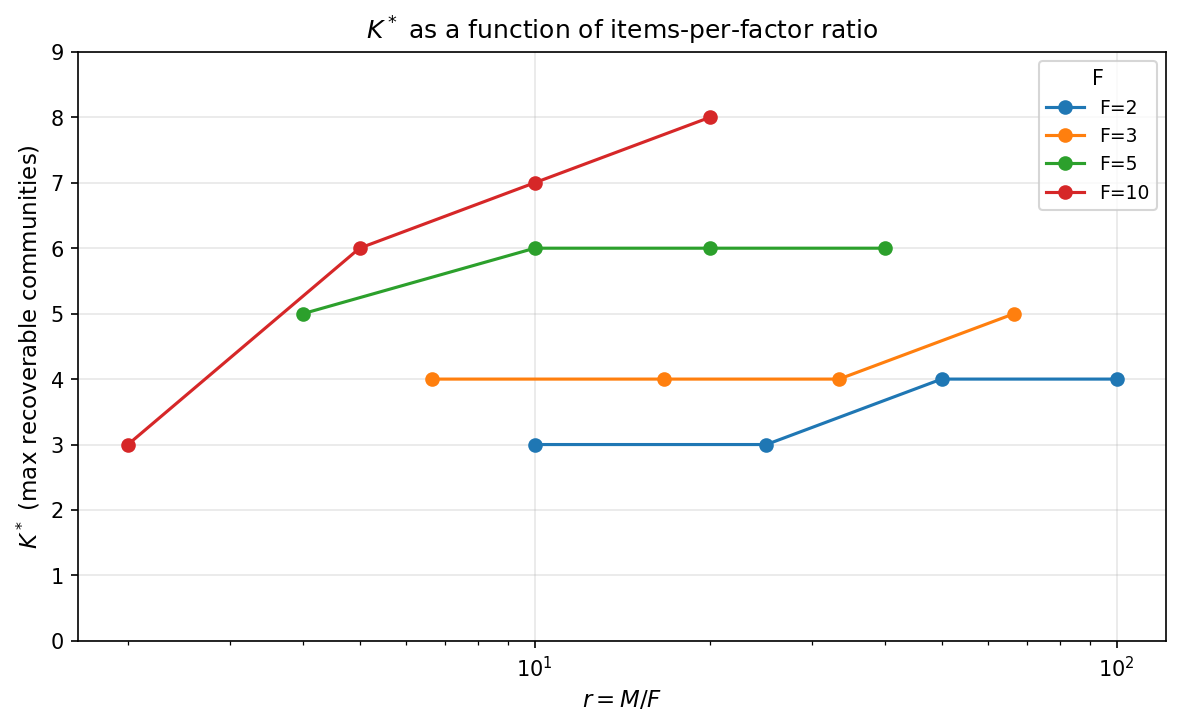}
    \caption{$K^*$ as a function of the items-per-factor ratio $r$, for $F \in \{2,3,5,10\}$. Each point aggregates two synthetic replicates per $(M, F, K_{\mathrm{true}})$ cell. The pipeline ceiling on $K$ grows with $F$ and saturates once $r$ is large enough to suppress the embedding noise.}
    \label{fig:Kstar}
\end{figure}

\noindent Results are shown in Figure~\ref{fig:Kstar} and Table~\ref{tab:Kstar}. Two clear patterns emerge: $K^*$ grows with both $F$ and $r = M/F$, and saturates at an $F$-dependent ceiling. For $F=2$ the ceiling is $K^* \approx 4$; for $F=3$, $K^*\approx 5$; for $F=5$, $K^*\approx 6$; for $F=10$ with sufficient $r$, $K^*\approx 8$. This is consistent with the geometric expectation that an $(F-1)$-dimensional embedding after market-mode removal can host at most $\sim F$ well-separated regions, modulated by the items-per-factor ratio that controls the noise floor of the embedding.

\noindent The implication for the real datasets analysed in the main text is direct: with $F \leq 5$ and $r \in [5, 12]$, the pipeline can recover at most $K \approx 4$--$5$ communities. The fact that single-construct scales such as RWAS, GCBS and PWE consistently recover $K=3$ is therefore not a saturation artifact: their detectable ceiling is around $K^*=4$, so $K_{\mathrm{cons}}=3$ reflects the structure in the data rather than the pipeline limit.

\begin{table*}[t]
\centering
\renewcommand{\arraystretch}{1.1}
\setlength{\tabcolsep}{8pt}
\begin{tabular}{c c c c c c}
\toprule
$F \backslash M$ & 20 & 50 & 100 & 200 \\
\midrule
2 & $K^*=3$ ($r=10$)  & 3 ($r=25$)  & 4 ($r=50$)  & 4 ($r=100$) \\
3 & 4 ($r=6.7$)        & 4 ($r=16.7$) & 4 ($r=33.3$) & 5 ($r=66.7$) \\
5 & 5 ($r=4$)          & 6 ($r=10$)   & 6 ($r=20$)   & 6 ($r=40$) \\
10 & 3 ($r=2$)         & 6 ($r=5$)    & 7 ($r=10$)   & 8 ($r=20$) \\
\bottomrule
\end{tabular}
\caption{Empirical maximum recoverable number of communities $K^*$ as a function of $(M, F)$, with $r = M/F$ in parentheses. Strong-signal mixtures with $\xi=0.9$, $N=800$. The ceiling on $K^*$ grows with $F$; the floor (small $r$) grows with $M/F$.}
\label{tab:Kstar}
\end{table*}

%=============================================================================
% APPENDIX: Market mode removal vs.\ Newman--Girvan null model
%=============================================================================
\section{Why market mode removal instead of Newman--Girvan}
\label{app:newman-failure}
\begin{figure*}[t]
  \centering
  \includegraphics[width=\linewidth]{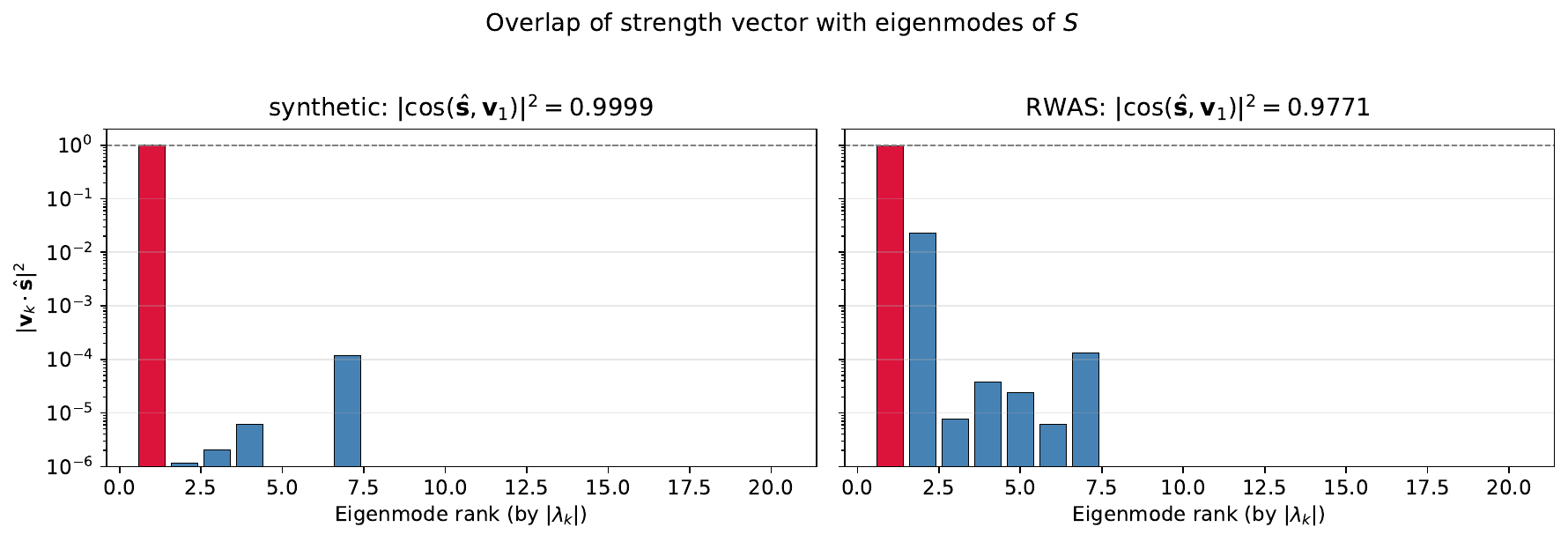}
  \caption{Squared projections $c_k^2 = (\mathbf{v}_k \cdot \hat{\mathbf{s}})^2$ of the normalised strength vector on the eigenmodes of $S$ (log scale). \\
  \emph{(a)} Synthetic data: essentially all weight lies on $\mathbf{v}_1$ ($c_1^2 = 0.9999$), so the Newman--Girvan and market-mode null models coincide. \\
  \emph{(b)} RWAS: $c_1^2 = 0.977$ with residual mass concentrated on $\mathbf{v}_2$ ($c_2^2 \simeq 0.022$). The misalignment causes the Newman--Girvan model to couple the market mode with the first informative eigenvector, contaminating the modularity landscape.}
  \label{fig:overlap}
\end{figure*}

\noindent Both null models subtract a rank-1 matrix from the similarity~$S$:
\begin{equation}
  J^{\mathrm{MM}} = S - \lambda_1\,\mathbf{v}_1\mathbf{v}_1^{\!\top}\,,
  \qquad
  J^{\mathrm{NG}} = S - \frac{\mathbf{s}\,\mathbf{s}^{\!\top}}{2m}
                   = S - A\,\hat{\mathbf{s}}\hat{\mathbf{s}}^{\!\top}\,,
  \label{eq:two_J}
\end{equation}
with $\mathbf{s} = S\mathbf{1}$, $2m = \mathbf{1}^{\!\top} S\,\mathbf{1}$ and $A = \|\mathbf{s}\|^2/(2m)$.  Expanding $\hat{\mathbf{s}}$ in the eigenbasis of~$S$,
\begin{equation}
  \hat{\mathbf{s}} \;=\; \sum_k c_k\,\mathbf{v}_k\,,
  \qquad
  c_k = \mathbf{v}_k\cdot\hat{\mathbf{s}}\,,
  \qquad
  \sum_k c_k^2 = 1\,,
  \label{eq:s_expansion}
\end{equation}
makes the difference between the two subtractions explicit.  MM is an orthogonal rank-1 deflation along~$\mathbf{v}_1$; NG is rank-1 along $\hat{\mathbf{s}}$, and is equivalent to MM only in the limit $c_1^2 = 1$. \\
\noindent On synthetic data (factor/mixture) with homogeneous subject norms, the strength vector is essentially collinear with $\mathbf{v}_1$, and $J^{\mathrm{NG}} \simeq J^{\mathrm{MM}}$ up to numerical noise (Fig.~\ref{fig:overlap}a). \\
\noindent On real Likert data this no longer holds. For example, on the RWAS dataset the residual mass is almost entirely concentrated on~$\mathbf{v}_2$ (Fig.~\ref{fig:overlap}b).
The asymmetry is driven by subject-level heterogeneity in $\lVert\mathbf{x}_i\rVert^2$: Likert-specific effects (floor/ceiling, indecisiveness and extreme response styles) make the norm distribution across subjects inhomogeneous, so~$\mathbf{s}$ acquires a component along the second mode.
Restricted to the two-dimensional subspace
$\mathrm{span}\{\mathbf{v}_1,\mathbf{v}_2\}$, $J^{\mathrm{NG}}$ reads
\begin{equation}
  J^{\mathrm{NG}}\big|_{2\times 2}
  \;=\;
  \begin{pmatrix}
    \lambda_1 - A\,c_1^2 & -A\,c_1 c_2 \\
    -A\,c_1 c_2 & \lambda_2 - A\,c_2^2
  \end{pmatrix}\,.
  \label{eq:2x2}
\end{equation}
The off-diagonal entry $-A c_1 c_2$ couples the market mode with the first informative mode and is not removed by either the scale factor~$A$ or the angle $c_2$ alone.  With the measured RWAS values the off-diagonal term is comparable in magnitude to $\lambda_2$ itself, so its diagonalisation produces a rotated top eigenmode:
\begin{equation}
  \mathbf{u}_{\mathrm{top}}^{\mathrm{NG}}
  \;=\;
  \alpha\,\mathbf{v}_1 + \beta\,\mathbf{v}_2\,,
  \label{eq:rotation}
\end{equation}
with eigenvalue $\lambda_{\mathrm{top}}(J^{\mathrm{NG}})$, inflated with respect to $\lambda_2(J^{\mathrm{MM}})$.
Empirically, the residual of $\mathbf{u}_{\mathrm{top}}^{\mathrm{NG}}$ outside the $(\mathbf{v}_1,\mathbf{v}_2)$-plane has norm close to zero, confirming that the interaction is confined to this two-dimensional subspace. \\
\noindent The consequence for modularity maximisation is direct: the dominant mode of $J^{\mathrm{NG}}$ is a mixture of the market mode and the first informative mode.
Any partition recovered from $J^{\mathrm{NG}}$ is therefore a mixture of the genuine community signal and a trivial ellipsoid bisection.  $J^{\mathrm{MM}}$, by removing exactly $\lambda_1\,\mathbf{v}_1\mathbf{v}_1^{\!\top}$, leaves $\mathbf{v}_2$ untouched by construction and recovers more informative partitions.
 
\begin{figure*}[t]
  \centering
  \includegraphics[width=0.9\linewidth]{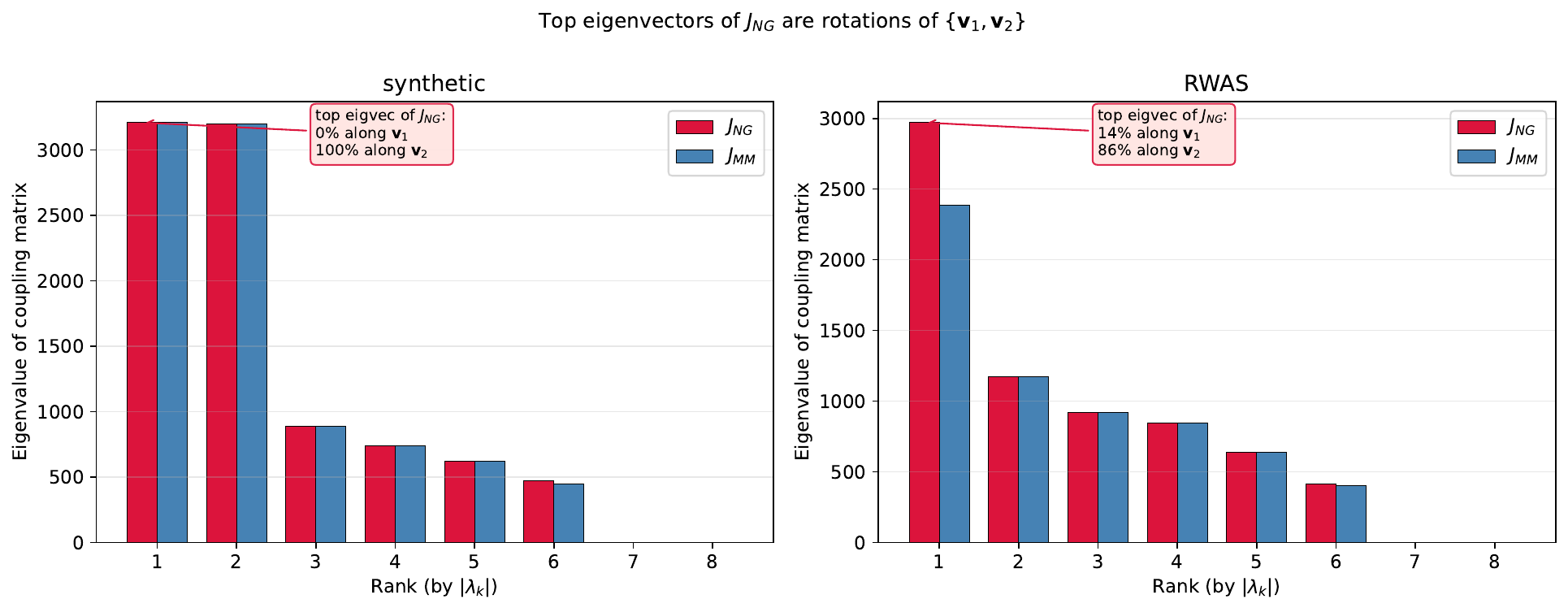}
  \caption{Top eigenvalues of $J^{\mathrm{NG}}$ (red) and $J^{\mathrm{MM}}$ (blue) for synthetic data (left) and RWAS (right). \\
  On synthetic data the leading eigenvector of $J^{\mathrm{NG}}$ is $100\%$ aligned with $\mathbf{v}_2$ of $S$; on RWAS it is a mixture of $\mathbf{v}_1$ and $\mathbf{v}_2$, with its eigenvalue inflated relative to $\lambda_2(J^{\mathrm{MM}})$.
  This inflation pushes a mixed, partially uninformative mode to the top of the $J^{\mathrm{NG}}$ spectrum, whereas $J^{\mathrm{MM}}$ leaves $\mathbf{v}_2$ intact by construction.}
  \label{fig:spectra}
\end{figure*}

%==========================================================
% APPENDIX: market mode vs MP distributions
%==========================================================
\section{Appendix: Spectral structure of the similarity matrix and limitations of the Marchenko--Pastur bulk criterion}
\label{app:spectral-nullmodel}

\noindent Our setting is closely related to the one tackled by \cite{garlaschelli-macmahon15} for correlation matrices used as weighted adjacency matrices.
In their approach the modularity matrix is cleaned, in addition to the market mode, of all eigencomponents whose eigenvalues are compatible with the Marchenko--Pastur (MP) bulk.

\noindent For $X_c$ with independent, zero-mean rows of covariance $\sigma^2 I_M$ (i.e., no item correlations), the MP law gives the upper edge of the bulk spectrum of $X_c X_c^\top$:
\begin{equation}
  \lambda_+^{\mathrm{MP}}
  = 2 N \hat{\sigma}^2 \left(1 + \sqrt{\frac{M}{N}}\right)^2,
  \label{eq:lambda_plus}
\end{equation}
where $\hat{\sigma}^2$ is the empirical item variance.
Eigenvalues of $S_{\mathrm{clean}}$ that exceed $\lambda_+^{\mathrm{MP}}$ are interpreted as carrying structured signal; under the standard random-matrix argument, $K$ communities produce $K - 1$ such eigenvalues \cite[cf.][]{fortunato2010-os,newman2006-dt}.

\begin{figure*}[t]
  \centering
  \includegraphics[width=0.9\linewidth]{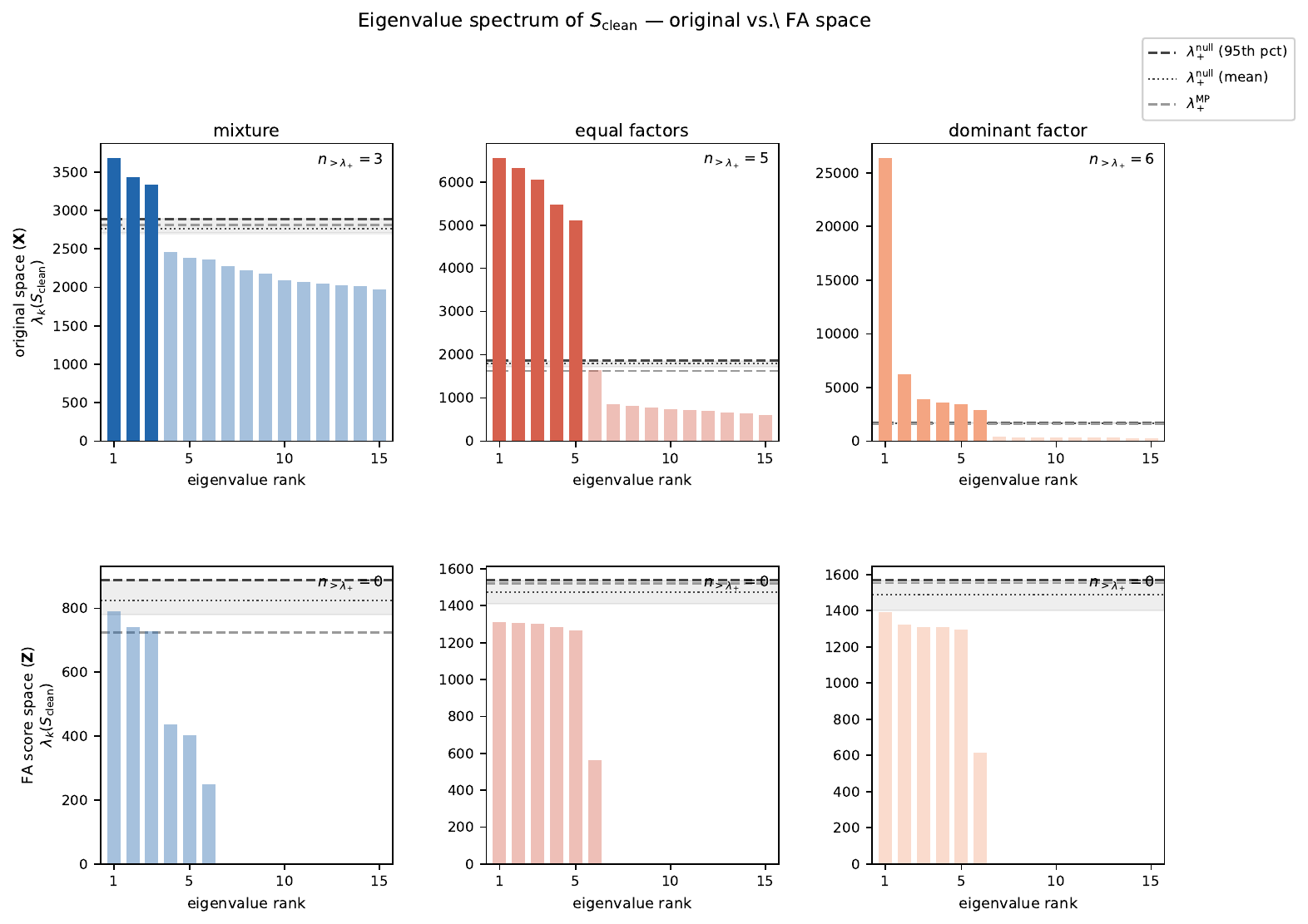}
  \caption{Eigenvalue spectrum of $S_{\mathrm{clean}}$ (market mode removed) for three synthetic datasets: a mixture model with $K=4$ communities, a homogeneous equal-factor model, and a homogeneous dominant-factor model ($N=700$, $M=50$, $F=5$).
  \textbf{Top row:} original item space $\mathbf{X}$.
  \textbf{Bottom row:} FA score space $\mathbf{Z}$ (similarity computed on factor scores, null obtained by independently reshuffling each factor column).
  Bars show the top ranked eigenvalues $\lambda_k(S_{\mathrm{clean}})$ in descending order; darker bars mark eigenvalues above the empirical bulk edge $\lambda_+^{\mathrm{null}}$ (dashed line, 95th percentile of the maximum eigenvalue across $n_{\mathrm{rep}}=30$ subject-reshuffled realisations).
  The grey dashed line shows the analytical estimate
  $\lambda_+^{\mathrm{MP}} = 2N\hat\sigma^2(1+\sqrt{d/N})^2$ (Eq.~\ref{eq:lambda_plus}, with $d=M$ or $d=F$ respectively).
  Top row: for both factor-only datasets 6 eigenvalues exceed the threshold (false positives), while the mixture dataset yields $K-1=3$ above-threshold eigenvalues only for strong signal.
  Bottom row: all datasets yield zero eigenvalues above the threshold.
  For the factor-only datasets this is the correct result; for the mixture dataset it is a false negative caused by the null model being inflated by the between-community variance absorbed into the FA scores.}
  \label{fig:spectrum_mp}
\end{figure*}

\noindent In practice, $\lambda_+^{\mathrm{MP}}$ can also be estimated empirically as the 95th percentile of the maximum eigenvalue of $S_{\mathrm{clean}}$ computed on $n_{\mathrm{rep}}$ subject-reshuffled realisations of $X$.
As shown in Fig.~\ref{fig:spectrum_mp} (top row), the analytical expression \eqref{eq:lambda_plus} and the empirical estimate agree closely.

\noindent However, for the MP distribution to be a good descriptor of the noise spectrum, the variables over which the correlations are computed need to be i.i.d.
This assumption fails whenever the data have a strong \emph{homogeneous} factor structure.
Consider a dataset generated by $F$ latent factors with no community structure.
Factor $f$ induces a rank-1 contribution to the inter-item covariance; after projection onto subject space, each factor contributes one eigenvalue of order $N$ to $S_{\mathrm{clean}}$.
For $F$ factors, this produces $F$ eigenvalues above the noise floor regardless of whether any community structure is present, and the dominant-factor variant produces an additional isolated spike on top of these, reflecting its hierarchical (one general factor on top of group-specific factors) loading pattern.
Concretely, for a dataset with $F=5$ factors ($N=700$, $M=50$) we observe up to 6 eigenvalues above $\lambda_+^{\mathrm{null}}$ in the dominant-factor case, while a four-community mixture dataset with the same parameters yields the expected $K-1=3$ (Fig.~\ref{fig:spectrum_mp}, top row).\\
The fundamental issue is that this procedure retains all directions of variance in $X_c$, not only those associated with block structure. 
Hence, removing only the market mode is enough to clean the signal of the general answer tendency, while it is not reliable a criterion that aims to isolate community-induced variance from factor-induced variance, since both appear as spikes above the bulk. \\
This motivates working in a lower-dimensional representation (e.g., factor-analysis scores) where the factor directions are explicitly projected out before similarity and community-detection are computed, as described in the main text.\\
The bottom row of Fig.~\ref{fig:spectrum_mp} shows that this resolves the false-positive problem: for both factor-only datasets the eigenvalue count above $\lambda_+^{\mathrm{null}}$ drops to zero.
However, a complementary failure mode appears for the mixture dataset, as all eigenvalues are now compatible with the reshuffled null model. 
The analytical value of $\lambda_+$ is less inflated than its empirical estimate, yet it remains too high to capture the significant eigenvalues; in practice this manifests as cases where the cluster separation is not sharp enough for the criterion to detect more than one eigenvalue above the threshold, which would imply only a binary split along the main axis of variation. \\
The result is that neither space yields a reliable spectral threshold for detecting communities, which motivates a milder cleaning of the matrix --- removing only the market mode --- that may induce false positives when the data exhibit a strong factor structure, but whose significance can be assessed post hoc with the approach described in the main text.

%==========================================================
% APPENDIX: eigenvector entropy 
%==========================================================
\section{The entropy of the first eigenvectors is not enough to characterize clustered data}
\label{app:eigenvector-entropy}

\begin{figure*}[t]
    \centering
    \begin{subfigure}[c]{0.8\linewidth}
        \includegraphics[width=\linewidth]{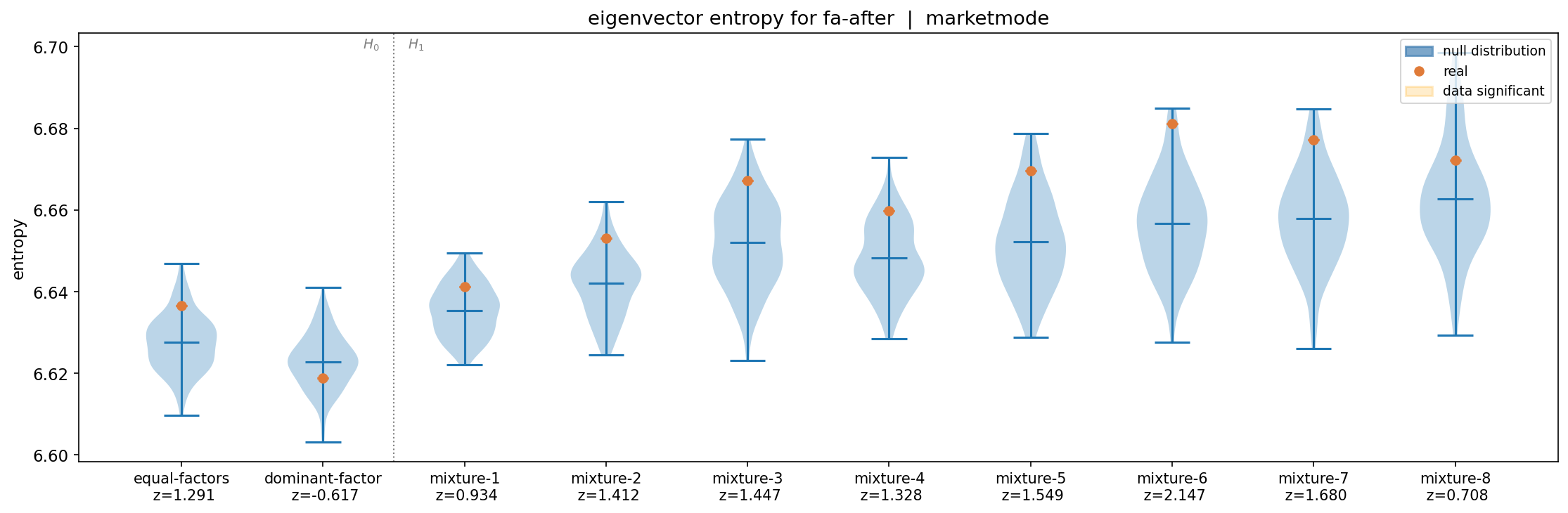}
    \end{subfigure}
    \begin{subfigure}[c]{0.8\linewidth}
        \includegraphics[width=\linewidth]{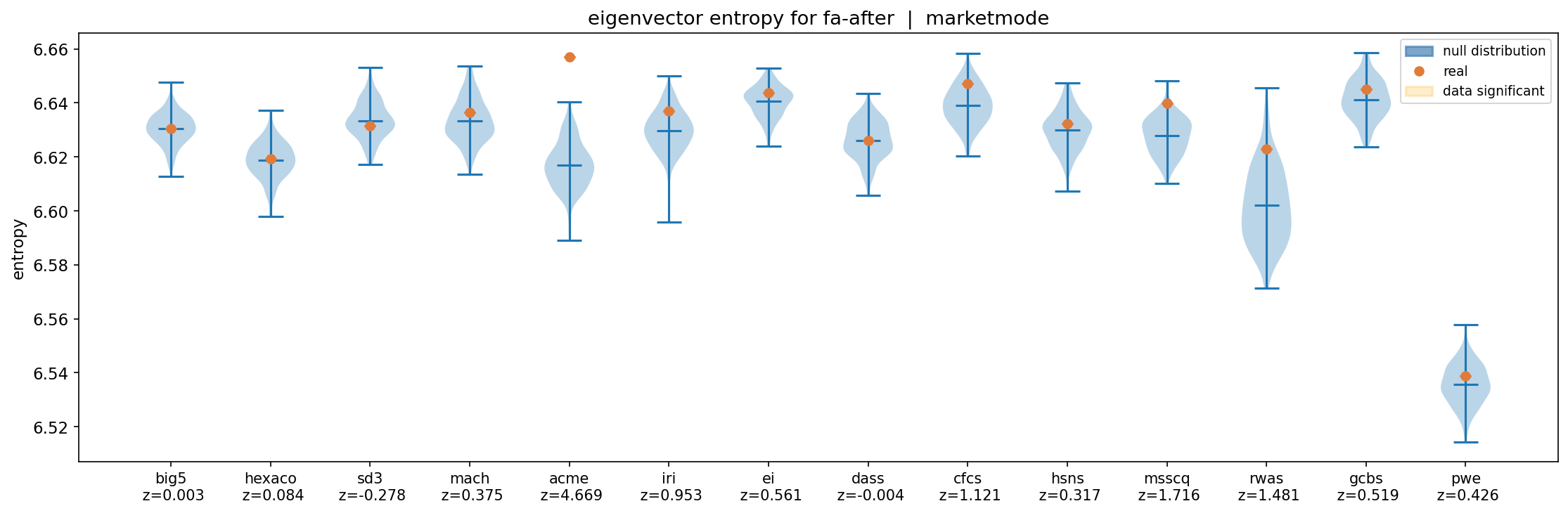}
    \end{subfigure}
    \caption{Average entropy of the first $K-1$ eigenvectors of the similarity matrix for each dataset (orange dot) vs.\ the resampling null distribution (blue violin plot), where $K$ is the number of communities detected in the original dataset. The top (bottom) panel shows the synthetic (real) datasets.
    The empirical values consistently fall within the bulk of the resampling null across both $H_0$ and mixture datasets, showing that the per-component entropy of the individual eigenvectors is not a reliable discriminator between $H_0$ and $H_1$ scenarios.}
    \label{fig:entropy-eigvec}
\end{figure*}

\noindent A natural alternative to the cloud entropy introduced in Section~\ref{sec:significance} would be to look at the entropy of the individual eigenvectors of the similarity matrix, computed component-by-component as a one-dimensional density estimate.
The intuition is that for clustered data the leading $K-1$ eigenvectors should exhibit a discretised, multimodal component distribution (one mode per community), so that their per-component entropy should be lower than under the resampling null.
We tested this idea on both synthetic and real datasets by computing the average entropy of the first $K-1$ eigenvectors of $S^{\mathrm{clean}}$ and comparing it to the entropy obtained on column-wise resampled versions of the data.
Figure~\ref{fig:entropy-eigvec} shows that the empirical values systematically fall well within the resampling-null distribution, both for $H_0$ datasets that are known to be unstructured and for mixture or real datasets where the cloud entropy in the joint eigenvector space (Sec.~\ref{sec:significance}) does instead detect significant structure.
This indicates that the marginal entropy of single eigenvectors is not informative about the presence of a clustered population: it is dominated by the smooth, unimodal envelope of each eigenvector's components and is essentially blind to the multi-dimensional concentration of points that signals genuine communities.
For this reason we do not include this observable in our significance pipeline and rely instead on the joint cloud entropy in the $(K-1)$-dimensional eigenvector embedding, which jointly exploits the geometry across the informative eigenvectors and is sensitive to cluster concentration.

%==========================================================
% APPENDIX: sensitivity to K-NN in cloud entropy
%==========================================================
\section{Sensitivity to the $k$-NN neighbourhood size in the cloud entropy estimator}
\label{app:knn-sensitivity}

\noindent The Kozachenko--Leonenko estimator (Eq.~\eqref{eq:KL_entropy}) requires specifying the number of nearest neighbours $k$.
This parameter controls the spatial scale over which the local density is estimated:
small $k$ resolves fine-grained local geometry but yields high variance, while large $k$ averages over broader neighbourhoods and reduces variance at the cost of smoothing out subtler cluster separations.
For community detection this translates into a direct sensitivity trade-off:
a small $k$ produces a liberal test sensitive to weaker separations, while a large $k$ yields a conservative test that flags only strongly concentrated clouds.\\

\noindent We choose two operating points for $k$ that bracket the informative regime. The geometric upper limit derived from the crossing-point analysis (see below) is at $\sim 0.9\,\overline{n}_{\mathrm{comm}}$; in the main text we adopt a slightly lower value (0.8) to keep the operating point safely below the transition for datasets with unbalanced community sizes:
\begin{equation}
    k_{\mathrm{low}} = \bigl\lfloor 0.1\,N \bigr\rfloor,
    \qquad
    k_{\mathrm{high}} = \bigl\lfloor c\,\overline{n}_{\mathrm{comm}} \bigr\rfloor, \quad c\in\{0.8, 0.9\},
    \label{eq:k-rules}
\end{equation}
where $\overline{n}_{\mathrm{comm}}$ is the size-weighted average community size across bootstrap replicates,
\begin{equation}
    \overline{n}_{\mathrm{comm}} \;=\; \frac{1}{n_{\mathrm{boot}}} \sum_{b=1}^{n_{\mathrm{boot}}} \sum_{k} \frac{n_{k,b}^2}{N},
    \label{eq:avg-comm}
\end{equation}
with $n_{k,b}$ the size of the $k$-th community at bootstrap $b$.
The interpretation is geometric. The loose criterion $k_{\mathrm{low}} = 0.1\,N$ probes a neighbourhood substantially smaller than any plausible community, so the local density estimate is sensitive to any fine-grained anisotropy in the embedding; this is the most permissive test and detects sub-threshold structure. The strict criterion $k_{\mathrm{high}}$ probes a neighbourhood comparable to the typical community: when $k$ matches the community size the density estimate is dominated by within-community geometry; pushing $k$ above $\overline{n}_{\mathrm{comm}}$ forces the estimator to average across communities and the signal disappears. The geometric limit is at $c\approx 0.9$, identified empirically by the crossing-point analysis below; in the main text we use the more conservative $c=0.8$ so that the operating point remains in the significant regime also for datasets with unbalanced community sizes, where the smallest community can be much smaller than $\overline{n}_{\mathrm{comm}}$ and the transition consequently shifts to lower $k$. The weighting in Eq.~\eqref{eq:avg-comm} gives larger communities a higher weight, in line with their larger contribution to the local-density estimate.\\

\begin{figure*}[t]
    \centering
    \includegraphics[width=\linewidth]{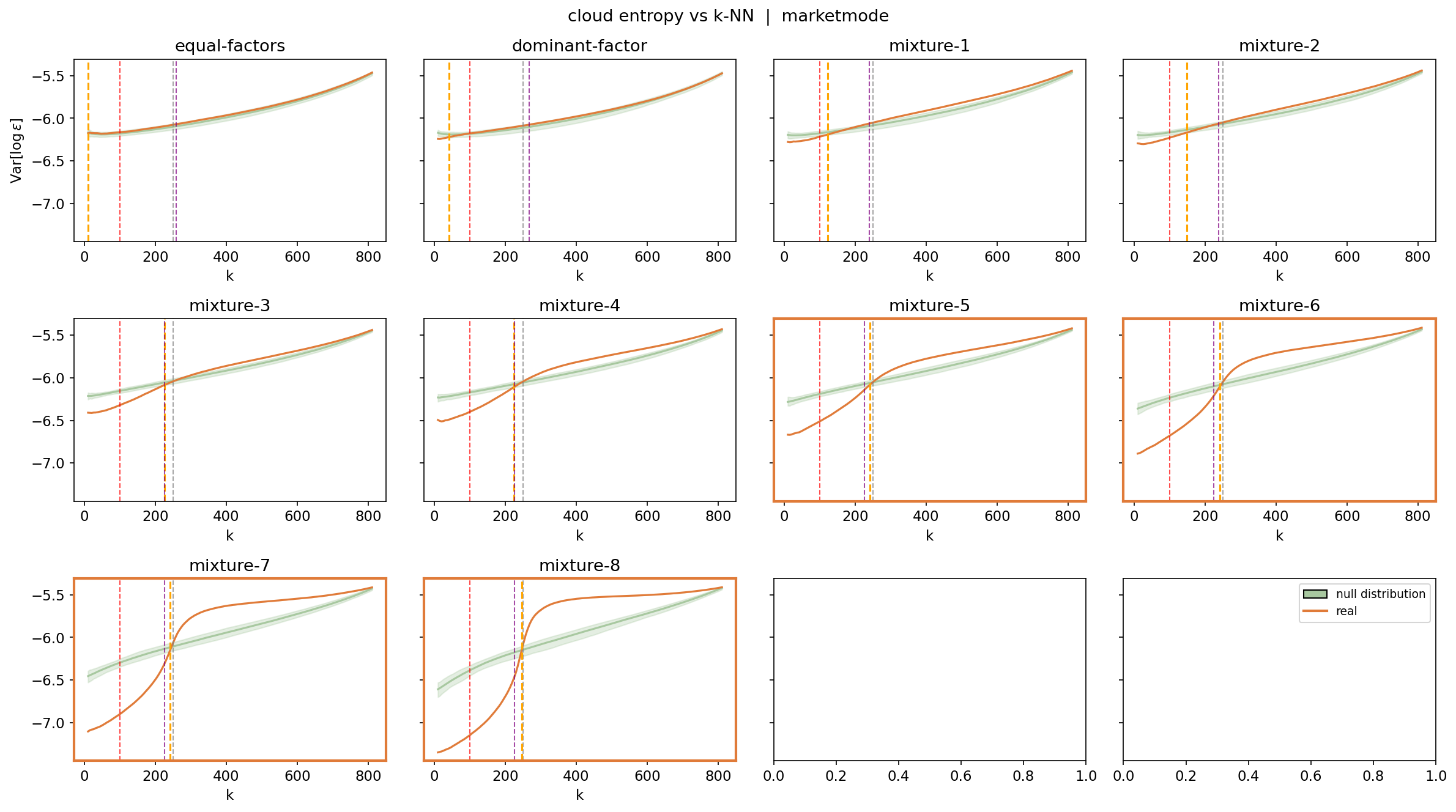}
    \caption{Cloud entropy of the synthetic datasets ($N=1000$) as a function of the $k$-NN neighbourhood size $k$, ranging from $k=10$ to $k=800$.
    The green band marks the cloud entropy under the column-wise resampling null (mean $\pm$ standard deviation across $n_{\mathrm{rep}}$ replicates); the empirical curve is shown in solid line.
    The vertical orange dashed line marks the empirical crossing point $k_{\mathrm{cross}}$, i.e.\ the smallest $k$ at which the empirical entropy enters the null envelope.
    For reference, the grey dashed line marks $k = N/K_{\mathrm{cons}}$ (uniform-community heuristic) and the purple dashed line marks $\overline{n}_{\mathrm{comm}}$, the size-weighted average community size (Eq.~\eqref{eq:avg-comm}); the two coincide when communities are equipopulated.
    This figure visualises the geometric meaning of the operating points $k_{\mathrm{low}}$ and $k_{\mathrm{high}}$ defined in Eq.~\eqref{eq:k-rules}: $k_{\mathrm{low}} = 0.1\,N$ sits deep in the significant regime, while the geometric upper bound $k_{\mathrm{high}} = 0.9\,\overline{n}_{\mathrm{comm}}$ tracks $k_{\mathrm{cross}}$ from below.}
    \label{fig:crossing-points}
\end{figure*}

\paragraph{Visualising the crossing point.}
Before validating the rules quantitatively, it is useful to look at how the significance of the cloud entropy depends on $k$ on a single dataset.
Figure~\ref{fig:crossing-points} reports the cloud entropy as a function of $k$ for the synthetic dataset series ($N=1000$), together with the permutation-null envelope.
For small $k$ the empirical entropy sits well below the null and the metric is significant; as $k$ grows the neighbourhood size becomes comparable to the typical community size and the empirical curve gradually meets the null band.
We define the \emph{crossing point} $k_{\mathrm{cross}}$ as the smallest $k$ at which the empirical curve enters the null envelope (equivalently, the smallest $k$ at which the permutation $p$-value of the cloud entropy exceeds $\alpha=0.05$, with $n_{\mathrm{rep}}=50$).
The two geometric scales used to define the operating points in Eq.~\eqref{eq:k-rules}---$N/K_{\mathrm{cons}}$ and the size-weighted average community size $\overline{n}_{\mathrm{comm}}$---are also shown for reference: they coincide when communities are equipopulated and bracket $k_{\mathrm{cross}}$ across all detected mixture datasets, providing a visual rationale for the geometric upper limit $k_{\mathrm{high}}\approx 0.9\,\overline{n}_{\mathrm{comm}}$. The main-text operating point $0.8\,\overline{n}_{\mathrm{comm}}$ trades a slightly less aggressive scale for robustness on unbalanced partitions.

\paragraph{Validation across $N$.}
To assess whether the geometric rules of Eq.~\eqref{eq:k-rules} track $k_{\mathrm{cross}}$ consistently as the sample size varies, we tested the two rules on synthetic data at $N \in \{400, 800, 1000, 1500, 2000\}$, generating one $H_0$ (dominant-factor, $F=4$, $M=50$) and one $H_1$ (mixture, $K_{\mathrm{true}}=4$, signal $\xi=0.6$) dataset per $N$. For each dataset we computed the cloud entropy at 25 log-spaced values of $k$ in $[5, N/2]$ and extracted $k_{\mathrm{cross}}$ as illustrated in Fig.~\ref{fig:crossing-points}. Table~\ref{tab:k-rules-validation} compares $k_{\mathrm{cross}}$ on the mixture dataset with the geometric upper bound $0.9\,\overline{n}_{\mathrm{comm}}$, and reports the ratio $k_{\mathrm{low}} / k_{\mathrm{cross}}$ to verify that $k_{\mathrm{low}}$ sits well within the significant regime.

\begin{table*}[!h]
\centering
\renewcommand{\arraystretch}{1}
\setlength{\tabcolsep}{5pt}
\begin{tabular}{c c c c c c c}
\toprule
$N$ & $K_{\mathrm{cons}}$ & $\overline{n}_{\mathrm{comm}}$ & $k_{\mathrm{cross}}$ & $k_{\mathrm{low}}$ & $k_{\mathrm{high}}$ & $k_{\mathrm{high}}/k_{\mathrm{cross}}$ \\
\midrule
 400 & 4 & 186 &  68 &  40 & 167 & 2.46 \\
 800 & 4 & 375 & 333 &  80 & 338 & 1.01 \\
1000 & 4 & 437 & 281 & 100 & 394 & 1.40 \\
1500 & 4 & 668 & 494 & 150 & 601 & 1.22 \\
2000 & 4 & 866 & 643 & 200 & 779 & 1.21 \\
\bottomrule
\end{tabular}
\caption{Empirical validation of $k_{\mathrm{low}} = 0.1\,N$ and $k_{\mathrm{high}} = 0.9\,\overline{n}_{\mathrm{comm}}$ on a mixture dataset ($\xi=0.6$, $K_{\mathrm{true}}=4$). $k_{\mathrm{cross}}$ is the observed crossing point (smallest $k$ at which the permutation $p$-value exceeds $0.05$). The ratio $k_{\mathrm{high}}/k_{\mathrm{cross}}$ is close to unity for $N\geq 800$, with $k_{\mathrm{high}}$ overshooting $k_{\mathrm{cross}}$ by 20--40\,\% on average; replacing the 0.9 factor with $\sim 0.75$ would yield a tighter fit but the present choice keeps the operating point geometrically interpretable as ``just inside the typical community''. The ratio $k_{\mathrm{low}}/k_{\mathrm{cross}}$ is always well below 1, confirming that $k_{\mathrm{low}}$ stays within the significant regime at all $N$.}
\label{tab:k-rules-validation}
\end{table*}

\paragraph{Caveat on $n_{\mathrm{rep}}$.}
Both the null mean and the null minimum/maximum stabilise only once $n_{\mathrm{rep}}$ is large enough; for $n_{\mathrm{rep}} \lesssim 50-80$ the estimate of the null tails is noisy and the apparent significance fluctuates from run to run. We recommend $n_{\mathrm{rep}} \geq 100$ for any application of the strict criterion, and we adopt $n_{\mathrm{rep}} = 150$ throughout the main results. Within this regime, the criteria in Eq.~\eqref{eq:k-rules} are insensitive to the choice of $n_{\mathrm{rep}}$ because they probe geometric, not statistical, scales.\\

\begin{figure*}[h]
    \centering
    \includegraphics[width=0.6\linewidth]{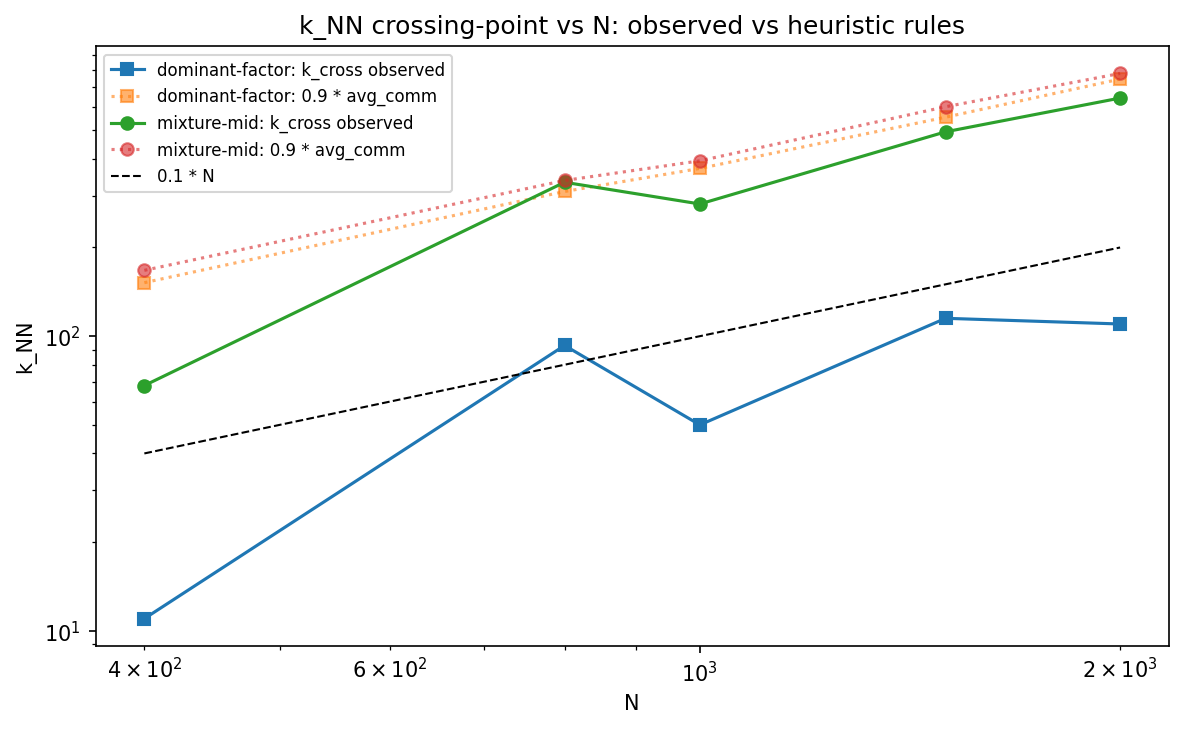}
    \caption{Observed crossing-point $k_{\mathrm{cross}}$ (solid markers) of the cloud entropy significance vs $N$, compared with the heuristic rules $k_{\mathrm{low}} = 0.1\,N$ (dashed) and $k_{\mathrm{high}} = 0.9\,\overline{n}_{\mathrm{comm}}$ (dotted, per dataset). The squares mark the $H_0$ dominant-factor dataset, the circles mark the $H_1$ mixture dataset ($\xi=0.6$). On $H_1$ the heuristic $k_{\mathrm{high}}$ tracks $k_{\mathrm{cross}}$ within a factor 1.0--1.4 for $N\geq 800$; on $H_0$ the crossing is essentially noise (the metric is never robustly significant).}
    \label{fig:knn-summary}
\end{figure*}

\end{document}